\documentclass{tcibook}
\usepackage{fancyhea}
\usepackage{work}
\usepackage{bm}       
\usepackage{graphicx}
\usepackage{hyperref}      
\usepackage{multirow}

\newcommand{\nc}{\newcommand}  

\def\beq{\begin{equation}}
\def\eeq#1{\label{#1}\end{equation}}
\def\eeqn{\end{equation}}

\newenvironment{Eqnarray}%
   {\arraycolsep 0.14em\begin{eqnarray}}{\end{eqnarray}}
\def\beqa{\begin{Eqnarray}}
\def\eeqa#1{\label{#1}\end{Eqnarray}}
\def\eeqan{\end{Eqnarray}}

\nc{\ra}{\rightarrow}  
\nc{\slsh}{\slash\hspace*{-0.22cm}}
\def\Re{{\cal R \mskip-4mu \lower.1ex \hbox{\it e}\,}}
\def\Im{{\cal I \mskip-5mu \lower.1ex \hbox{\it m}\,}}

\nc{\vev}[1]{ \left\langle {#1} \right\rangle }
\nc{\bra}[1]{ \langle {#1} | }
\nc{\ket}[1]{ | {#1} \rangle }
\nc{\fb}{\,{\rm fb}^{-1}}
\nc{\ev}{{\rm eV}}
\nc{\kev}{{\rm keV}}
\nc{\Mev}{{\rm MeV}}
\nc{\gev}{{\rm GeV}}
\nc{\tev}{{\rm TeV}}
\nc{\mev}{{\rm MeV}}

\def\D{{\cal D}}

\def\O{{\cal O}}

\def\del{\partial}
\def\Dslash{\not{\hbox{\kern-4pt $D$}}}
\def\dslash{\not{\hbox{\kern-2pt $\del$}}}
\def\pslash{\not{\hbox{\kern-2pt $p$}}}
\def\ETmiss{ \not{\hbox{\kern-4pt $E$}}_T }

\def\msb{{\bar{\ssstyle M \kern -1pt S}}}

\def\babar{\mbox{\sl B\hspace{-0.4em} {\small\sl A}\hspace{-0.37em} \sl B\hspace{-0.4em} {\small\sl A\hspace{-0.02em}R}}}

\begin{document}

\def\bibname{References}
\bibliographystyle{plain}

\raggedbottom

\pagenumbering{roman}

\parindent=0pt
\parskip=8pt
\setlength{\evensidemargin}{0pt}
\setlength{\oddsidemargin}{0pt}
\setlength{\marginparsep}{0.0in}
\setlength{\marginparwidth}{0.0in}
\marginparpush=0pt

\pagenumbering{arabic}

\renewcommand{\chapname}{chap:intro_}
\renewcommand{\chapterdir}{.}
\renewcommand{\arraystretch}{1.25}
\addtolength{\arraycolsep}{-3pt}


\catcode`\@=11

\renewcommand\contentsname{Table of Contents}
\renewcommand\thesection {\@arabic\c@section}
\renewcommand\thesubsection {\thesection.\@arabic\c@subsection}
\renewcommand\thesubsubsection{\thesubsection.\@arabic\c@subsubsection}
\renewcommand\theparagraph{\thesubsubsection.\@arabic\c@paragraph}
\renewcommand\thesubparagraph {\theparagraph.\@arabic\c@subparagraph}

\renewcommand\tableofcontents{%
    \if@twocolumn
      \@restonecoltrue\onecolumn
    \else
      \@restonecolfalse
    \fi
    \section*{\contentsname
        \@mkboth{%
           \MakeUppercase\contentsname}{\MakeUppercase\contentsname}}%
    \@starttoc{toc}%
    \if@restonecol\twocolumn\fi
    }

\renewcommand \thetable
     {\@arabic\c@table}
\renewcommand \thefigure
     {\@arabic\c@figure}

\renewcommand{\sectionmark}[1]{ \markright{#1}{} }

\catcode`@=12

\newcommand{\Bbar}{\,\overline{\!B}{}}
\newcommand{\Dbar}{\,\overline{\!D}{}}
\newcommand{\Kbar}{\,\overline{\!K}{}}
\def\B0bar{\Bbar{}^0}
\def\D0bar{\Dbar{}^0}
\def\K0bar{\Kbar{}^0}
\def\rhobar{\bar\rho}
\def\etabar{\bar\eta}
\newcommand{\nn}{\nonumber}

\def\lsim{\mathrel{\mathpalette\vereq<\!\!}}
\def\gsim{\mathrel{\mathpalette\vereq>\!\!}}
\def\vereq#1#2{\lower3pt\vbox{\baselineskip.5pt\lineskip.5pt
    \ialign{\\$#1\hfill##\hfil\\$\crcr#2\crcr\sim\crcr}}}

\def\ProjectX{\mbox{\sl P\hspace{-0.35em} r\hspace{-0.35em} o\hspace{-0.35em} j\hspace{-0.35em} e\hspace{-0.35em} c\hspace{-0.35em} t\hspace{-0.25em}   X}}
\def\ProjectX{Project~X}
\def\ProjectX{{\slshape Project~X}} 

\newcommand{\Kplus}{\ensuremath{K^+\rightarrow \pi^+ \nu \overline{\nu}}}
\newcommand{\Kzero}{\ensuremath{K^0_L\rightarrow \pi^0 \nu \overline{\nu}}}

\chapter*{Report of the Quark Flavor Physics Working Group}
\label{chap:qfp}


\begin{center}

\begin{center}

{Conveners: J.N.~Butler,$^{13}$ Z.~Ligeti,$^{22}$ J.L.~Ritchie$^{\,38}$}

{Task Force leaders:\\
V.~Cirigliano,$^{23}$ S.~Kettell$^{\,3}$ (Kaons); \
R.~Briere,$^{6}$ A.A.~Petrov$^{\,13}$ (Charm); \\
A.~Schwartz,$^{8}$ T.~Skwarnicki,$^{37}$ J.~Zupan$^{\,8}$ ($B$~physics);\\
N.~Christ,$^{10}$ S.R.~Sharpe,$^{41}$ R.S.~Van~de~Water$^{\,13}$~(Lattice~QCD)}

\smallskip
W.~Altmannshofer,$^{13}$
N.~Arkani-Hamed,$^{20}$
M.~Artuso,$^{37}$
D.M.~Asner,$^{32}$
C.~Bernard,$^{42}$
A.J.~Bevan,$^{34}$
M.~Blanke,$^{12}$
G.~Bonvicini,$^{13}$
T.E.~Browder,$^{17}$
D.A.~Bryman,$^{2}$
P.~Campana,$^{14}$
R.~Cenci,$^{24}$
D.~Cline,$^{4}$
J.~Comfort,$^{1}$ 
D.~Cronin-Hennessy,$^{26}$
A.~Datta,$^{27}$
S.~Dobbs,$^{29}$
M.~Duraisamy,$^{27}$
A.X.~El-Khadra,$^{18}$
J.E.~Fast,$^{32}$
R.~Forty,$^{12}$
K.T.~Flood,$^{5}$
T.~Gershon,$^{40}$
Y.~Grossman,$^{11}$
B.~Hamilton,$^{24}$
C.T.~Hill,$^{13}$
\\
R.J.~Hill,$^{7}$
D.G.~Hitlin,$^{5}$
D.E.~Jaffe,$^{3}$
A.~Jawahery,$^{24}$
C.P.~Jessop,$^{28}$
A.L.~Kagan,$^{8}$
D.M.~Kaplan,$^{19}$
M.~Kohl,$^{16}$
P.~Krizan,$^{21}$
A.S.~Kronfeld,$^{13}$ 
K.~Lee,$^{4}$
L.S.~Littenberg,$^{3}$
D.B.~MacFarlane,$^{35}$
P.B.~Mackenzie,$^{13}$
B.T.~Meadows,$^{8}$
J.~Olsen,$^{33}$
M.~Papucci,$^{22}$
Z.~Parsa,$^{3}$
G.~Paz,$^{13}$
G.~Perez,$^{12,44}$
L.E.~Piilonen,$^{39}$
K.~Pitts,$^{18}$
M.V.~Purohit,$^{36}$
B.~Quinn,$^{27}$
B.N.~Ratcliff,$^{35}$
D.A.~Roberts,$^{24}$
J.L.~Rosner,$^{7}$
P.~Rubin,$^{15}$ 
J.~Seeman,$^{35}$
K.K.~Seth,$^{29}$
B.~Schmidt,$^{12}$ 
A.~Schopper,$^{12}$
M.D.~Sokoloff,$^{8}$
A.~Soni,$^{3}$
K.~Stenson,$^{9}$
S.~Stone,$^{37}$
R.~Sundrum,$^{24}$
R.~Tschirhart,$^{13}$
A.~Vainshtein,$^{26}$
Y.W.~Wah,$^{7}$
G.~Wilkinson,$^{31}$
M.B.~Wise,$^{5}$
\\
E.~Worcester,$^{3}$
J.~Xu,$^{25}$
T.~Yamanaka$^{\,30}$

\smallskip

$^{1}$Arizona State University, Tempe, AZ 85287-1504, USA\\ 
$^{2}$University of British Columbia, Vancouver, BC V1V 1V7, Canada\\ 
$^{3}$Brookhaven National Laboratory, Upton, NY 11973-5000, USA\\ 
$^{4}$University of California, Los Angeles, Los Angeles, CA 90095, USA\\ 
$^{5}$California Institute of Technology, Pasadena, CA 91125, USA\\ 
$^{6}$Carnegie Mellon University, Pittsburgh, PA 15213, USA\\
$^{7}$University of Chicago, Enrico Fermi Institute, Chicago, IL 60637, USA\\ 
$^{8}$University of Cincinnati, Cincinnati, OH 45221, USA\\ 
$^{9}$University of Colorado, Boulder, CO 80309, USA\\ 
$^{10}$Columbia University, New York, NY 10027, USA\\ 
$^{11}$Cornell University, Ithaca, NY 14853, USA\\
$^{12}$European Organization for Nuclear Research (CERN), Geneva, Switzerland\\
$^{13}$Fermi National Accelerator Laboratory, Batavia, IL 60510, USA\\
$^{14}$Laboratori Nazionali di Frascati dell'INFN, I-00044 Frascati, Italy\\
$^{15}$George Mason University, Fairfax, VA 22030, USA\\
$^{16}$Hampton University, Hampton, VA 23668, USA\\
$^{17}$University of Hawaii, Honolulu, Hawaii 96822, USA\\
$^{18}$University of Illinois, Urbana, IL 61801, USA\\ 
$^{19}$Illinois Institute of Technology, Chicago, IL 60616, USA\\ 
$^{20}$Institute for Advanced Study, Princeton, NJ 08540, USA\\
$^{21}$Jozef Stefan Institute, 1000 Ljubljana, Slovenia\\
$^{22}$Lawrence Berkeley National Laboratory, Berkeley, CA 94720, USA\\ 
$^{23}$Los Alamos National Laboratory, Los Alamos, NM 87545, USA\\ 
$^{24}$University of Maryland, College Park, MD 20742, USA\\ 
$^{25}$University of Michigan, Ann Arbor, MI 48109, USA\\
$^{26}$University of Minnesota, Minneapolis, MN 55455 USA\\ 
$^{27}$University of Mississippi, Oxford, MS 38677, USA\\
$^{28}$University of Notre Dame, Notre Dame, IN 46556, USA\\
$^{29}$Northwestern University, Evanston, IL  60208 USA\\ 
$^{30}$Osaka University, Toyanaka, Osaka 560-0043, Japan\\
$^{31}$University of Oxford, Oxford, OX1 3RH, United Kingdom\\ 
$^{32}$Pacific Northwest National Laboratory, Richland, WA  99352\\
$^{33}$Princeton University, Princeton, NJ 08544, USA\\ 
$^{34}$Queen Mary University of London, London, E1 4NS, United Kingdom\\ 
$^{35}$SLAC National Accelerator Laboratory, Menlo Park, CA 94025, USA\\ 
$^{36}$University of South Carolina, Columbia, SC 29208, USA\\ 
$^{37}$Syracuse University, Syracuse, NY 13244-5040, USA\\ 
$^{38}$University of Texas at Austin, Austin, TX 78712-0587, USA\\ 
$^{39}$Virginia Polytechnic Institute and State University, Blacksburg, VA 24061, USA\\ 
$^{40}$University of Warwick, Coventry CV4 7AL, United Kingdom\\
$^{41}$University of Washington, Seattle, WA 98195 USA\\ 
$^{42}$Washington University, St.\ Louis, MO 63130, USA\\
$^{43}$Wayne State University, Detroit, MI 48201, USA\\
$^{44}$Weizmann Institute of Science, Rehovot 76100, Israel

\end{center}


\end{center}


\bigskip
\begin{quote}
\centerline{\bf Abstract}

This report represents the response of the Intensity Frontier Quark Flavor Physics Working
Group to the Snowmass charge. We summarize the current status of quark flavor
physics and identify many exciting future opportunities for studying the
properties of strange, charm, and bottom quarks.   The ability of these studies to
reveal the effects of new physics at high mass scales make them an essential
ingredient in a well-balanced experimental particle physics program.

\end{quote}


\newpage

\tableofcontents

\lhead[\fancyplain{}{\bf\thepage}]%
      {\fancyplain{}{}}

\newpage

\rhead[\fancyplain{}{\bf Quark Flavor Physics Working Group}]%
      {\fancyplain{}{\bf\thepage}}
\lhead[\fancyplain{}{\bf\thepage}]%
      {\fancyplain{}{\bf\boldmath\rightmark}}


\section{Introduction}
\label{sec:intro}

This report, from the Quark Flavor Physics working group,
describes the physics case for precision studies of flavor-changing
interactions of bottom, charm, and strange quarks, and it discusses the 
experimental program needed to exploit these physics opportunities.
It also discusses the role of theory and the importance of lattice QCD
to future progress in this field.
The report is the result of a
process that began before Snowmass, in the fall of 2011 with the DOE-sponsored
workshop on Fundamental Physics at the Intensity Frontier (Rockville, MD).   The Heavy Quarks
working group from that workshop continued into the Snowmass process, albeit with
a change of name to Quark Flavor Physics to better reflect our emphasis on 
quark flavor mixing.  The Heavy Quarks report~\cite{Hewett:2012ns} from that workshop 
provided a starting point for our Snowmass efforts.  
\phantom{\cite{Hewett:2012ns}}%

With the initiation of the Snowmass process, our working group grew.  Also, four
Task Forces were organized to focus on four closely related, but distinct, areas
of effort in quark-flavor physics:  kaons, charm, $B$-physics, and lattice QCD.
Our working group had physical meetings during the Community Planning Meeting
at Fermilab (October, 2012), at the Intensity Frontier Workshop at Argonne (April, 2013),
and at Snowmass itself at the University of Minnesota (July, 2013).  Consequently, 
this report is the culmination of discussions that were 
conducted over a period of almost two years.   

This report describes the physics case for quark-flavor physics, and it represents
the aspirations of a substantial community of physicists in the U.S.\ who are
interested in this physics.  This report is not a review of
quark-flavor physics, and no attempt has been made to
provide complete references to prior work.  Rather,  it focuses on the 
opportunities for spectacular discoveries during the remainder of this decade
and during the next decade, made possible by the extraordinary reach to high mass
scales that is possible in quark-flavor physics experiments.

Nevertheless, before looking forward,  
it provides useful context to briefly review some history.
In the 1990's, the U.S.\ was the leader both on the Energy Frontier and in
quark flavor-physics experiments at the Intensity Frontier.  $B$ physics
was still dominated by the CLEO experiment for most of that decade.  
The most sensitive rare $K$
decay experiments performed to date were then underway at the Brookhaven AGS,
including an experiment that made the first observation of the extremely rare 
$K^+ \to \pi^+ \nu \overline{\nu}$ decay,
and a fixed-target experiment using the Tevatron at Fermilab
was underway that observed 
direct CP violation in $K_L^0 \to \pi \pi$ decays. 
Toward the end of that decade, the asymmetric $e^+e^-$ 
$B$ factories began running at SLAC and KEK, leading to increases in  the
size of $B$ meson data sets by two orders of magnitude and also opening
the door to measurements of time-dependent CP asymmetries,
which provided the experimental basis for the 2008 Nobel Prize.
In the midst of this success,  a number of new and ambitious
quark-flavor initiatives were put forward in the U.S.  These included the BTeV
proposal which would have used the Tevatron collider for 
$B$ physics, the CKM proposal which would have made the first high-statistics
measurement of $K^+ \to \pi^+ \nu \overline{\nu}$ using the Fermilab Main Injector,
and the RSVP proposal which included an experiment (KOPIO) to measure
$K_L^0 \to \pi^0 \nu \overline{\nu}$ at the Brookhaven AGS.
After lengthy consideration in an enviroment characterized by flat budgets and 
a predilection for a fast start on the International Linear Collider
on U.S.\ soil, all of these initiatives were ultimately terminated.
Also, as accelerator breakthroughs capable of increasing $B$-factory luminosity
by more than another order of magnitude were made,
the opportunity to upgrade the PEP-II $B$ factory at SLAC was not pursued.  
This history is relevant in order to stress that the U.S.\ has been a leader in 
flavor-physics experiments --- involving a vigorous community ---
until very recently.  Nonetheless, 
this sequence of events  inevitably encouraged many in the 
flavor-physics community in the U.S.\ to migrate elsewhere, 
most often to ATLAS or CMS at the LHC.

In spite of these developments in the U.S.,  strong physics imperatives have
motivated 
a rich quark flavor physics
program that is flourishing around the world.  Kaon experiments, $B$-physics
experiments, and charm experiments are running and under construction in Asia
and Europe. Indeed, CERN --- the laboratory that now owns the Energy
Frontier~--- is also the home of a running $B$-physics experiment (LHCb), which
has a clear upgrade path, and a rare $K$ decay experiment (NA62), focusing
on $K^+ \to \pi^+ \nu \overline{\nu}$, which will begin
taking data near the end of 2014.  
This reflects the world-wide consensus that  flavor-physics
experiments are critical to progress in  particle physics.


Looking forward, it is clear that there continues to be strong
interest and a potentially substantial community in the U.S.\ for an 
experimental 
flavor-physics program. 
The motivation for this program can be described very simply.
If the LHC observes
new high-mass states, it will be necessary to distinguish among models
proposed to explain them.  This will require tighter constraints from the
flavor sector, which can come from more precise experiments using
strange, charm, and bottom quark systems.  
If the LHC does not make such discoveries, then the ability of 
precision flavor-physics experiments to probe mass scales far above LHC, 
through virtual effects, is the best hope to see signals that may point toward
the next energy scale to~explore.

In the following sections of this report, we describe the general physics case
for quark-flavor physics, followed by the reports of each of the Task Forces. 
The Task Forces were in communication with each other, but worked independently
on these reports.  Finally, this report concludes with a discussion of how the
U.S.\ high-energy physics program can, at relatively modest cost compared to
most other initiatives, participate in critical flavor-physics experiments
offshore and regain some of its leadership status by executing a program of
rare kaon decay experiments at Fermilab.

\section{Quark Flavor as a Tool for Discovery}

An essential feature of flavor physics experiments is their ability to probe
very high mass scales, beyond the energies directly accessible in collider experiments.  
In addition, flavor physics can teach us about properties of TeV-scale new
physics, which cannot be learned from the direct production of new particles at
the LHC.  This is because quantum effects allow virtual particles to modify the
results of precision measurements in ways that reveal the underlying physics.
(The determination of the $t$--$s$ and $t$--$d$ couplings in the standard model
(SM) exemplifies how measurements of some properties of heavy particles may only
be possible in flavor physics.) Even as the LHC embarks on probing the TeV
scale, the ongoing and planned precision flavor physics experiments are
sensitive to beyond standard model (BSM) interactions at mass scales which are
higher by several orders of magnitude.  These experiments will provide essential
constraints and complementary information on the structure of models put forth
to explain any discoveries at LHC, and they have the potential to reveal new
physics that is inaccessible to the LHC.

Throughout the history of particle physics discoveries made in studies of rare
processes have led to new and deeper understanding of nature.  A classic example
is beta decay, which foretold the electroweak mass scale and the ultimate
observation of the $W$ boson.  Results from kaon decay experiments were crucial
for the development of the standard model:  the discovery of CP violation in
$K_L^0 \to \pi^+ \pi^-$ decay ultimately pointed toward the three-generation CKM
model~\cite{Kobayashi:1973fv, Cabibbo:1963yz}, the absence of
strangeness-changing neutral current decays (i.e., the suppression of $K_L^0 \to
\mu^+ \mu^-$ with respect to $K^+ \to \mu^+ \nu$) led to the prediction of a
fourth quark~\cite{GIM} (charm), and the measured value of the $K_L$\,--\,$K_S$
mass difference made it possible to predict the charm quark
mass~\cite{Gaillard-Lee, Vainshtein} before charm particles were directly
detected.  The larger than expected $B_H$\,--\,$B_L$ mass
difference foretold the high mass of the top quark.  Precision measurements of
time-dependent CP-violating asymmetries in $B$-meson decays in the \babar\ and
Belle experiments firmly established the CKM phase as the dominant source of CP
violation observed to date in flavor-changing processes --- leading to the 2008
Nobel Prize for Kobayashi and Maskawa.  At the same time, corrections to the SM
at the tens of percent level are still allowed, and many extensions of the SM
proposed to solve the hierarchy problem are likely to give rise to changes in
flavor physics that may be observed in the next generation of experiments.

\subsection{Strange, Charm, and Bottom Quarks as Probes for New Physics}

\begin{figure}[t]
\centerline{\includegraphics*[height=5.35cm]{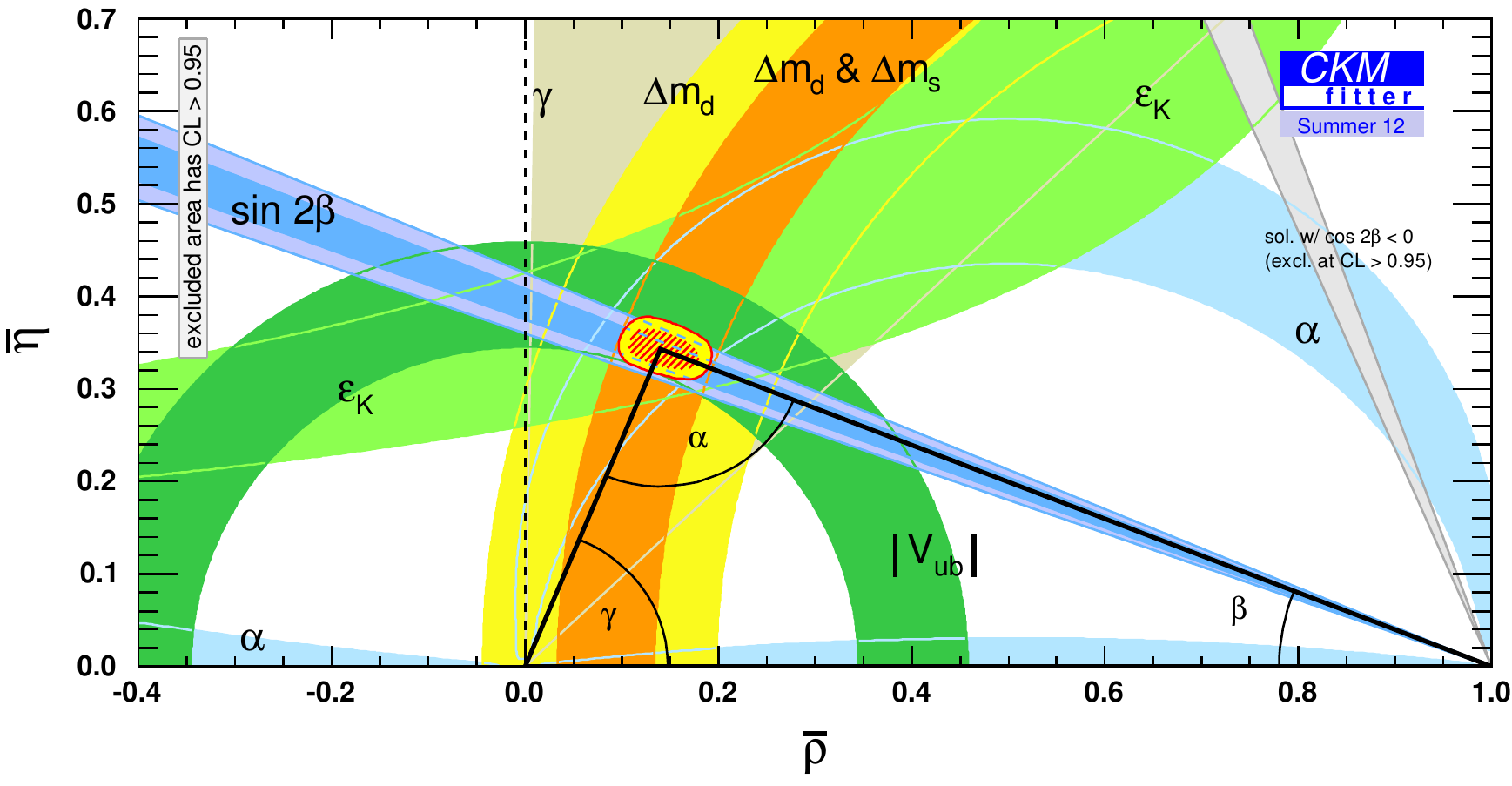} \hfill
\raisebox{5pt}{\includegraphics*[height=5.35cm]{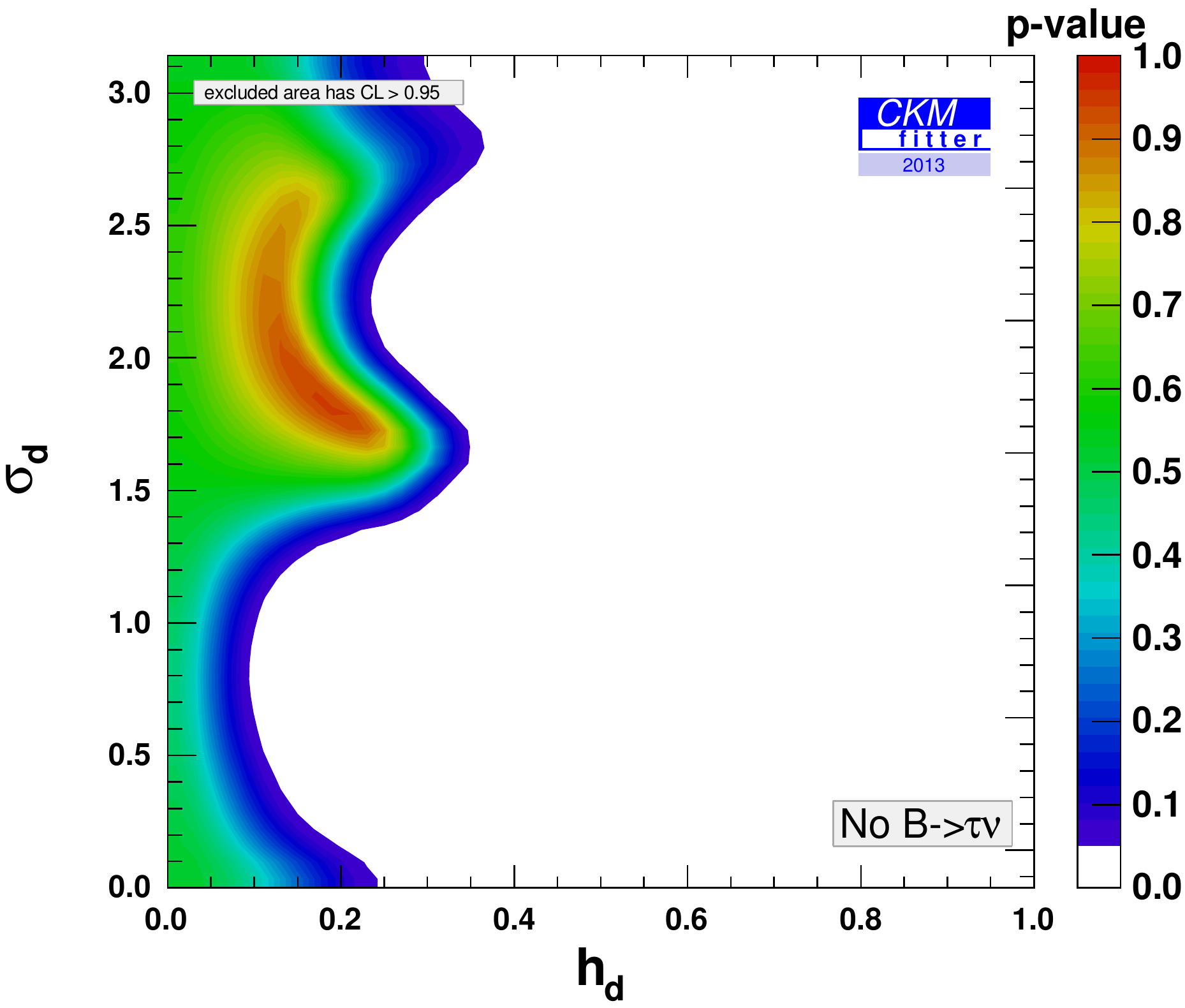}}}
\caption{Left: Constraints on the apex of the unitarity triangle in the
$\rhobar-\etabar$ plane (at 95\% CL)~\cite{ckmfitter,Charles:2004jd}. 
Right: the allowed $h_d-\sigma_d$ new physics parameter space (see text) in
$B^0$--$\B0bar$ mixing.}
\label{fig:hdsd}
\end{figure}

In the past decade our understanding of flavor physics has improved
significantly due to the $e^+e^-$ $B$ factories, \babar, Belle, CLEO, the
Tevatron experiments, and most recently LHCb.  While kaon physics was crucial
for the development of the SM, and has provided some of the most stringent
constraints on BSM physics since the 1960s, precision tests of the CKM picture
of CP violation in the kaon sector alone have been hindered by theoretical
uncertainties in calculating direct CP violation ($\epsilon_K'$).  The $B$
factories and LHCb provided many stringent tests by precisely measuring numerous
CP-violating and CP-conserving quantities, which in the SM are determined in
terms of just a few parameters, but are sensitive to different possible BSM
contributions.  The consistency of the measurements and their agreement with CP
violation in $K^0$--$\K0bar$ mixing, $\epsilon_K$, and with the SM predictions
(shown in the left plot in Fig.~\ref{fig:hdsd}) strengthened the ``new physics
flavor problem."  It is the tension between the relatively low (TeV) scale
required to stabilize the electroweak scale, and the high scale that is
seemingly required to suppress BSM contributions to flavor-changing processes. 
This problem arises because the SM flavor structure is very special, containing
small mixing angles, and because of additional strong suppressions of
flavor-changing neutral-current (FCNC) processes.  Any TeV-scale new physics
must preserve these features, which are crucial to explain the observed pattern
of weak decays.

The motivation for a broad program of precision flavor physics measurements has
gotten even stronger in light of the first LHC run.  With the discovery of a new
particle whose properties are similar to the SM Higgs boson, but no sign of
other high-mass states, the LHC has begun to test naturalness as a guiding
principle of BSM research.  If the electroweak scale is unnatural, we have
little information on the next energy scale to explore (except for a hint at the
TeV scale from dark matter, a few anomalous experimental results, and neutrinos
most likely pointing at a very high scale).  The flavor physics program will
explore much higher scales than can be directly probed.  However, if the
electroweak symmetry breaking scale is stabilized by a natural mechanism, new
particles should be found at the LHC.  Since the largest quantum correction to
the Higgs mass in the SM is due to the top quark, the new particles will likely
share some properties of the SM quarks, such as symmetries and interactions. 
Then they would provide a novel probe of the flavor sector, and flavor physics
and the LHC data would provide complementary information.  Their combined study
is our best chance to learn more about the origin of both electroweak and flavor
symmetry breaking. 

Consider, for example, a model in which the only suppression of new
flavor-changing interactions comes from the large masses of the new particles
that mediate them (at a scale $\Lambda\gg m_W$).  Flavor physics, in particular
measurements of meson mixing and CP violation, put severe lower bounds on
$\Lambda$.  For some of the most important four-quark operators contributing to
the mixing of the neutral $K$, $D$, $B$, and $B_s$ mesons, the bounds on the
coefficients $C/\Lambda^2$ are summarized in Table~\ref{tab:DF2}.  For $C=1$,
they are at the scale $\Lambda \sim (10^2-10^5)$\,TeV. Conversely, for $\Lambda
= 1$\,TeV, the coefficients have to be very small. Therefore, there is a
tension.  The hierarchy problem can be solved with new physics at $\Lambda \sim
1$\,TeV.  Flavor bounds, however, require much larger scales, or tiny
couplings.  This tension implies that TeV-scale new physics must have special
flavor structures, e.g., possibly sharing some of the symmetries that shape the
SM Yukawa interactions. The new physics flavor puzzle is thus the question of
why, and in what way, the flavor structure of the new physics is non-generic. As
a specific example, in a supersymmetric extension of the SM, there are box
diagrams with winos and squarks in the loops.  The size of such contributions
depends crucially on the mechanism of SUSY breaking, which we would like to
probe.  

\begin{table}[t!]
\centerline{\begin{tabular}{c|cc|cc|c}
\hline\hline
\multirow{2}{*}{Operator} &
  \multicolumn{2}{c|}{Bounds on $\Lambda$~[TeV]~($C=1$)} &
  \multicolumn{2}{c|}{Bounds on $C$~($\Lambda=1$\,TeV) } & 
  \multirow{2}{*}{Observables}\\
&   Re  & Im  &  Re  &  Im  &  \\
\hline 
$(\bar s_L \gamma^\mu d_L )^2$  &~$9.8 \times 10^{2}$& $1.6 \times 10^{4}$
  & $9.0 \times 10^{-7}$& $3.4 \times 10^{-9}$ & $\Delta m_K$; $\epsilon_K$ \\
($\bar s_R\, d_L)(\bar s_L d_R$)   & $1.8 \times 10^{4}$& $3.2 \times 10^{5}$
  & $6.9 \times 10^{-9}$& $2.6 \times 10^{-11}$ &  $\Delta m_K$; $\epsilon_K$ \\
\hline
$(\bar c_L \gamma^\mu u_L )^2$  &$1.2 \times 10^{3}$& $2.9 \times 10^{3}$
  & $5.6 \times 10^{-7}$& $1.0 \times 10^{-7}$ & $\Delta m_D$; $|q/p|, \phi_D$ \\
($\bar c_R\, u_L)(\bar c_L u_R$)   & $6.2 \times 10^{3}$& $1.5 \times 10^{4}$
  & $5.7 \times 10^{-8}$& $1.1 \times 10^{-8}$ &  $\Delta m_D$; $|q/p|, \phi_D$\\
\hline
$(\bar b_L \gamma^\mu d_L )^2$    &  $6.6 \times 10^{2}$ 
  &  $9.3 \times 10^{2}$  &  $2.3 \times 10^{-6}$
  &  $1.1 \times 10^{-6}$  &  $\Delta m_{B_d}$; $S_{\psi K_S}$  \\
($\bar b_R\, d_L)(\bar b_L d_R)$  &   $2.5 \times 10^{3}$ & $3.6
  \times 10^{3}$ &  $3.9 \times 10^{-7}$ &   $1.9 \times 10^{-7}$
  & $\Delta m_{B_d}$; $S_{\psi K_S}$ \\
\hline 
$(\bar b_L \gamma^\mu s_L )^2$    &  $1.4 \times 10^2$ & $2.5 \times 10^2$  &
  $5.0\times10^{-5}$  &  $1.7\times10^{-5}$  & $\Delta m_{B_s}$; $S_{\psi\phi}$ \\
($\bar b_R \,s_L)(\bar b_L s_R)$  &  $4.8 \times 10^2$ & $8.3 \times 10^2$ &
  $8.8\times10^{-6}$ &  $2.9\times10^{-6}$  & $\Delta m_{B_s}$; $S_{\psi\phi}$ \\ 
\hline\hline
\end{tabular}}
\caption{Bounds on some $\Delta F=2$ operators of the form $(C/\Lambda^2)\,
{\cal O}$, with ${\cal O}$ given in the first column. The bounds on $\Lambda$
assume $C=1$, the bounds on $C$ assume $\Lambda=1$\,TeV.
(From Ref.~\cite{Isidori:2010kg}.)}
\label{tab:DF2}
\end{table}

To be sensitive to BSM contributions to FCNC processes (where the SM is
suppressed, but not absent), many measurements need to be done, and it is only
their combination that can reveal a signal.  (There are some exceptions, mainly
processes forbidden in the SM, but   considering only those would reduce the
sensitivity of the program to BSM physics.)  To visualize the constraints from
many measurements, it is convenient to use the Wolfenstein
parameterization~\cite{Wolfenstein:1983yz} of the CKM matrix (for a review,
see~\cite{Hocker:2006xb}),
\begin{equation}\label{ckmdef}
V_{\rm CKM} = \left( \begin{array}{ccc}
  V_{ud} & V_{us} & V_{ub} \\
  V_{cd} & V_{cs} & V_{cb} \\
  V_{td} & V_{ts} & V_{tb} \end{array} \right)
= \left( \begin{array}{ccc}
  1-\frac{1}{2}\lambda^2 & \lambda  &  A\lambda^3(\rhobar-i\etabar) \cr
  -\lambda  &  1-\frac{1}{2}\lambda^2  &  A\lambda^2 \cr
  A\lambda^3(1-\rhobar-i\etabar)  &  -A\lambda^2  &  1 \end{array} \right)
  + {\cal O}(\lambda^4) \,.
\end{equation}
It exhibits the hierarchical structure of the CKM matrix by expanding in a small
parameter, $\lambda \simeq 0.23$.  The unitarity of this matrix in the SM
implies many relations, such as that defining the ``unitarity triangle" shown in
Fig.~\ref{fig:hdsd}, which arises from rescaling $V_{ud}\, V_{ub}^* +
V_{cd}\, V_{cb}^* + V_{td}\, V_{tb}^* = 0$ by $V_{cd}\,V_{cb}^*$ and
choosing two vertices of the resulting triangle to be $(0,0)$ and $(1,0)$.

As a result of second-order weak interaction processes, there are transitions
between the neutral meson flavor eigenstates, so the physical mass eigenstates
are their linear combinations, denoted as $|B_{H,L}\rangle = p |B^0\rangle \mp q
|\Bbar^0\rangle$.  (The $p$ and $q$ parameters differ for the four neutral
mesons, but the same notation is commonly used without distinguishing indices.) 
In a large class of models, the BSM physics modifies the mixing amplitude of
neutral mesons, and leaves tree-level decays unaffected.  This effect can be
parameterized by just two real parameters for each mixing amplitude.  For $B^0 -
\B0bar$ mixing, writing $M_{12} = M_{12}^{\rm SM}\, \big(1 + h_d\,
e^{2i\sigma_d}\big)$, the constraints on $h_d$ and $\sigma_d$ are shown in the
right plot in Fig.~\ref{fig:hdsd}.  (Evidence for $h_d \neq 0$ would rule out
the SM.)  Only in 2004, after the first significant constraints on $\gamma$ and
$\alpha$ from \babar\ and Belle, did we learn that the BSM contribution to
$B^0$--$\B0bar$ mixing must be less than the SM amplitude~\cite{Ligeti:2004ak,
Charles:2004jd}.  The right plot in Fig.~\ref{fig:hdsd} shows that order $20\%$
corrections to $|M_{12}|$ are still allowed for (almost) any value of the phase
of the new physics contribution, and if this phase is aligned with the SM
($\sigma_d = 0$ mod $\pi/2$), then the new physics contribution does not yet
have to be much smaller than the SM one.  Similar conclusions apply for $B_s^0$
and $K^0$ mixings~\cite{Bona:2007vi, Lenz:2012az}, as well as many other
$\Delta F=1$ FCNC transition amplitudes.

The fact that such large deviations from the SM are not yet excluded gives very
strong motivations to continue flavor physics measurements in order to observe
deviations from the SM predictions or establish an even stronger hierarchy
between the SM and new physics contributions.

In considering the future program, the following issues~\cite{Grossman:2009dw}
are of key importance:

\begin{enumerate}\vspace*{-10pt}\itemsep 2pt

\item
What are the expected deviations from the SM predictions induced by
new physics at the TeV scale? \\
As explained above, TeV-scale new physics with generic flavor structure is ruled
out by many orders of magnitude.  However, sizable deviations from the SM are
still allowed by the current bounds, and in many scenarios
observable effects are expected.

\item
What are the theoretical uncertainties? \\
These are highly process dependent.  Some measurements are limited by
theoretical uncertainties (due to hadronic, strong interaction, effects), but in
many key processes the theory uncertainties are very small, below the expected
sensitivity of future experiments.

\item
In which processes will the sensitivity to BSM physics increase the most?\\
The useful data sets can increase by a factor of order 100 (in most cases
10--1000), and will probe effects predicted by fairly generic BSM scenarios.

\item
What will the measurements reveal, if deviations from the SM are [not] seen?\\
The flavor physics data will be complementary with the high-$p_T$  part of
the LHC program. The synergy of measurements can reveal a lot about what the
new physics at the TeV scale is, and what it is not.

\end{enumerate}\vspace*{-10pt}

This report concentrates on the physics and prospects of a subset of
measurements, for which the answers to these questions are the clearest, both in
terms of theoretical cleanliness and experimental feasibility.  The experiments
will enable many additional measurements which are not discussed here, some due
to lack of space, and some because they will be more important than can now be
anticipated. (Recall that the best measurements of the CKM angles $\alpha$ and
$\gamma$ at \babar\ and Belle were not in formerly expected decay modes.)
While future theory progress is important, the value of more sensitive
experiments is not contingent on it.

\subsection{The Role of Theory}

To find a convincing deviation from the SM, a new physics effect has to be
several times larger than the experimental uncertainty of the measurement and
the theoretical uncertainty of the SM prediction.  One often distinguishes two
kinds of theoretical uncertainties, perturbative and nonperturbative (this
separation is not unambiguous).  Perturbative uncertainties come from the
truncation of expansions in small (or not-so-small) coupling constants, such as
$\alpha_s$ at a few GeV scale.  There are always higher order terms that have
not been computed.  Nonperturbative effects arise because QCD becomes strongly
interacting at low energies, and these are often the limiting uncertainties. 
There are, nevertheless, several possibilities to get at the fundamental physics
in certain cases.

\begin{itemize}\vspace*{-12pt}\itemsep 0pt

\item For some observables the hadronic parameters (mostly) cancel, or can be
extracted from data (e.g., using the measured $K\to\pi\ell\nu$ form factor to
predict $K\to\pi\nu \overline{\nu}$, several methods to extract $\gamma$, etc.).

\item In many cases, CP invariance of the strong interaction implies that the
dominant hadronic physics cancels, or is CKM suppressed (e.g., measuring $\beta$
from $B\to\psi K_S$, and some other CP asymmetries).

\item In some cases one can use symmetries of the strong interaction which arise
in certain limits, such as the chiral or the heavy quark limit, to establish
that nonperturbative effects are suppressed by small parameters, and to estimate
or extract them from data (e.g., measuring $|V_{us}|$ and $|V_{cb}|$, inclusive
rates).

\item 
Lattice QCD is a model-independent method to address nonperturbative phenomena. 
The most precise results to date are for matrix elements involving at most one
hadron in the initial and the final state (allowing, e.g., extractions of
magnitudes of CKM elements).

\end{itemize}\vspace*{-12pt}
All of these approaches use experimental data from related processes to fix some
parameters, constrain the uncertainties, and cross-check the methods.  Thus,
experimental progress on a broad program will not only reduce the uncertainties
of key measurements, but also help reduce theoretical uncertainties.

As an example, consider extracting $\gamma$ from $B \to D K$. This is one of the
cleanest measurements in terms of theoretical uncertainties, because all the
necessary hadronic quantities can be measured.  All $B \to D K$ based analyses
consider decays of the type $B \to D^0 (\D0bar)\, K\, (X) \to F_D\, K\, (X)$,
where $F_D$ is a final state that is accessible in both $D^0$ and $\D0bar$
decay, allowing for interference, and $X$ represents possible extra particles in
the final state.  Using several $B \to DKX$ decays modes (say, $n$ different $X$
states and $k$ different $D^0$ and $\D0bar$ decay modes), one can perform $n k$
measurements, which depend on $n+k$ decay amplitudes.  Thus, one can determine
all hadronic parameters, as well as the weak phase $\gamma$, with very little
theoretical uncertainty.

The main reason why many CP asymmetry measurements have small theoretical
uncertainties is because they involve ratios of rates, from which the leading
amplitudes cancel, so the uncertainties are suppressed by the relative magnitude
of the subleading amplitudes.  This is the case for the time dependent CP
asymmetry in $B\to \psi K_S$, in which case the subdominant amplitude is
suppressed by a factor $\sim50$ due to CKM elements and by the ratio of the
matrix element of a loop diagram compared to a tree diagram.  However, it is not
simple to precisely quantify the uncertainties below the percent level.  In
other modes (e.g., $B\to \phi K_S$, $\eta' K_S$, etc.) the loop suppression of
the hadronic uncertainty is absent, and the theoretical understanding directly
impacts at what level new physics can be unambiguously observed.

Symmetries of the strong interaction that occur for hadrons containing light
quarks ($m_{u,d,s} < \Lambda_{\rm QCD}$) or for hadrons containing a heavy quark
($m_{b,c} > \Lambda_{\rm QCD}$) have played critical roles in understanding
flavor physics.  Chiral perturbation theory has been very important for kaon
physics, and isospin symmetry is crucial for the determination of $\alpha$ in $B
\to \pi\pi$, $\rho\rho$, and $\rho\pi$ decays.  For $B$ and $D$ mesons, extra
symmetries of the Lagrangian emerge in the $m_{b,c} \gg \Lambda_{\rm QCD}$
limit, and these heavy quark spin-flavor symmetries imply, for example, that
exclusive semileptonic $B\to D^{(*)}\ell \overline{\nu}$ decays are described by a
single universal Isgur-Wise function in the symmetry limit.  For inclusive
semileptonic $B$ decays, an operator product expansion can be used to compute
sufficiently inclusive rates; applications include the extraction of
$|V_{cb}|$.  As is often the case, after understanding the symmetry limit and
its implications, it is the analysis of subleading effects where many
theoretical challenges lie.  The theoretical tools to make further progress are
well-developed, but much work remains to be done to reach the ultimate
sensitivities.

Lattice QCD has become an important tool in flavor physics, and significant
improvements are expected.  As substantial investment in computational
infrastructure is required, a separate section discusses it in this report. 
Lattice QCD allows first-principles calculations of some nonperturbative
phenomena.  In practice, approximations have to be made due to finite computing
power, which introduce systematic uncertainties that can be studied (e.g.,
dependence on lattice spacing, spatial volume, etc.). Due to new algorithms and
more powerful computers, matrix elements which contain at most one hadron in the
final state should soon be calculable with percent level uncertainties.  Matrix
elements involving states with sizable widths, e.g., $\rho$ and $K^*$, are more
challenging.  So are calculations of matrix elements containing more than one
hadron in the final state, and it will require further developments to obtain
small uncertainties for those.  Thus, lattice QCD errors are expected to become
especially small for leptonic and semileptonic decays, and meson mixing.

In summary, there are many observables with theoretical uncertainties at the few
percent level, matching the expected experimental sensitivity, which is
necessary to allow a discovery of small new physics contributions.  The full
exploitation of the experimental program requires continued support of
theoretical developments.



\section{Report of the Kaon Task Force}

Kaon decays have played a pivotal role in shaping the standard model
(SM).  Prominent examples include the introduction of internal
``flavor'' quantum numbers (strangeness), parity violation ($K \to
2 \pi, 3 \pi$ puzzle), quark mixing, meson-antimeson oscillations,
discovery of CP violation, suppression of flavor-changing neutral
currents (FCNC), discovery of the GIM (Glashow-Iliopoulos-Maiani)
mechanism and prediction of charm.  Now and looking ahead, kaons
continue to have high impact in constraining the flavor sector of
possible extensions of the SM.

In the arena of kaon decays, a key role is played by the FCNC
modes mediated by the quark-level processes $s \to d (\gamma,\, \ell^+
\ell^-,\, \nu \overline{\nu})$, and in particular the four theoretically
cleanest modes
$K^+ \to \pi^+ \nu \overline{\nu}$,
$K_L \to \pi^0 \nu \overline{\nu}$, 
$K_L \to \pi^0  e^+ e^-$, 
$K_L \to \pi^0  \mu^+ \mu^-$.  
Because of the peculiar suppression of the SM amplitude (top-quark loop
suppressed by $|V_{td}V_{ts}| \sim \lambda^5$) which in general is not present 
in SM extensions, kaon FCNC modes offer a unique window on the flavor
structure of such extensions.  This argument by itself provides a strong
and model-independent motivation to study these modes in the LHC era.
Rare kaon decays can elucidate the flavor structure of
SM extensions, information that is in general not accessible from high-energy
colliders.

The actual ``discovery potential" depends on the precision of the
prediction for these decays in the SM, the level of
constraints from other observables, and how well we can measure their
branching ratios.

\begin{table}[tb]
\centerline{\begin{tabular}{c|c|c|l}
\hline\hline
{Observable} &
  {SM Theory} &
  {Current Expt.} & 
  {Future Experiments}\\
\hline 
${\cal B}(K^+ \to \pi^+ \nu \overline{\nu})$ 
&$7.81(75)(29)\times 10^{-11}$
& $1.73^{+1.15}_{-1.05} \times 10^{-10}$ 
& $\sim$10\% at  NA62 \\
& & E787/E949 & $\sim$5\% at  ORKA \\
& &  & $\sim$2\%  at \ProjectX\  \\
\hline
${\cal B}(K^0_L \to \pi^0 \nu \overline{\nu})$ 
& $2.43(39)(6)\times 10^{-11}$
&$<2.6 \times 10^{-8}$  \  E391a  &$ 1^{\rm st}$ observation at  KOTO \\
&   &  & $\sim$5\% at  \ProjectX\  \\
\hline
${\cal B}(K^0_L \to \pi^0 e^+ e^-)$ 
& $(3.23^{+0.91}_{-0.79})\times 10^{-11}$ 
& $<2.8 \times 10^{-10}$ \ KTeV  & $\sim$10\% at  \ProjectX\  \\
\hline
${\cal B}(K^0_L \to \pi^0 \mu^+ \mu^-)$ 
& $(1.29^{+0.24}_{-0.23})\times 10^{-11}$
& $<3.8 \times 10^{-10}$  \ KTeV  &  $\sim$10\% at \ProjectX\  \\
\hline
$|P_T|$ 
& $ \sim 10^{-7}$ & $<0.0050$  & $<0.0003$ at  TREK  \\
in $K^+ \to \pi^0 \mu^+ \nu$   &   &  & $<0.0001$ at  \ProjectX\  \\
\hline 
$ \Gamma(K_{e2})/\Gamma(K_{\mu2})$  
& $2.477 (1)  \times 10^{-5} $
& $2.488(10) \times 10^{-5}$  
 &  $\pm 0.0054 \times 10^{-5}$ at TREK \\
   &  &   (NA62, KLOE)  & $\pm 0.0025 \times 10^{-5}$  at \ProjectX\  \\ 
\hline
${\cal B}(K^0_L \to \mu^\pm e^\mp)$ 
& $< 10^{-25}$  &$< 4.7 \times 10^{-12}$ & $< 2 \times 10^{-13}$ at  \ProjectX\  \\
\hline\hline
\end{tabular}}
\vspace{0.2cm}
\caption{A summary of the reach of current and proposed experiments for some key rare kaon decay measurements, in comparison to standard model theory and the current best experimental results. In the SM   predictions for the $K \to \pi \nu \overline{\nu}$
and $K \to \pi \ell^+ \ell^-$   the first error is parametric, the second denotes the  intrinsic theoretical uncertainty. 
}
\label{tab:exptKsummary}
\end{table}

\subsection{Rare Kaon Decays in the Standard Model: Status and Forecast}

State-of-the-art predictions (see Ref.~\cite{brod-projectx}
and references therein) are summarized in 
Table~\ref{tab:exptKsummary} along with current and expected experimental results.
The predictions show our current  knowledge of the theoretical branching ratio uncertainties: 
$K^+ \to \pi^+ \nu \overline{\nu}$ at the 10\% level, 
$K_L \to \pi^0 \nu \overline{\nu}$ at the 15\% level, 
and $K_L \to \pi^0  e^+ e^-$ and 
$K_L \to \pi^0  \mu^+ \mu^-$ at the 25--30\% level. 
In the neutrino modes, the intrinsic 
theoretical uncertainty is a small
fraction of the total, which is currently dominated by the uncertainty in
CKM parameters.
In the charged lepton modes, the uncertainty is dominated by long
distance contributions which can be parametrized in terms of the
rates of other decays (such as $K_S \to \pi^0 \ell^+ \ell^-$).  
It is expected that in the next decade progress
in lattice QCD and in $B$ meson measurements (LHCb
and Belle~II) will reduce the uncertainty on both
$K \to \pi \nu \overline{\nu}$ modes to the 5\% level.  Substantial
improvements in $K _L\to \pi^0 \ell^+ \ell^-$ will have to rely on
lattice QCD computations, requiring evaluation of bi-local operators.
Exploratory steps exist, but involve new techniques, making it hard to
forecast the level of uncertainty that can be achieved.  Therefore,
from a theory perspective, the golden modes remain the
$K \to \pi \nu \overline{\nu}$ decays, because they have small
long-distance contamination (negligible in the CP-violating $K_L$
mode).  The $K \to \pi \nu \overline{\nu}$ decay rates, especially in
the $K_L$ mode, can be predicted with smaller theoretical uncertainties
than other FCNC decay rates involving quarks.

\subsection{Beyond the Standard Model Physics Reach}

The beyond the standard model (BSM) reach of rare FCNC kaon decays has received significant
attention in the literature, through both explicit model analyses and
model-independent approaches based on effective field theory (EFT).  In the
absence of a clear candidate for the TeV extension of the SM, 
the case for
discovery potential and model-discriminating power can be presented
very efficiently in terms of an EFT approach
to BSM physics.  
In this approach, one parametrizes the effects of new heavy particles
in terms of local operators whose coefficients are suppressed by inverse
powers of the heavy new physics mass scale.
The important point is that the EFT approach allows us to make
statements that apply to classes of models, not just any specific SM
extension.  In this context, one can ask two important questions: 
(i) how large a deviation from the SM can we expect in rare decays from
existing constraints?  (ii) if a given class of operators dominates,
what pattern of deviations from the SM can we expect in various rare
kaon decays?  
\phantom{\cite{Kronfeld:2013uoa}}%

Our discussion here parallels the one given in Ref.~\cite{sebastian}, to which we refer for more 
details. To leading order in $v/\Lambda$ (where $v \sim 200$~GeV and 
$\Lambda$ is the scale of new physics),  six operators can affect the $K \to \pi \nu \overline{\nu}$ decays. 
Three of  these are four-fermion operators and affect the  $K \to \pi \ell^+ \ell^-$ decays  as well 
(one of these operators  contributes to $K \to \pi \ell \nu$  by $SU(2)$  gauge invariance). 
The coefficients of these operators are largely unconstrained by other observables, and therefore one can expect  sizable 
deviations from the SM in  $K \to \pi \nu \overline{\nu}$  (both modes) and $K \to \pi \ell^+ \ell^-$, 
depending on the flavor structure of the BSM scenario.

\begin{figure}[t!] 
\centering
\includegraphics[width=0.45\textwidth]{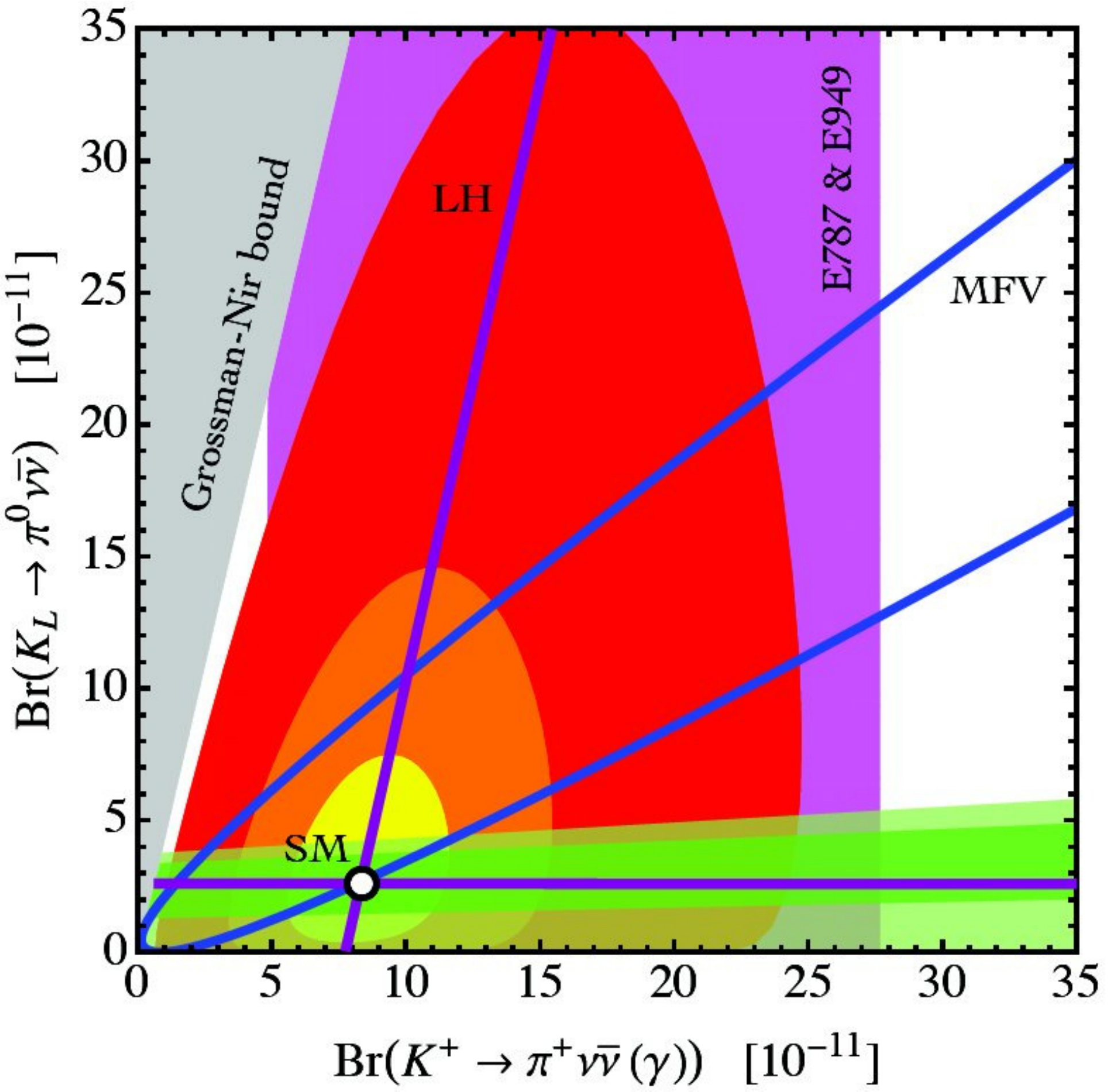}
\caption{\label{fig:constraints} Predictions for the $K \to \pi \nu
  \overline{\nu}$ branching ratios assuming dominance of the 
 $Z$-penguin operators,  
  for different choices of 
 the effective couplings  $C_{\rm NP},\tilde{C}_{\rm NP}$~\cite{uli-projectx}. 
The SM point is indicated by a white dot with black border.  
The yellow, orange, and red shaded contours correspond to
 $|C_{\rm NP},\tilde{C}_{\rm NP}| \leq \left \{0.5, 1, 2 \right \} |\lambda_t  \hspace{0.5mm}  C_{\rm SM}|$, 
the magenta band indicates the 68\% confidence level~(CL) constraint
on ${\cal B} (K^+ \to \pi^+ \nu \overline{\nu} \hspace{0.5mm} (\gamma))$
from experiment~\cite{Adler:2008zza}, and the gray area is
theoretically inaccessible~\cite{Grossman:1997sk}.  The blue parabola represents the subspace
accessible to MFV models.  The purple straight lines represent the
subspace accessible in models that have only LH currents, due to the
constraint from $\epsilon_K$~\cite{Blanke:2009pq}.  The green band
represents the region accessible after taking into account the
correlation of $K_L \to \pi^0 \nu \overline{\nu}$ with
$\epsilon^\prime_K/\epsilon_K$: the (light) dark band corresponds to
predictions of $\epsilon^\prime_K/\epsilon_K$ within a factor of (5) 2
of the experimental value, using central values for the hadronic
matrix elements as reported in~\cite{Bauer:2009cf}  and references therein. }
\end{figure}

The other  three  leading operators contributing to  $K \to \pi \nu \overline{\nu}$ 
involve the Higgs field and reduce,  after 
electroweak symmetry breaking,  to effective flavor-changing $Z$-boson interactions, with both left-handed (LH) and 
right-handed (RH) couplings to quarks.  
These  ``$Z$-penguin'' operators   (both LH and RH) are the leading effect in many SM extensions, and 
affect  a large number of kaon observables 
($K \to \pi \ell^+ \ell^-$, $\epsilon_K$,  $\epsilon_K'/\epsilon_K$, and in the case of one operator 
$K \to \pi \ell \nu$ through $SU(2)$ gauge invariance). 
Focusing on this  class of operators, the relevant part of the effective Lagrangian reads  
\begin{equation} \label{eq:3}
{\cal L}_{\rm eff} \propto 
\left ( \lambda_t \hspace{0.5mm} C_{\rm SM} + C_{\rm NP} \right )  
\bar d_L \gamma_\mu s_L Z^\mu  + \widetilde C_{\rm NP} \, d_R \gamma_\mu s_R Z^\mu \,,
\end{equation}
where $\lambda_q = V_{qs}^\ast V_{qd}$ with $V_{ij}$ denoting elements of
the CKM matrix, and $C_{\rm SM} \approx 0.8$ encodes the SM contribution
to the LH $Z$-penguin  (the RH $Z$-penguin  is highly suppressed in the SM by
small quark masses).
Assuming dominance of the $Z$-penguin operators,  one can study  the 
expectations  for the $K \to \pi \nu  \overline{\nu}$ branching ratio   for different choices of 
the effective couplings  $C_{\rm NP},\ \tilde{C}_{\rm NP}$, and address the correlations 
with other observables.  This is  illustrated in Fig.~\ref{fig:constraints}.  
In this framework,   
$\epsilon^\prime_K / \epsilon_K$ 
provides
the strongest constraint on the CP violating mode $K_L \to \pi^0 \nu \overline{\nu}$~ \cite{Buras:2000qz,Bauer:2009cf,Blanke:2007wr,Buras:1998ed,Buras:1999da}. 
This is illustrated by the green bands in Fig.~\ref{fig:constraints}, where
one can see that the requirement $\epsilon^\prime_K / \epsilon_K  \in [0.2, 5]
(\epsilon^\prime_K/\epsilon_K)_{\rm exp}$ limits deviations in the $K_L \to \pi^0 \nu
\overline{\nu}$ to be of ${\cal O}(1)$, while leaving room for larger deviations 
in the CP conserving mode $K^+ \to \pi^+ \nu \overline{\nu}$.  
The  correlation between $\epsilon^\prime_K/\epsilon_K$ and $K_L \to \pi^0 \nu \overline{\nu}$ 
can be evaded only if there is a cancellation 
among  the $Z$-penguin and  other contributions to $\epsilon^\prime_K/\epsilon_K$. 
Moreover, we stress that  this conclusion holds in all models in which the $Z$-penguin
provides the dominant contribution to $K \to \pi \nu \overline{\nu}$   decays.  
While this is not true in general,  we think  this constraint should be  
one of the drivers of the design sensitivity for 
$K_L \to \pi^0 \nu \overline{\nu}$ experiments.

\begin{figure}[t!]
\centering
\includegraphics[width=0.45\textwidth]{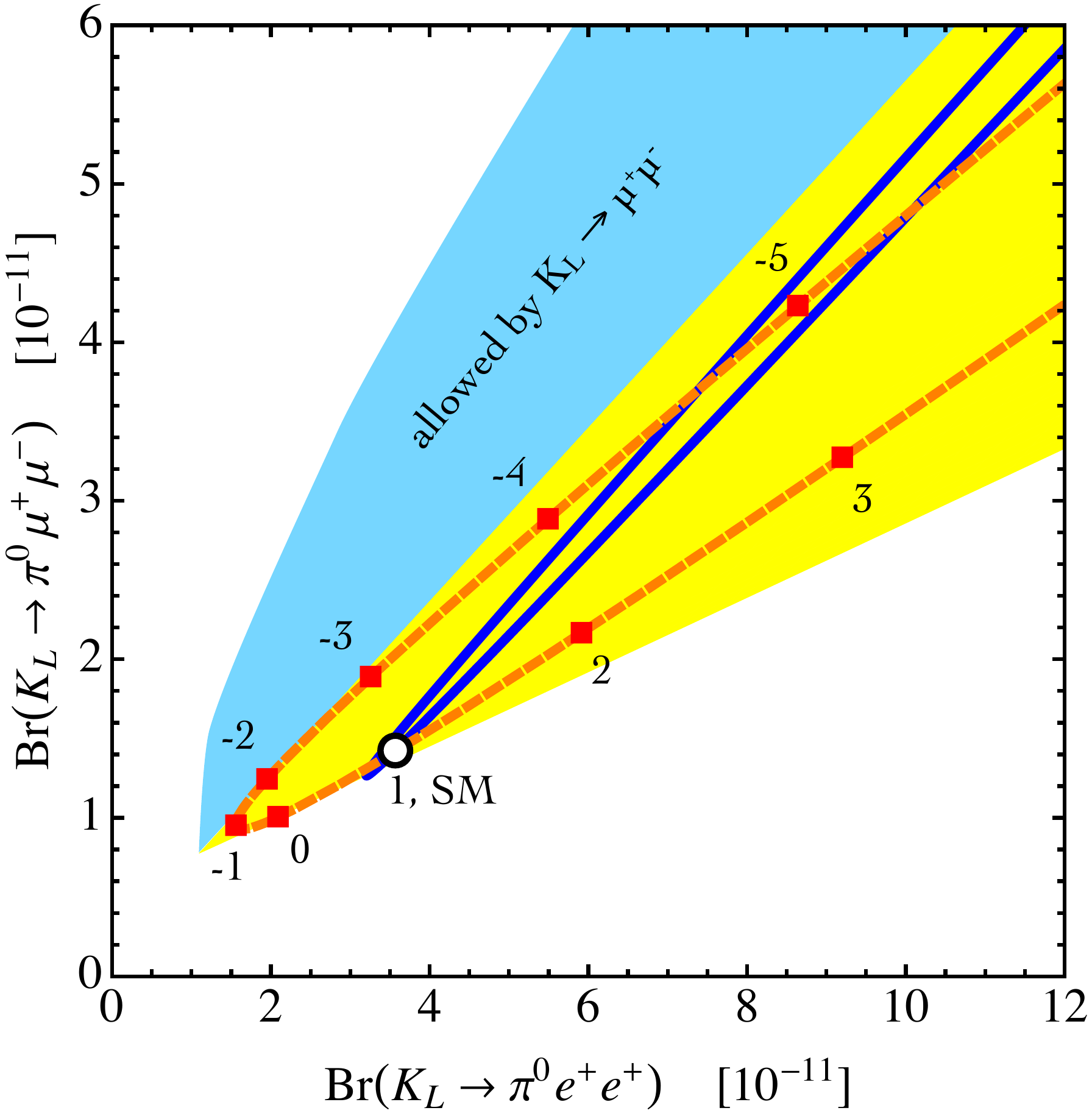}
\vspace{0.2cm}
\caption{Predictions for the $K_L \to \pi^0 \ell^+ \ell^-$  branching ratios assuming different types of NP contributions~\cite{uli-projectx}.
The SM point is indicated by a white dot with black border.
The blue parabola  represents the region accessible  by allowing only for $C_{\rm NP}$ with arbitrary modulus and phase. 
The subspace accessible when $C_{V,A} \neq 0$ is represented by 
the dashed orange parabola (common rescaling of $C_{A,V}$)  and the yellow shaded region (arbitrary values of $C_{A,V}$).
 The subspace accessible when $C_{S,P} \neq 0$  (compatibly with $K_L \to \mu^+ \mu^-$) is represented by the light blue shaded region.
 }
\label{fig:1bis}
\end{figure}

The number of operators that affect the $K_L \to \pi^0 \ell^+ \ell^-$
($\ell = e,\mu)$ decays is larger than the case of
$K \to \pi \nu \overline{\nu}$. Besides (axial-)vector operators resulting
from $Z$- and photon-penguin diagrams, (pseudo-)scalar operators
associated with Higgs exchange can play a role~\cite{Mescia:2006jd}. In
a model-independent framework:
\begin{equation}
\label{eq:5}
{\cal L}_{\rm eff} \supset C_A Q_A  + C_V Q_V + C_P Q_P + C_S Q_S ~, 
\end{equation}
with
\begin{equation} \label{eq:6}
Q_A =  (\bar d \gamma^\mu s ) (\bar \ell \gamma_\mu \gamma_5 \ell) \,, \quad 
Q_V =  (\bar d \gamma^\mu s ) (\bar \ell \gamma_\mu \ell) \,, \quad 
Q_P =  (\bar d  s ) (\bar \ell \gamma_5 \ell) \,, \quad 
Q_S =  (\bar d s ) (\bar \ell \ell) \,.
\end{equation}
In Figure~\ref{fig:1bis} we depict the accessible parameter space
corresponding to various classes of NP. The blue parabola illustrates
again the predictions obtained by allowing only for a contribution
$C_{\rm NP}$ with arbitrary modulus and phase. We see that in models
with dominance of the LH $Z$-penguin the deviations in
$K_L \to \pi^0 \ell^+ \ell^-$ are strongly correlated. A large
photon-penguin can induce significant corrections in $C_V$, which
breaks this correlation and opens up the parameter space as
illustrated by the dashed orange parabola and the yellow shaded
region. The former predictions are obtained by employing a common
rescaling of $C_{A,V}$, while in the latter case the coefficients
$C_{A,V}$ are allowed to take arbitrary values. If besides $Q_{A,V}$,
$Q_{P,S}$ can also receive sizable NP corrections, then a further relative
enhancement of ${\rm Br} (K_L \to \pi^0 \mu^+ \mu^-)$ compared to
${\rm Br} (K_L \to \pi^0 e^+ e^-)$ is possible. This feature is
exemplified by the light blue shaded region that corresponds to the
parameter space that is compatible with the constraints on $C_{P,S}$
arising from $K_L \to \mu^+ \mu^-$.
Finally, we note that  $K_L \to \mu^+\mu^-$ itself is
another FCNC mode of interest,  as it is sensitive to 
different combinations of new physics couplings. 
The constraining power of $K_L \to \mu^+\mu^-$  is  limited by the 
current  understanding of the dispersive part of the amplitude. 
Despite this,  the mode already provides useful diagnostic power,  
as  in combination with $K \to \pi \nu \overline{\nu}$ can help  
distinguish among LH or RH coupling of $Z$ and $Z^\prime$ to 
quarks~\cite{ Blanke:2008yr, Blanke:2009am, Buras:2012jb}.

Rare kaon decays have been extensively studied within well motivated
extensions of the SM, such as supersymmetry (SUSY)~\cite{Buras:2004qb}
and warped extra dimensions (Randall-Sundrum)
models~\cite{Bauer:2009cf, Blanke:2008yr}.  In all cases, deviations
from the SM can be sizable and perhaps most importantly the
correlations between various rare $K$ decays are essential in
discriminating among models.  Rare
$K \to \pi \nu \overline{\nu}$ experiments can also probe the existence of
light states very weakly coupled to the SM appearing in various dark
sector models~\cite{Kamenik:2011vy}, through the experimental
signature $K \to \pi +$\,(missing energy) and distortions to the pion
spectrum.

\subsubsection*{Other modes}

Besides the FCNC modes, kaon decays also provide exquisite probes of the
charged-current sector of SM extensions, probing  scales of TeV or above. 
Theoretically, the cleanest probes are (i) the ratio $R_K \equiv \Gamma (K \to e
\nu) / \Gamma (K \to \mu \nu)$, which tests lepton universality, scalar, and
tensor charged-current interactions; (ii)~the transverse muon polarization
$P_\mu^T$ in the semi-leptonic decay $K^+ \to \pi^0 \mu^+ \nu_\mu$, which is
sensitive to BSM sources of CP violation in scalar charged-current operators. 
In both cases there is a clean discovery window provided by the precise SM
theoretical prediction~\cite{Cirigliano:2007xi} ($R_K$) and by the fact that in
the SM $P_\mu^T$ is generated only by very small and known final state
interactions~\cite{Efrosinin:2000yv}. 
Table~\ref{tab:exptKsummary} provides a summary of  SM predictions for these
processes, along with current and projected experimental sensitivities  at
ongoing or planned experiments.

\subsection{Experimental Program}

Following the termination of a world-class kaon program in the U.S.\ by
2002, leadership in kaon physics shifted to Europe and Japan,
where a program of experiments aiming for orders of
magnitude improvements in reach for new physics is now in progress.

The NA62 experiment at CERN~\cite{NA62}  uses, for the first time, 
the in-flight technique to search for \Kplus.
NA62
will finish commissioning at the end of 2013 and start physics running
toward the end of 2014.  The NA62 goal is
to measure the \Kplus\ branching ratio with 10\% precision along with
a robust and diverse kaon physics program.

The KOTO experiment at JPARC~\cite{KOTO} is an in-flight measurement of \Kzero.
Significant experience and a better understanding of the backgrounds
were obtained in its predecessor, E391a.
The anticipated experimental sensitivity is a few SM
signal events in three years of running with 300~kW of beam power.
Commissioning runs were undertaken in 2012 and 2013 and physics running started in 2013, but the longer
term performance of the experiment will depend upon beam power
evolution of the JPARC accelerator.

The TREK experiment at JPARC~\cite{TREK} will search for T~violation in stopped charged
kaon decays by measuring the polarization asymmetry in $K^+ \to \pi^0
\mu^+ \nu_\mu$ decays.  TREK needs at least 100~kW (the proposal assumed
270~kW) for this measurement.  While the accelerator is running at
lower power, collaborators have proposed a search for lepton flavor
universality violation through the measurement of $\Gamma (K \to e
\nu)/\Gamma (K \to \mu \nu)$ at the 0.2\% level, which will use much
of the TREK apparatus and requires only 30~kW of beam power and will
be ready to run in 2015.  
At the same time, this configuration allows for sensitive searches for a
heavy sterile neutrino ($N$) in $K^+ \to \mu^+ N$, and for light bosons (heavy
photons from the dark sector, $A^\prime \to e^+ e^-$) in the 
$K^+ \to \mu^+ \nu_\mu\, e^+ e^-$ and $K^+ \to \pi^+  e^+ e^-$ decays,
where the new particles would be identified as narrow peaks in the respective
momentum and  $e^+ e^-$ invariant mass spectra.
The uncertainty of the JPARC beam power
profile and potential conflicts for beamline real estate make the long
term future of the TREK experiment unclear.

The KLOE-2 experiment~\cite{KLOE2}  will extend the results of KLOE to improve
neutral kaon interference measurements, CPT and quantum mechanics
tests, and a wide range of measurements of non-leptonic and radiative
kaon decays.

The ORKA experiment is proposed to measure \Kplus\ with 1000 event
sensitivity at the Fermilab Main Injector
(MI)~\cite{Comfort:2011zz}. After a five year run ORKA will reach a
precision of 5\% on the branching ratio, which is the expected level of
theoretical precision.  This high-precision measurement would be one
of the most incisive probes of quark flavor physics in the coming
decade. ORKA is a stopped kaon experiment that builds on the
experience of the E787/949 experiments at Brookhaven that observed
seven candidate events. Backgrounds, primarily from other kaon decays
at branching fractions as much as 10 orders of magnitude larger, have
similar signatures to the signal. ORKA takes advantage of the
extensive knowledge of background rates and characteristics from
E787/E949 by using the same proven experimental techniques.  The
methods for suppressing backgrounds are well known, as are the
background rates and experimental acceptance. Improvements in detector
performance are possible due to significant advances in detector
technology in the 25 years since E787 first ran. The new ORKA detector
with beam supplied by the MI running at 95~GeV with moderate duty
factor presents an opportunity to extend the E787/E949 approach by two
orders of magnitude in sensitivity. The first order of magnitude
improvement comes from the substantially brighter source of low energy
kaons and the second arises from incremental improvements to the
experimental techniques firmly established at BNL. ORKA will observe
210 SM events per year and will make a wide variety of measurements in
addition to the \Kplus\ mode. ORKA will search for and study a range
of important reactions involving kaon and pion decays, such as tests
of lepton universality, symmetry violations, hidden sector particles,
heavy neutrinos and other topics. ORKA will be a world-leading kaon
physics experiment, train a new generation of kaon physicists and
position the U.S. to move forward to a \ProjectX\ kaon program. It is
an essential step in developing a robust intensity frontier program in
the U.S.\ at \ProjectX.

The U.S.\ has an opportunity through ORKA to re-establish a leadership
position in kaon physics.  

\subsubsection*{\ProjectX}

A flagship experiment of the \ProjectX\ physics program will
measure the \Kzero\ branching ratio with 5\% precision.
This effort will build on the KOTO experience, benefit from
the KOPIO initiative~\cite{KOPIO}, and take advantage of the beam power and
flexibility provided by Stage 2 of \ProjectX.

KOPIO proposed to measure \Kzero\ with a
SM sensitivity of 100 events at the BNL Alternating Gradient
Synchrotron (AGS) as part of the RSVP (Rare Symmetry Violating Processes)
project.  The experimental technique and sensitivity were
well-developed and extensively reviewed.  KOPIO was designed to use a
neutral beam at a $42^\circ$ targeting angle produced by 24~GeV
protons from the AGS. The neutral kaons would have an average
momentum of 800~MeV$/c$ with a range of 300--1500~MeV$/c$.  A low
momentum beam was critical for the time-of-flight (TOF) strategy of
the experiment.

The TOF technique is even better matched to the kaon momentum produced by the
3\,GeV proton beam at \ProjectX\ where the higher momentum tail present in the AGS beam is suppressed. Performance of the TOF strategy was
limited by the design bunch width of 200\,ps at the AGS. The \ProjectX\
beam pulse timing, including target time slewing, is expected to be
less than 50\,ps and would substantially improve the momentum
resolution and background rejection capability of the
\Kzero\ experiment driven with the \ProjectX\ beam.

The AGS $K_L$ yield per proton is 20 times the \ProjectX\ yield;
however, the 0.5~mA \ProjectX\ proton flux is
150 times the RSVP goal of $10^{14}$ protons every 5 seconds.  Hence
the neutral kaon flux at \ProjectX\ will be 8 times the AGS flux goal into the
same beam acceptance.  The \ProjectX\ neutral beam will contain 
about a factor of three more neutrons, but neutron interactions will
be highly suppressed by the evacuated beamline and detector volume.
The nominal five-year \ProjectX\ run is
2.5 times longer than the KOPIO initiative at the AGS and hence the
reach of a \ProjectX\ \Kzero \ experiment would be 20 times greater than RSVP.

A TOF-based \Kzero\ experiment driven by \ProjectX\ would be
re-optimized for the \ProjectX\ $K_L$ momentum spectrum, TOF resolution,
and corresponding background rejection. It is likely that this
optimization would result in a smaller neutral beam solid angle which
would simplify the detector design, increase the acceptance and relax
the requirement to tag photons in the fierce rate environment of the
neutral beam. Optimizing the performance will probably require a
proton pulse train frequency of 20--50 MHz and an individual proton
pulse timing width of $\sim\!20$~ps.  Based on the E391a and KOTO
experience, a careful design of the target and neutral beam channel is
required to minimize the neutron halo and to assure target survival in
the intense proton beam.  The high $K_L$ beam flux and the potential of
break-through improvements in TOF performance and calorimeter
technology support the viability of a \Kzero\ experiment
with $\sim\!1000$ SM event sensitivity.

If ORKA~\cite{Comfort:2011zz} observes a significant non-SM result, the
\Kplus \ decay mode could be studied with higher statistics with a
$K^+$ beam driven by \ProjectX. The high-purity, low-momentum $K^+$
beam designed for ORKA could also serve experiments to precisely
measure the polarization asymmetry in $K^+ \to \pi^0 \mu^+ \nu_\mu$
decays and to continue the search for lepton flavor universality
violation through the measurement of $\Gamma (K \to e \nu)/\Gamma (K
\to \mu \nu)$ at high precision.

Depending upon the outcome of the TREK experiment at JPARC, a T
violation experiment would be an excellent candidate for \ProjectX, as
would a multi-purpose experiment dedicated to rare modes that involve
both charged and neutral particles in the final state.  This
experiment might be able to pursue $K_L\to \pi^0 \ell^+ \ell^-$ as
well as many other radiative and leptonic modes. The kaon physics
program at \ProjectX\ could be very rich indeed.

\subsection{Conclusions}

Kaon decays are extremely sensitive probes of the flavor and CP-violating sector
of any SM extension.  The $K \to \pi \nu \overline{\nu}$ golden modes have great
discovery potential: (i)~sizable, ${\cal O}(1)$, deviations from the SM are 
possible; (ii)~even small deviations can be detected due to the precise
theoretical predictions.  Next generation searches should aim for a sensitivity
level of $10^3$ SM events (few \% uncertainty) in both $K^+$ and $K_L$ modes, in
order to maximize discovery potential.

We foresee searches for both $K \to \pi \nu \overline{\nu}$ modes as flagship
measurements of a reinvigorated U.S.-led kaon program.  As summarized in
Table~\ref{tab:exptKsummary},  through ORKA and \ProjectX\ this program has the
opportunity to pursue a broad set of measurements, exploring the full discovery
potential and model-discrimination power of kaon physics.

\section[Report of the $B$-Physics Task Force]{Report of the {\boldmath $B$}-Physics Task Force}


\subsection{Physics Motivation}

\subsubsection*{Searches for BSM Physics}

  
Rare $B$ physics processes are sensitive to new physics (NP) because the  heavy
particles can contribute through virtual corrections to the effective weak
Hamiltonian. 
This makes it possible, for example, to
probe  extended Higgs sectors and to test for
the presence of new gauge interactions or for extended matter content such as
the ones encountered in supersymmetric models.  The sensitivity to NP depends on
how large the flavor violating couplings are.  For instance,  in the most
conservative case of Minimal Flavor Violation (MFV)  with new particles only
contributing in the loops, the rare $B$ processes  can probe mass scales of
roughly $\sim{\mathcal O}({\rm TeV})$  with the next generation experiments.
In the case of general flavor violation with ${\mathcal O}(1)$
off-diagonal couplings, on the other hand, one probes mass scales of ${\mathcal
O}(10^3\,{\rm TeV})$~\cite{Hewett:2012ns}.  Because  the dependence on new
particle masses and (flavor-violating) couplings is different than in the
on-shell production,  the NP searches at LHCb and Belle~II are also
complementary to the high $p_T$ NP searches at ATLAS and CMS. 

Observables that are especially interesting for the 
future $B$ physics program are those
that have small or systematically improvable theoretical uncertainties.  An
important input is provided by measurements of the standard CKM unitarity
triangle.  The angle $\gamma$ and modulus $|V_{ub}|$ are determined from
tree-level processes and thus less prone to contributions from NP.  They
provide the SM ``reference'' determination of the CKM unitarity triangle (in
effect its apex, the values of $\overline{\rho}$ and $\overline{\eta}$).  $|V_{ub}|$ is
measured from inclusive and exclusive $b\to u \ell \nu$ processes.  There is an
on-going effort to improve the theory predictions using both the continuum
methods and lattice QCD, and a factor of a few improvements on the errors
seem feasible.   For instance, the present theory error on $|V_{ub}|$
from exclusive $B\to \pi \ell \nu$ can be reduced from present $8.7\%$ to $2\%$ by
2018~\cite{LatticeWhitePaper} (see Table~\ref{tab:error}).  The theoretical uncertainties in the measurement
of $\gamma$ from $B\to DK$ decays are even smaller.   All the required hadronic
matrix elements can be measured, because of the cascade nature of the $B\to DK,\
D\to f$ decay, if enough final states $f$ are taken into account.  The
irreducible 
theoretical errors thus enter only at the level of one-loop
electroweak corrections and are below ${\mathcal
O}(10^{-6})$~\cite{Zupan:2011mn}.  The present experimental errors are $\pm
12^\circ$ from the average of \babar\ and Belle measurements.   LHCb has
recently matched this precision. The errors are statistics-limited and  will be
substantially decreased in the future.

The tree-level determinations of $\overline{\rho}$ and $\overline{\eta}$ can then be
compared with the  measurements from loop-induced FCNCs, for instance with the
time-dependent CP asymmetry in $B\to J/\psi\, K_S$ and related modes determining
the angle $\beta$. With improved theoretical control BSM physics can be
constrained or even discovered.  NP could also enter in the $B_s-\Bbar_s$
mixing. In the SM the mixing phase is small, suppressed by $\lambda^2$ compared
to~$\beta$. Thus, in the SM, the corresponding time-dependent CP asymmetry in
the $b\to c \overline{c} s$ dominated decays, such as $B_s \to J/\psi\, \phi$, is
predicted very precisely, $\beta_s^{\rm (SM)} = 0.0182 \pm 0.0008$.  The LHCb
result, $\beta_s = -0.035\pm0.045$~\cite{LHCb:2013oba}, is consistent with the
SM expectation, but the statistical uncertainty is much greater than that of the
SM prediction. Since the uncertainty of the SM prediction is very small, future
significant improvements of the measurement of $\beta_s$ will directly translate
to a better sensitivity to BSM physics.


Another important search for NP is to compare the time-dependent CP asymmetries
of penguin-dominated $b\to q \overline{q} s$ processes with the tree dominated $b\to
c \overline{c} s$ decays. Observables that probe this are the differences of CP
asymmetries $S_{J/\psi K_S} - S_{\phi K_S}$, $S_{\psi K_S} - S_{\eta' K_S}$,
etc., in $B_d$ decay, and $S_{J/\psi\phi} - S_{\phi\phi}$ in $B_s$ decay.

The list of interesting observables in $B$ physics is very long. One
could emphasize in particular the rare $B$ decays with leptons in the final
state.  The $B_s\to \ell^+\ell^-$ decay is especially interesting for SUSY
searches in view of the fact that these are $(\tan\beta)^6$ enhanced.   LHCb
presented first evidence of this decay, with ${\cal B}(B_s\to
\mu^+\mu^-)=(3.2^{+1.5}_{-1.2})\times 10^{-9}$~\cite{Aaij:2012nna} consistent
with the SM prediction $(3.54 \pm 0.30) \times 10^{-9}$~\cite{Buras:2012ru, DeBruyn:2012wk}. 
Also, CMS has very recently reported a measurement of
${\cal B}(B_s\to \mu^+\mu^-)=(3.0^{+1.0}_{-0.9})\times 10^{-9}$~\cite{Chatrchyan:2013bka}
consistent with the earlier LHCb result, and LHCb has also updated its
value to ${\cal B}(B_s\to \mu^+\mu^-)=(2.9^{+1.1}_{-1.0})\times 10^{-9}$~\cite{Aaij:2013aka}.
This puts strong constraints on the large $\tan\beta$ region of MSSM,
favored by the measured Higgs mass for the case of TeV scale squarks.  The
theoretical errors on the SM prediction are still several times smaller than the
experimental ones, making more precise measurements highly interesting.
With the LHCb upgrade, the search for $B_d\to \ell^+\ell^-$ will also get near the
SM level.  Rare decays involving a $\nu \overline{\nu}$ pair are theoretically very
clean, and 
Belle~II should reach the SM level in
$B\to K^{(*)}\nu \overline{\nu}$; the current constraints are an order of magnitude
weaker.  There is also a long list of interesting measurements in $b\to s
\gamma$ and $b\to s \ell^+\ell^-$ mediated inclusive and exclusive decays, CP
asymmetries, angular distributions, triple product correlations, etc., which
will be probed much better in the future.  The $s\leftrightarrow d$ processes,
with lower SM rates, will provide many other challenging measurements and
opportunities to find NP. Rare $B$ decays can  also be used as probes for
``hidden sector" particle searches, for lepton flavor violation, and for baryon
number violating processes.


There are also some intriguing deviations from the SM in the current data. The
D0 collaboration  measured the CP-violating dilepton asymmetry to be $4\sigma$
away from zero, $A_{\rm SL}^b = (7.87 \pm 1.96) \times 10^{-3} \approx 0.6\,
A_{\rm SL}^d + 0.4\, A_{\rm SL}^s$~\cite{Abazov:2011yk}.  The measured
semileptonic asymmetry is a mixture of $B_d$ and $B_s$ ones, where $A_{\rm SL}
\simeq 2(1-|q/p|)$ in each case measures the mismatch of the CP and mass
eigenstates.  The quantity $(1-|q/p|)$ is model-independently suppressed by
$m_b^2/m_W^2$, with an additional $m_c^2/m_b^2$ suppression in the SM, which NP
may violate~\cite{Laplace:2002ik}. Since the D0 result allows plenty
of room for NP, it will be important for LHCb and Belle~II to clarify the
situation. LHCb has recently measured $A_{\rm SL}^s = (-2.4\pm5.4\pm3.3)
\times10^{-3}$ \cite{Xing:2012mt}  which complements $A_{\rm SL}^d$ measured at
$e^+e^-$ $B$ factories. Further improvement in experimental errors on both
quantities is needed.  



Another interesting anomaly is the hint of the flavor universality violation in
$B\to D^{(*)}\tau \nu$ decays observed by \babar~\cite{Lees:2013uzd} which differ from the SM prediction expected from the $B\to D^{(*)}\ell \nu$ rates by $3.4\sigma$. Combined with the slight excess of ${\cal B}(B\to \tau\nu)$ over the SM the measurements can be explained using charged Higgs exchange, e.g., in the two Higgs doublet model, but with nontrivial flavor structure \cite{Fajfer:2012jt}. The MFV 
hypothesis is not preferred. To settle the case
it will require larger data sets at the future
$e^+e^-$ $B$ factories (and measuring the $B\to \mu \overline{\nu}$ mode as well).

Any of the above measurements could lead to a discovery of new physics. In addition,
a real strength of the $B$ physics program is that a pattern of modifications
in different measurements can help to
zoom in on the correct NP model. Further information will also be provided by
rare kaon decay experiments and searches for lepton flavor violation in charged lepton decays
such as $\mu \to e \gamma$, $\mu$-to-$e$ conversion on a nucleus, 
$\tau\to \mu \gamma$, and $\tau\to 3\mu$.   This program will provide complementary information to
the on-shell searches at the LHC.



\subsubsection*{Bottomonium Spectroscopy}


Recent observations of the spin-singlet states
$h_b(1P)$, $h_b(2P)$, and $\eta_b(2S)$ have contributed greatly to our
understanding of bottomonium multiplets below the open-bottom threshold.
The unexpected discovery of narrow, manifestly exotic 
$b \overline{b}$ mesons
above the open-bottom threshold has revealed a gap in our
understanding of the QCD spectrum.
Future data will probably reveal
additional $b \overline{b}$ threshold states.
Understanding the spectrum and decays of these states
are major challenges for lattice QCD and phenomenology.
Comparisons of these bottomonium states
with the $XYZ$ states in the charmonium system
should provide additional clues. 
The challenge of bottomonium spectroscopy 
is discussed in detail in a Snowmass White Paper~\cite{Bodwin:2013nua}.

\subsection[Physics Potential of $e^+e^-$ Experiments: Belle~II]{\boldmath 
Physics Potential of $e^+e^-$ Experiments: Belle~II}

The spectacular successes of the $B$-factory experiments
Belle and \babar\ highlight the advantages of $e^+e^-$ collider
experiments:
\begin{itemize}\vspace*{-10pt}\itemsep 2pt
\item Running on the $\Upsilon(4S)$ resonance produces an
especially clean sample of $B^0\B0bar$ pairs in a quantum correlated 
$1^{--}$ state. The low background level allows reconstruction 
of final states containing $\gamma$'s and particles decaying 
to $\gamma$'s: $\pi^0$, $\rho^\pm$, $\eta$, $\eta'$, etc. 
Neutral $K^0_L$ mesons are also efficiently reconstructed.
Detection of the decay products of one $B$ allows the flavor of
the other $B$ to be tagged.
\item Due to low track multiplicities and detector occupancy, the
reconstruction efficiency is high and the trigger bias is low. 
This substantially reduces corrections and systematic uncertainties
in many types of measurements, e.g., Dalitz plot analyses.
\item By utilizing asymmetric beam energies, the Lorentz boost $\beta$ 
of the $e^+e^-$ system can be made large enough such that a
$B$ or $D$ meson travels an appreciable distance before decaying.
This allows precision measurements of lifetimes, mixing parameters,
and CP violation (CPV). Note that measurement of the $D$ lifetime 
provides a measurement of the mixing parameter $y^{}_{CP}$, while 
measurement of the $B$ lifetime (which is already well measured) 
allows one to determine the decay time resolution function from data.
\item Since the absolute delivered luminosity is measured
with Bhabha scattering, an $e^+e^-$ experiment measures
{\it absolute\/} branching fractions. These are complementary
to {\it relative\/} branching fractions measured at
hadron colliders, and in fact are used to normalize
the relative measurements.
\item Since the initial state is completely known, one can 
perform ``missing mass'' analyses, i.e., infer the existence
of new particles via energy/momentum conservation rather
than reconstructing their final states.
By fully reconstructing a $B$ decay in one hemisphere 
of the detector, inclusive decays such as 
$B\to X^{}_s\ell^+\ell^-\!,\,X^{}_s\gamma$ 
can be measured in the opposite hemisphere. 
\item In addition to producing large samples of
$B$ and $D$ decays, an $e^+e^-$ machine produces
large sample of $\tau$ leptons. This allows one to
measure rare $\tau$ decays and search for forbidden 
$\tau$ decays with a high level of background rejection.
\end{itemize}\vspace*{-10pt}

To extend this physics program beyond the Belle and \babar\
experiments, the KEKB $e^+e^-$ accelerator at the KEK laboratory 
in Japan will be upgraded to SuperKEKB, and the Belle 
experiment will be upgraded to Belle~II.
The KEKB accelerator achieved a peak luminosity of
$2.1 \times 10^{34} \, {\rm cm^{-2} s^{-1}}$, and the
Belle experiment recorded a total integrated luminosity 
of 1040~fb$^{-1}$ (just over 1.0~ab$^{-1}$). The 
SuperKEKB accelerator plans to achieve a luminosity
of $8 \times 10^{35} \, {\rm cm^{-2} s^{-1}}$, and the
Belle~II experiment plans to record 50~ab$^{-1}$ 
of data by 2022. As $\sigma(e^+e^-\to b\overline{b})\approx 1.1$~nb 
at the $\Upsilon(4S)$ resonance, this data sample will 
contain $5\times 10^{10}$ $B\Bbar$ pairs. Such a
large sample will improve the precision 
of time-dependent CPV measurements and the sensitivity 
of searches for rare and forbidden decays. Systematic 
errors should also be reduced, as control samples from 
which many are calculated will substantially increase. 


A discussion of the complete physics program of 
Belle~II is beyond the scope of this summary. 
Here we touch upon only a few highlights. More
complete writeups can be found in Refs.~\cite{Aushev:2010bq}
and \cite{Bona:2007qt}; the latter was written in the context 
of the proposed --- but declined --- SuperB experiment in Italy.
The expected sensitivity of Belle~II in 50~fb$^{-1}$ of data
for various topical $B$ decays is listed in Table~\ref{tab:eeBsummary}.

\begin{table}[t]
\centerline{\begin{tabular}{c|c|c|c}
\hline\hline
\multirow{2}{*}{Observable} &  \multirow{2}{*}{SM theory} 
  &  Current measurement  &  Belle~II \\
  &    &  (early 2013)  &  ($50\, {\rm ab^{-1}}$) \\
\hline 
$S(B \to \phi K^0)$  &  $0.68$  &  $0.56 \pm 0.17$ & $ \pm 0.03$ \\
$S(B \to \eta^\prime K^0)$  & $0.68$  & $0.59 \pm 0.07$  & $\pm 0.02$ \\
$\alpha$ from $B \to \pi\pi,\,\rho\rho$  &  &  $\pm 5.4^\circ$ &  $\pm 1.5^\circ$ \\
$\gamma$ from $B \to D K$  &  &  $\pm 11^\circ$ &  $\pm 1.5^\circ$ \\
$S(B \to  K_S \pi^0 \gamma)$ & $< 0.05$  & $-0.15 \pm 0.20$ & $\pm 0.03$ \\
$S(B \to  \rho \gamma)$ & $<0.05$  & $-0.83 \pm 0.65$ & $\pm 0.15$  \\
$A_{\rm CP}(B \to  X_{s+d}\,\gamma)$  
  &  $<0.005$ &  $0.06 \pm 0.06$ & $\pm 0.02$  \\
$A_{\rm SL}^d$ & $-5 \times 10^{-4}$ & $-0.0049 \pm 0.0038$  & $\pm 0.001$ \\
\hline
${\cal B}(B \to  \tau \nu)$ 
& $1.1 \times 10^{-4}$   & $(1.64 \pm 0.34) \times 10^{-4}$ & $\pm 0.05 \times 10^{-4}$ \\
${\cal B}(B \to  \mu \nu)$ 
& $4.7 \times 10^{-7}$   & $< 1.0 \times 10^{-6}$ & $\pm 0.2 \times 10^{-7}$ \\
${\cal B}(B \to  X_s \gamma)$ 
& $3.15 \times 10^{-4}$  & $(3.55 \pm 0.26)\times 10^{-4}$ & $\pm 0.13 \times 10^{-4}$  \\
${\cal B}(B \to  K \nu \overline{\nu})$ 
& $3.6 \times 10^{-6}$  & $<1.3 \times 10^{-5}$ & $\pm 1.0 \times 10^{-6}$ \\
${\cal B}(B \to X_s  \ell^+ \ell^- )$  ($1 < q^2 < 6$\,GeV$^2$)
& $1.6 \times 10^{-6}$   & $(4.5 \pm 1.0) \times 10^{-6}$ & $\pm 0.10 \times 10^{-6}$  \\
$ A_{\rm FB}(B^{0}\to K^{*0}\ell^{+}\ell^{-})$ zero crossing & 7\% & 18\% & 5\%  \\ 
$|V^{}_{ub}|$ from $B\to\pi\ell^+\nu$ ($q^2>16$\,GeV$^2$) & 
$9\%\to 2\%$ & 11\% & 2.1\% \\
\hline\hline
\end{tabular}}
\caption{The expected reach of Belle~II with 50~ab$^{-1}$ of data
for various topical $B$ decay measurements. Also 
listed are the SM expectations and the current 
experimental results.  
For ${\cal B}(B \to X_s  \ell^+ \ell^- )$, the quoted 
measurement~\cite{Beringer:1900zz} 
covers the full $q^2$ range.
For $|V^{}_{ub}|$ and the $A_{\rm FB}$ zero crossing, we 
list the fractional errors.}
\label{tab:eeBsummary}
\end{table}

\begin{figure}[t]
\hspace*{1.5cm}
\includegraphics[width=0.32\textwidth, clip]{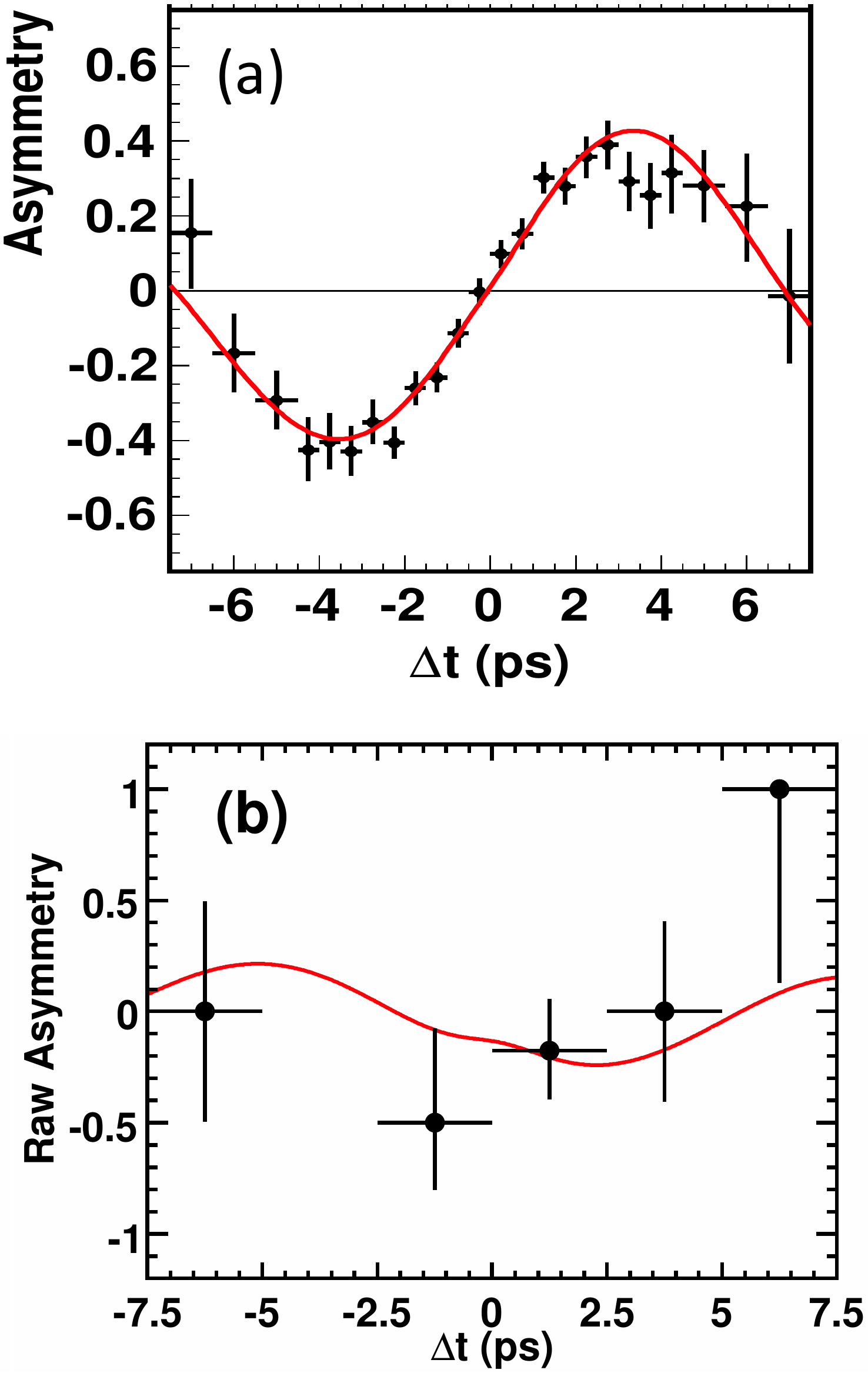}
\hspace*{1cm}
\vbox{
\includegraphics[width=0.46\textwidth, clip]{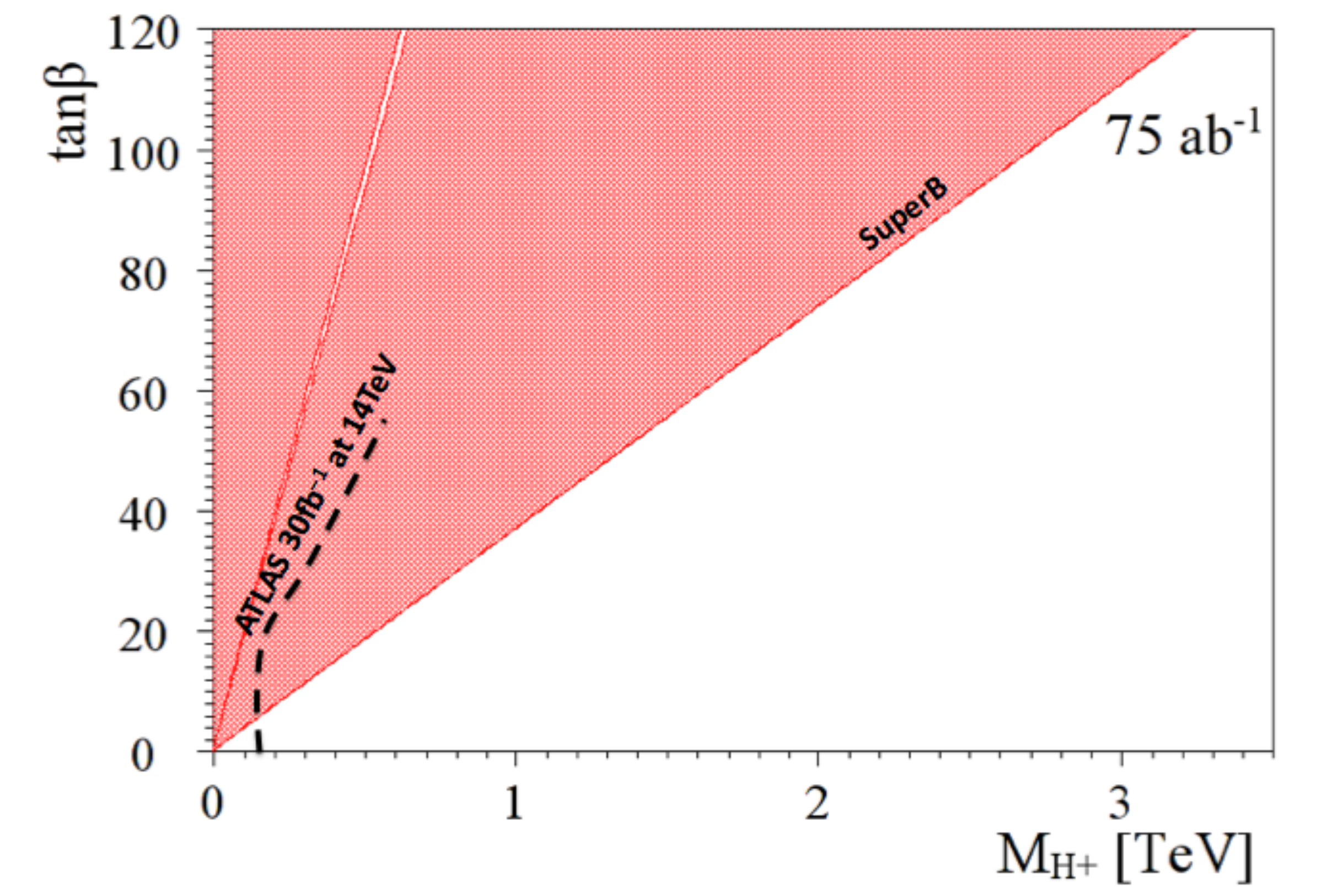}
\vspace*{1.5cm}
}
\caption{Left: Belle measurements of the 
time-dependent CP asymmetry versus $\Delta t$ for 
(a) $B \to J/\psi K^0$ and (b) $B \to K^0_S K^0_S$. 
The parameter $\sin(2\beta)$ is determined from the 
amplitude of the oscillations. Belle~II should obtain
statistics for  $B \to K^0_S K^0_S$ (and other 
loop-dominated modes) comparable to those obtained 
by Belle for $B \to J/\psi K^0$. 
Right: The expected constraint in $m^{}_H$ vs. $\tan\beta$ 
parameter space for a Type~II Higgs doublet model that would
result from 75~ab$^{-1}$ of data at a super-$B$-factory.
For comparison, also shown is the expected constraint 
from ATLAS in 30~fb$^{-1}$ of data.
}
\label{fig:belleII_prospects}
\end{figure}

As mentioned above, a main strength of a $B$ factory
experiment is the ability to make precision measurements
of CP violation, and this capability will be 
exploited to search for NP sources of CPV.
The difference between $B^0$ and $\B0bar$ decay rates 
to a common self-conjugate state
is sensitive to both direct CPV (i.e., occurring in 
the $B^0$ and $\B0bar$ decay amplitudes), and indirect CPV 
from interference between the 
$B\to f$ decay and 
$B\to\B0bar\to f$ mixing amplitudes. The indirect CPV 
was originally measured at Belle and \babar\ for all-charged final 
states such as $J/\psi K^0$~\cite{Adachi:2012et,Aubert:2009aw} 
(see Fig.~\ref{fig:belleII_prospects}, left) and 
$\pi^+\pi^-$~\cite{Ishino:2006if,Lees:2012kx};
at Belle~II, this measurement will be extended with
good statistics to more challenging final states such
as $B^0\to K^0_S K^0_S$ (Fig.~\ref{fig:belleII_prospects}, 
left, shows a first measurement by Belle), $B^0\to K^0\pi^0$,
and $B^0\to X_{s+d}\,\gamma$. The last mode proceeds via
electromagnetic $b \to s \gamma$ and $b \to d \gamma$ penguin amplitudes, where
$X_{s+d}$ represents the hadronic system in these decays.
In a fully inclusive measurement, the $\gamma$ is measured
but $X_{s+d}$ is not reconstructed.
In the SM there is a 
robust expectation that direct CP violation is negligible, i.e., 
the decay rates for $B$ and $\Bbar$ to  $X_{s+d}\,\gamma$ 
are equal. A measured difference would be a strong 
indication of NP, and differences of up to 10\% appear 
in some non-SM  scenarios. The best measurement with 
existing $B$-factory data is consistent with no difference 
and has a 7\% absolute error~\cite{Beringer:1900zz}. Belle~II 
should reduce this uncertainty to below~1\%.

Both Belle and \babar\ used the 
$b\to c\overline{c}s$ ``tree'' mode $B^0\to J/\psi\,K^0$ to 
measure the phase $\beta$ of the CKM unitary triangle 
to high precision: 
$\sin(2\beta) = 0.665 \pm 0.022$~\cite{Amhis:2012bh}.
However, this phase can also be measured in 
$b\to s\overline{s} s$ ``loop'' decays 
such as $B^0 \to \phi K^0$ and $B^0 \to \eta^\prime K^0$.
Since virtual NP contributions could compete with the SM loop
diagrams, these modes are sensitive to NP.
Comparing the values of $\sin(2\beta)$
measured in $b\to c\overline{c}s$ 
and in $b\to s\overline{s} s$ processes thus provides 
a way 
to search for NP. The decay $B^0 \to \eta^\prime K^0$ is 
the most precisely measured $b\to s\overline{s} s$ mode;
the value of $\sin(2\beta)$ obtained is
$0.59 \pm 0.07$~\cite{Amhis:2012bh}, about 
$1.2\sigma$ lower than that measured in 
$B^0\to J/\psi\,K^0$ decays. Belle~II is expected
to reduce this error by almost an order of magnitude,
making the test much more sensitive.

The $B^0\to K^0\pi^0$ CP asymmetry  is an important component of a sum rule
which holds in the isospin limit~\cite{Gronau:2005kz}
\begin{equation}
{\cal A}^{}_{K^+\pi^-}\, \frac{{\cal B}_{K^+\pi^-}}{\tau^{}_{B^0}}
+ 
{\cal A}^{}_{K^0\pi^+}\, \frac{{\cal B}_{K^0\pi^+}}{\tau^{}_{B^+}}
= 2 {\cal A}^{}_{K^+\pi^0}\, \frac{{\cal B}_{K^+\pi^0}}{\tau^{}_{B^+}}
+ 
2 {\cal A}^{}_{K^0\pi^0}\, \frac{{\cal B}_{K^0\pi^0}}{\tau^{}_{B^0}}\,,
\label{eqn:kpi_test}
\end{equation}
where ${\cal A}$ denotes a CP asymmetry, ${\cal B}$ 
a branching fraction, and $\tau$ 
a lifetime.
This sum rule is thought to be accurate 
to a few percent precision and provides a robust test 
of the SM. The limitation of the test is the precision 
of ${\cal A}^{}_{K^0\pi^0}$, which is difficult to 
measure and currently known to only
$\sim$\,14\% precision~\cite{Fujikawa:2008pk}. 
At Belle~II this is expected to be reduced to 
$\sim$\,3\% precision, greatly improving 
the sensitivity of Eq.~(\ref{eqn:kpi_test}) 
to~NP.

Numerous rare $B$ decays that were observed with low statistics
by Belle and \babar\ or not at all will become accessible
at Belle~II. One example is $B^+\to\tau^+\nu$, which in 
the SM results from a $W$-exchange diagram and has an 
expected branching fraction of 
$(0.76\,^{+0.10}_{-0.06}) \times 10^{-4}$~\cite{Charles:2011va}.
This mode is sensitive to supersymmetric models and others 
that predict the existence of a charged Higgs.
The final state contains multiple neutrinos and thus
is feasible to study only at an $e^+ e^-$ experiment.
The current average branching fraction from Belle and
\babar\ is $(1.15 \pm 0.23) \times 
10^{-4}$~\cite{Aubert:2009wt,Hara:2010dk,Lees:2012ju,Adachi:2012mm},
somewhat higher than the SM expectation. Belle~II should 
reduce this error to about $0.04 \times 10^{-4}$.  
The contribution of a charged Higgs boson within the
context of a Type II Higgs doublet model (e.g., which is also the tree level Higgs sector of the 
Minimal Supersymmetric Model) would increase the branching
fraction above the SM prediction by a factor
$1-(m^2_B/m^2_H)\tan^2\beta$, where $m^{}_H$ is the mass
of the charged Higgs and $\tan\beta$ is the ratio of
vacuum expectation values of up-type and down-type Higgses.
This relation can be used in conjunction with the measured
value of the branching fraction to constrain $m^{}_H$ and
$\tan\beta$. The expected constraint from a $B$-factory
experiment with 75~ab$^{-1}$ of data is shown in 
Fig.~\ref{fig:belleII_prospects} (right). One sees that 
a large region of phase space is excluded. For 
$\tan\beta \gsim 60$, the range $m^{}_H<2$~TeV/$c^2$ 
is excluded.


Other interesting processes include 
$b\to s\ell^+\ell^-$ and $b\to d\ell^+\ell^-$, 
with $\ell = e$ or $\mu$. These are also
sensitive to NP via loop diagrams. Belle~II will 
reconstruct a broad range of exclusive final states 
such as $B\to K^{(*)}\ell^+\ell^-$, from which one
can determine CP asymmetries, forward-backward asymmetries,
and isospin asymmetries (i.e., the asymmetry between
$B^+\to K^{(*)+}\ell^+\ell^-$ and
$B^0\to K^{(*)0}\ell^+\ell^-$). Belle~II will also measure 
inclusive processes such as $B\to X_{s+d}\,\ell^+\ell^-$,
for which theoretical predictions have less uncertainty
than those for exclusive processes. 
By running on the $\Upsilon(5S)$ resonance, Belle~II can
study $B^0_s$ decays. Topical decay modes include
$B^0_s\to D_s^{*+}D_s^{*-}$, $D_s^{*+}\rho^-$, and 
$B^0_s\to\gamma\gamma$, all of which are
challenging in a hadronic environment. 


The SuperKEKB project at KEK is well underway.  
Commissioning of the accelerator is expected to begin in 2015.
The high luminosity 
($8 \times 10^{35}$~cm$^{-2}$s$^{-1}$, 40 times larger than KEKB)
results mainly from a smaller $\beta^*$ function and reduced emittance.
As a result, the vertical beam spread at the interaction point
will shrink from $\sim$\,2~$\mu$m at KEKB to $\sim$\,60~nm 
at SuperKEKB. In addition, the beam currents will be approximately
doubled, and the beam-beam parameter will be increased by~50\%.

The Belle~II detector will be an upgraded version of the
Belle detector that can handle the increased backgrounds associated 
with higher luminosity.  The inner vertex detector will
employ DEPleted Field Effect (DEPFET) pixels located 
inside a new silicon strip tracker employing the APV25 
ASIC (developed for CMS) to handle the large rates.
There will also be a 
new small-cell drift chamber.  The particle identification
system will consist of an ``imaging-time-of-propagation'' (iTOP)
detector in the barrel region, and an aerogel-radiator-based 
ring-imaging Cherenkov 
detector in the forward endcap region.  The iTOP operates
in a similar manner as \babar's DIRC detector, except that 
the photons are focused with a spherical mirror onto a 
finely segmented array of multi-channel-plate (MCP) PMTs.
These MCP PMTs provide precise timing, which significantly
improves the discrimination power between pions and kaons
over that provided by imaging alone. The CsI(Tl) calorimeter
will be retained but instrumented with waveform sampling 
readout. The innermost layers of the barrel $K^0_L$/$\mu$~detector,
and all layers of the endcap $K^0_L$/$\mu$~detector,
will be upgraded to use scintillator in order to accommodate the 
higher rates.  Belle~II should be ready to roll in by the
spring of 2016 after commissioning of SuperKEKB is completed.
The U.S.\ groups on Belle~II are focusing their efforts on
the iTOP and $K^0_L$/$\mu$ systems.

\subsection{Physics Potential of Hadronic Experiments}

\subsubsection*{LHCb and its Upgrade}

The spectular successes of LHCb have realized some of the great potential
for studying the decays of particles containing $c$ and $b$ quarks at hadron 
colliders.
The production cross sections are quite
large.  
More than 100~kHz of
$b$-hadrons within the detector acceptance can be produced
even at reduced LHC luminosities (4$\times$10$^{32}$ cm$^{-2}$s$^{-1}$).
This is a much higher production rate
than can be achieved even in the next generation $e^{+}e^{-}$ $B$ factories. 
All species of $b$-flavored hadrons, including $ B_{s} $ and $ B_{c} $ mesons, and $b$
baryons, are produced.  However, compared to $e^{+}e^{-}$ colliders,
the environment is much more harsh for experiments. At hadron
colliders, the $b$ quarks are accompanied by a very high rate of background
events; they are produced over a very large range of momenta and angles; and
even in $b$-events of interest there is a complicated underlying event. The
overall energy of the center of mass of the hard scatter that produces the $b$
quark, which is usually from the collision of a gluon from each beam particle,
is not known, so the overall energy constraint that is so useful in
$e^{+}e^{-}$ colliders is not available. These features translate into
challenges in triggering, flavor tagging, and photon detection 
and they limit the overall efficiency. 

The CDF and D0 experiments at the Fermilab Tevatron demonstrated that these
problems could be successfully addressed using precision silicon vertex
detectors and specialized triggers.  While these experiments were mainly
designed for high-$p_{T}$ physics, they made major contributions to
bottom  and charm physics~\cite{cdf_b_results, d0_b_results}.  

The LHC produced its first collisions at 7 TeV center of mass energy at the end
of March 2010. The $b$ cross section at the LHC is  $\sim300\mu$b, a factor of three
higher than at the Tevatron and approximately 0.5\% of the inelastic cross
section. When the LHC reaches its design center of mass energy of 14 TeV in
2015, the cross section will be a factor of two higher.

The LHC program features for the first time at a hadron collider a  dedicated
$B$-physics experiment, LHCb~\cite{Alves:2008zz}.  LHCb covers the forward
direction from about 10 mr to 300 mr with respect to the beam line. $B$ hadrons
in the forward direction are produced by collisions of gluons of unequal energy,
so that the center of mass of the collision is Lorentz boosted in the direction
of the detector. Because of this, the $b$-hadrons and their decay products are
produced at small angles with respect to the beam and have momenta ranging from
a few GeV/$c$ to more than a hundred GeV/$c$. Because of the Lorentz boost, even though
the angular range of LHCb is small, its coverage in pseudorapidity is between 
about 2 -- 5, and both $b$~hadrons travel in the same direction, making $b$
flavor tagging possible.  With the small angular coverage, LHCb can stretch out
over a long distance along the beam without becoming too large transversely. A
silicon microstrip vertex detector (VELO) 
only 8~mm from the beam
provides precision tracking that enables LHCb to separate
weakly decaying particles from particles produced at the interaction vertex.
This allows the measurement of lifetimes and oscillations due to flavor mixing.
A 4 Tm dipole magnet downstream of the collision region, in combination with the
VELO, large area silicon strips (TT) placed downstream of the VELO but upstream
of the dipole,  and a combination of silicon strips (IT) and straw tube chambers
(OT) downstream of the dipole provides a magnetic spectrometer with excellent
mass resolution. There are two  ring imaging Cherenkov (RICH) counters, one upstream of
the dipole and one downstream, that together provide $K$--$\pi$ separation from
2 to 100 GeV/$c$.  An electromagnetic calorimeter (ECAL) follows the tracking
system and provides electron triggering and $\pi^{0}$ and $\gamma$
reconstruction. This is followed by a hadron calorimeter (HCAL) for triggering
on hadronic final states. A  muon detector at the end of the system provides muon
triggering and identification. 

LHCb has a very sophisticated trigger system \cite{Aaij:2012me}
that uses hardware at the lowest
level (L0) to process the signals from the ECAL, HCAL and muon systems. The L0
trigger reduces the rate to $\sim$1 MHz followed by  the High Level Trigger
(HLT), a large computer cluster, that reduces the rate to $\sim$3 kHz for
archiving to tape for physics analysis. LHCb is able to run at a luminosity of
4.0$\times$10$^{32}$ cm$^{-2}$s$^{-1}$. This is much smaller than the current peak
luminosity achieved by the LHC and only a few percent of the LHC design luminosity. 
The luminosity that LHCb can take efficiently  is currently limited by the 1 MHz
bandwidth between the Level 0 trigger system and the trigger cluster. Therefore,
the physics reach of LHCb is determined by the detector capabilities and not by
the machine luminosity. In fact, LHC implemented a ``luminosity leveling''
scheme in the LHCb collision region so that LHCb could run at its desired
luminosity throughout the store  while the other experiments, CMS and ATLAS,
could run at higher luminosities.  This mode of running will continue until 2017
when a major upgrade~\cite{Bediaga:2012py,lhcb_upgrade_tdr} 
of the LHCb trigger and parts of the detector and front end
electronics will increase the bandwidth to the HLT, increase archiving rate to 20 kHz,
and permit operation at a factor of 10 higher luminosity. 
Several subdetectors will be rebuilt for more robust performance at higher 
luminosities, including VELO (pixels), TT (finer strips), 
IT+OT (technology to be soon decided) and RICH (redesigned optics, MaPMTs).   

There have been three runs of the LHC. In the first ``pilot" run in 2010, LHCb
recorded 35 pb$^{-1}$, which was enough to allow it to surpass in precision many
existing measurements of $B$ decays. In 2011, the LHC delivered more than 5
fb$^{-1}$  to CMS and ATLAS. Since this luminosity was more than LHCb was
designed to handle, the experiment ran at a maximum luminosity that was 10\% of
the LHC peak luminosity. The total integrated luminosity was about 1 fb$^{-1}$.
In 2012 LHC delivered 20 fb$^{-1}$ to CMS and ATLAS with additional  2 fb$^{-1}$
collected by LHCb.  Until the LHCb upgrade is installed in the long shutdown
planned in 2018, LHCb plans to run at a luminosity of 4.0 $\times
10^{32}$cm$^{-2}$s$^{-1}$.  Between now and then, LHCb will accumulate about
1--2~fb$^{-1}$ per operating year, so a total of about 6.5 fb$^{-1}$ will be
obtained. The sensitivity will increase by more than this because the LHC will
run  at 14 TeV, with about a factor of two higher $B$ cross section.  After the
upgrade is installed, LHCb will integrate about 5 fb$^{-1}$ per year, so that
about 50 fb$^{-1}$ will be obtained over the decade following the upgrade
installation. 


The decay $B_{s}\to J/\psi \phi$ has been used to measure the CKM angle
$\phi_{s} (\equiv - 2\beta_s)$~\cite{LHCb:2013oba}. The result, using also the
decay mode  $B_{s}\to J/\psi f_{0}$ \cite{Stone:2008ak} 
first established by LHCb~\cite{LHCb:2012ae}, is
$\phi_{s} = 0.01 \pm 0.07 \pm 0.01 \ {\rm rad}$~\cite{LHCb:2013oba}. The
difference in the width of the CP-even and CP-odd $B_{s}$ mesons is $\Delta
\Gamma_{s} = (0.106 \pm 0.011 \pm 0.007)$ ps$^{-1}$. These results are
consistent with the SM, resolving a slight tension with earlier measurements
from the Tevatron, which deviated somewhat from the SM predictions. 
However, the experimental uncertainty on $\phi_s$ is still a factor of 40 larger than
that on the SM prediction, so improved measurements will probe higher mass scales of 
possible NP contributions.  

\begin{figure}[tb]
\centerline{\includegraphics[width=0.65\textwidth, clip]{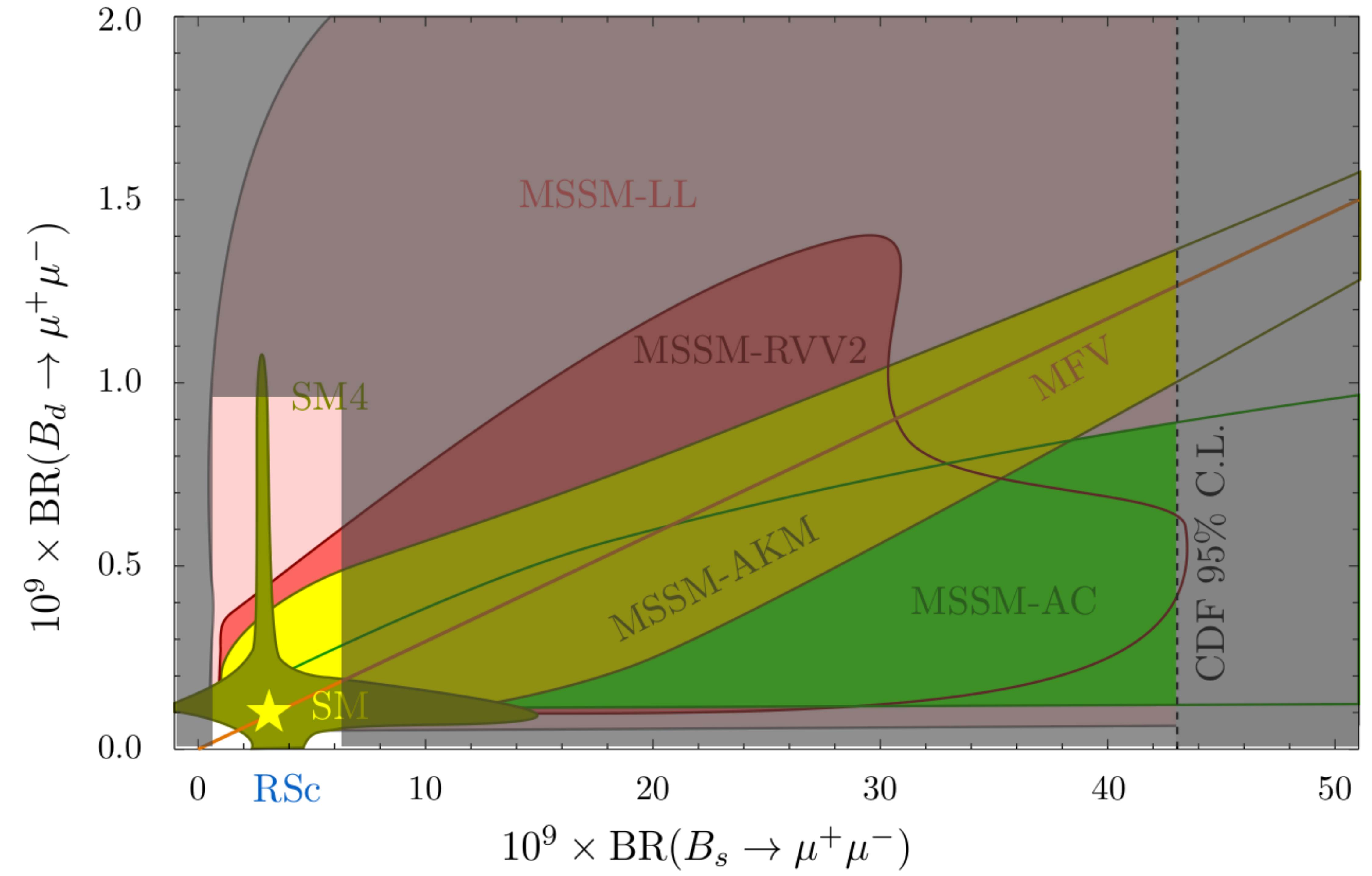}}
\caption{Correlation between the branching ratios of $B_s^0\to\mu^+\mu^-$ and
$B_d^0\to\mu^+\mu^-$ in various models. The SM point is marked by a star. From
Ref.~\cite{Straub:2012jb} with the $2.1 \, {\rm fb^{-1}}$ 
correction LHCb result~\cite{Aaij:2012nna} superimposed.}
\label{fig:bsmmbdmm}
\end{figure}

The rare decay $B_{s} \to \mu^{+}\mu^{-}$ is predicted in the SM  to
have a branching fraction $(3.54 \pm 0.30) \times 10^{-9}$~\cite{Buras:2012ru, DeBruyn:2012wk}. 
A higher or lower branching fraction would be an indicator for NP. 
LHCb presented first evidence of this decay based on 
$2.1 \, {\rm fb^{-1}}$ of data,  
with ${\cal B}(B_s\to \mu^+\mu^-)=(3.2^{+1.5}_{-1.2})\times 10^{-9}$ 
\cite{Aaij:2012nna} consistent with the SM prediction. 
LHCb has recently updated its result for $3 \, {\rm fb^{-1}}$ of data
to  ${\cal B}(B_s\to \mu^+\mu^-)=(2.9^{+1.1}_{-1.0})\times 10^{-9}$~\cite{Aaij:2013aka}.
(CMS has also recently reported a measurement of
${\cal B}(B_s\to \mu^+\mu^-)=(3.0^{+1.0}_{-0.9})\times 10^{-9}$~\cite{Chatrchyan:2013bka}.)
LHCb has also set an upper limit on ${\cal B}(B_d\to \mu^+\mu^-)< 0.74\times 10^{-9}$ (95\%\ C.L.)
with the $3 \, {\rm fb^{-1}}$ data set.
These measurements impose stringent constraints on SUSY models 
as illustrated in Fig.~\ref{fig:bsmmbdmm}.   
Further increase in statistics will probe even higher energy scales. 

LHCb has also produced results on the key decay $B^{0}\to K^{*0}\mu^{+}\mu^{-}$
(1.1 fb$^{-1}$) \cite{Aaij:2013iag} that could reveal evidence for NP.   One of
the interesting observables is the forward-backward asymmetry of the $\mu^{-}$
relative to the direction of the parent $B^{0}$ meson in the dimuon center of
mass vs.\  $q^{2}$ (dimuon invariant mass).  The SM prediction crosses zero
within a well-determined narrow region of $q^2$, due to the interference between
the SM box and electroweak penguin diagrams. NP can remove the crossover or
displace its location. Indications from low statistics at Belle, \babar, and CDF
seemed to indicate that this might be happening. The LHCb results are the
most precise so far, and are in good agreement with the SM within errors, which
however can be significantly reduced with the LHCb upgrade.  Many other
observables sensitive to NP have also been investigated. The CMS (5.2 fb$^{-1}$
\cite{CMSKstarmumu}) and ATLAS (4.9 fb$^{-1}$ \cite{ATLASKstarmumu})
Collaborations have also performed such studies. The results agree with the SM
and the previous measurements, but have larger errors than LHCb.  


Many other decays are being studied, including all-hadronic decays such 
as $B_{s}\to \phi\phi$~\cite{Aaij:2013qha} 
($\beta_s^{\rm eff}$ via interference of mixing and decay via gluonic penguin)
$B\to D\pi$, $B\to DK$ (determination of $\gamma$ from tree processes), and
states with photons such as $B_{s}\to \phi \gamma$ (search for right-handed
currents).  The expected sensitivity to selected important $B$ decays during the
present and upgraded phases of the LHCb experiment is shown in
Table~\ref{hadronB_tab_1}.
In addition, LHCb has also demonstrated the capability for a rich program of
studies of $B_c$ meson decays.

The physics output of LHCb also extends beyond its $B$ and charm (see next
section) core programs. Examples of other topics include measurements of the
production of electroweak gauge bosons in the  forward kinematic region covered
by the LHCb acceptance \cite{Aaij:2012vn}, studies of double parton scattering
\cite{Aaij:2012dz}, measurements of the properties of exotic hadrons
\cite{Aaij:2012pz,Aaij:2013zoa}, searches for lepton number and lepton flavor
violations \cite{Aaij:2012zr,Aaij:2013fia} and for long-lived new particles
\cite{Heijne:2012qia}.

\begin{table}[tb]
\centering
\begin{tabular}{c|c|c|c|c}   \hline\hline
\multirow{2}{*}{Observable} & Current SM  & Precision  & LHCb & LHCb Upgrade \\
	& theory uncertainty  & as of 2013 & (6.5 fb$^{-1}$) & (50 fb$^{-1}$) \\
\hline
$2\beta_s(B_{s}\to J/\psi \phi)$ &   $\sim0.003$ & 0.09 & 0.025 & 0.008 \\
$\gamma(B\to D^{(*)}K^{(*)})$ & $<1^\circ$ & $8^\circ$ & $4^\circ$ & $0.9^\circ$  \\
$\gamma(B_s\to D_{s}K)$ & $<1^\circ$ & --- &  $\sim11^\circ$ & $2^\circ$ \\ 
$\beta(B^{0}\to J/\psi K^{0}_S)$ &  small & $0.8^\circ$ & $0.6^\circ$ & $0.2^\circ$  \\ 
\hline
$2\beta_s^{\rm eff}(B_{s}\to \phi\phi)$ & 0.02 & 1.6  & 0.17  & 0.03 \\
$2\beta_s^{\rm eff}(B_{s}\to K^{*0}\bar{K}^{*0})$  & $<0.02$ & --- & 0.13 & 0.02  \\
$2\beta_s^{\rm eff}(B_{s}\to \phi\gamma)$ & 0.2\% & --- & 0.09 & 0.02 \\ 
$2\beta^{\rm eff}(B^{0}\to \phi K^{0}_{S})$ & 0.02 & 0.17 & 0.30 & 0.05 \\
$A_{\rm SL}^s$ & $0.03\times10^{-3}$ & $6\times10^{-3}$ & $1\times10^{-3}$ & $0.25\times10^{-3}$ \\
\hline 
${\cal B}(B_{s}\to \mu^{+}\mu^{-}) $ & 8\%  & 36\% & 15\% & 5\% \\
${\cal B}(B^{0}\to \mu^{+}\mu^{-}) / {\cal B}(B_{s}\to \mu^{+}\mu^{-})$
  & 5\%  & --- & $\sim$100\% &  $\sim$35\% \\
$A_{\rm FB}(B^{0}\to K^{*0}\mu^{+}\mu^{-})$ zero crossing & 7\% & 18\% & 6\% & 2\% \\ 
\hline\hline
\end{tabular}
\caption{Sensitivity of LHCb to key observables. The current sensitivity (based
on 1--3\,fb$^{-1}$, depending on the measurement) is compared to that expected
after 6.5 fb$^{-1}$ and that achievable with 50 fb$^{-1}$ by the upgraded
experiment assuming $\sqrt{s} = 14$\,TeV.  Note that at the upgraded LHCb, the
yield per fb$^{-1}$, especially in hadronic $B$ and $D$ decays, will be higher
on account of the software trigger. (Adapted from Ref.~\cite{Bediaga:2012py}.)}
\label{hadronB_tab_1}
\end{table}



\subsubsection*{ATLAS and CMS}

Two LHC detectors, CMS and ATLAS, are designed to explore high mass and
high-$p_{T}$ phenomena to look for new physics at the LHC. 
They must operate at luminosities of up to $10^{34}\,$cm$^{-2}$s$^{-1}$, which
implies the need to handle an average event pileup of $\sim$20. 
%
Both experiments can implement muon triggers with relatively low thresholds of a
few GeV/$c$.
However, the rate of low-$p_{T}$ muons from $B$ decays competes for scarce resources with 
the many other trigger signatures that could contain direct evidence of new physics.
Thus in practice, only $B$ final states containing dimuons are well preserved through the trigger pipelines.
The trigger efficiency is lower than in LHCb but at higher luminosity.
One example of this, discussed above, is the rare
decay $B_{d,s}\to \mu^{+}\mu^{-}$.
CMS has very recently reported a measurement of
${\cal B}(B_s\to \mu^+\mu^-)=(3.0^{+1.0}_{-0.9})\times 10^{-9}$~\cite{Chatrchyan:2013bka}.
If ATLAS and CMS can maintain their
trigger efficiency as the LHC luminosity and energy increase,
they can 
be competitive in this study. The decay $B^{0}\to
K^{*}\mu^{+}\mu^{-}$ presents more problems. The muons are softer and more
difficult to trigger on and the limited $K$--$\pi$ separation increases the
background to the $K^{*}$. However, as illustrated by their preliminary results 
these two experiments can play a
confirming role to LHCb in this study. Despite their limitations, these two experiments
will collect large numbers of $B$ decays and should be able to observe many new
decay modes and new particles containing $b$ and charm
quarks.

\section{Report of the Charm Task Force}


\subsection{Introduction to Charm Physics}

Studies of charm quarks can be split into two broad categories.  
First, in indirect searches for new physics affecting decays 
and oscillations, charm quarks furnish a unique probe of flavor physics 
in the up-quark sector, complementing strange and bottom physics.  
Second, as a probe of quantum chromodynamics (QCD)
charm aids our understanding of nonperturbative physics, 
since it is not much heavier than the characteristic scale 
$\Lambda \sim 1$\,GeV of QCD.  
Overall, charm adds much to the core new physics thrusts in heavy 
flavor physics while also adding significant breadth to the program.  
 
Charm physics measurements allow for direct determination of 
the Cabibbo-Kobayashi-Maskawa (CKM) matrix elements 
$|V_{cs}|$ and $|V_{cd}|$, can also help improve 
the accuracy of $|V_{cb}|$ and $|V_{ub}|$ determined from $B$ decays,
and $|V_{ts}|$ and $|V_{td}|$ from $B^0$ and $B_s^0$ mixing.  
Part of this richness is due to the usefulness of charm data in 
verifying lattice QCD (LQCD) results.  

Indirect searches for new physics with charm quarks provide competitive 
as well as complementary constraints to the results of direct searches 
at the Energy Frontier.  
One can classify searches in three broad categories, according to their 
``standard model background."  

\begin{enumerate}\vspace*{-10pt}\itemsep2pt
\item 
Searches in the processes that are allowed in the standard model. 

New physics contributions may often be difficult to discern in this case, 
except in cases of sufficient theoretical precision 
(e.g., leptonic decays of $D$ mesons, $D_q \to \ell \overline{\nu}$).   
Alternatively, testing relations that are only valid 
in the standard model, but not in BSM models, may prove advantageous; 
e.g., CKM triangle relations.  

\item
Searches in the processes that are forbidden in the standard model 
at tree level.

Flavor-changing neutral current (FCNC) interactions occur in the standard 
model only through loops and are therefore suppressed.  
New physics contributions can enter both at tree-level and from one-loop 
corrections.  
Examples include $D^0-\D0bar$ mixing, or inclusive and exclusive 
transitions mediated by $c \to u \gamma$ or  $c \to u \ell \overline{\ell}$. 

Searches for CP violation in charm decays and oscillations should be 
included here as well, as they require at least two different pathways 
to reach the final state, at least one of which is FCNC transition.  

\item
Searches in the processes that are forbidden in the standard model.

While these processes are generally very rare even in NP models, 
their observation, however, would constitute a high-impact discovery.  
Examples include searches for lepton- and baryon-number-violating transitions, 
such as $D^0 \to e^+\mu^-$, $D^0 \to \overline{p} e^+$, etc.
\end{enumerate}\vspace*{-10pt}

The QCD side of charm physics is also very vibrant.  
Recently, there has been much activity in XYZ state spectroscopy, 
in addition to continued studies of conventional charmonium.  
This provides a rich source of results 
in hadronic physics and radiative transitions.

\subsection{Current and Future Experiments}

Over the past decade, charm results have been dominated by results from 
detectors at the $e^+e^-$ flavor factories: BaBar, Belle, and CLEO-c.  
Currently, the BESIII experiment is running at charm threshold 
and Belle~II, which will run at and near the $\Upsilon(4S)$, 
is under construction;  both experiments have excellent capabilities 
in charm~\cite{Asner:2008nq, Aushev:2010bq}.  
While charm statistics are lower at threshold, the data are unique 
in their ability to measure strong phases and also excel at modes 
with neutrinos in the final state.  
The Belle~II detector should begin physics running in 2017;  
charm from continuum fragmentation at $B$-factory energies 
is complementary to threshold data.  

Hadron machines also provide important results. 
CDF was able to contribute due to displaced-vertex and 
muon triggers, producing notable results on $D^0-\D0bar$ oscillations.  
While muon triggers have produced some charm results from ATLAS and CMS, 
the current and future charm program at hadron colliders lies almost 
exclusively with the dedicated flavor experiment, LHCb.  
Many areas of charm physics are accessible at LHCb and the 2018 
upgrade will enhance opportunities even  more.  
Their physics reach~\cite{Bediaga:2012py, Gersabeck:2012hz} 
is an important addition to the $e^+e^-$ program.  

The BESIII program should continue at least until the end of the decade, 
and Belle~II and LHCb will carry charm physics well into the 2020's.  
One major decision point is the future of threshold charm after BESIII.  
Currently, the Cabibbo Lab near Rome is preparing a threshold  tau-charm 
factory proposal.  Interest has also been expressed by BINP at Novosibirsk 
and institutions in Turkey.  

Another source of information on charm could come from fixed target experiments,
of which the only currently approved example is PANDA~\cite{panda_fair} at the
FAIR facility at Darmstadt,  which will collide antiprotons in a storage ring
with gas, solid, or liquid targets. The ability of that experiment to contribute
will depend on the cross section for charm production by low energy antiprotons,
a quantity that has not been measured and whose theoretical estimates vary from
1$\mu$b to 10$\mu$b, and the amount of time dedicated to the charm program,
which competes with other aspects of the program that require the machine to
operate below or close to the bare charm production threshold.

\subsubsection*{Leptonic and Semileptonic Decays and CKM triangle relations}

In leptonic and semileptonic decays, all of the uncertainties from 
strong-interaction effects may be conveniently parametrized as 
decay constants and form factors, respectively.  
The remainder of the theory is straightforward weak-interaction physics.  
Indeed, comparing decay constants and form factors to LQCD predictions allows 
one to exclude large portions of parameter space for NP models with charged 
scalars.  

Leptonic decay rates depend on the square of both decays constants 
and CKM matrix elements.  
If one uses LQCD as in input, then $|V_{cq}|$ may be extracted.  
If the CKM matrix elements are taken from elsewhere (possibly unitarity 
constraints), then we can test LQCD results.  
In fact, by taking ratios of leptonic and semileptonic decays, 
one can cancel $|V_{cq}|$ to obtain pure LQCD tests.  

The Cabibbo-suppressed leptonic decay $D^+ \to \mu \nu$ is only 
measurable at threshold charm machines.  
Currently, it is essentially determined via one CLEO-c 
result~\cite{Eisenstein:2008aa}, 
although BESIII has a preliminary result based on a dataset 
3.5 times larger~\cite{Rong:2012pb}.  
This result, $f_D = (203.91 \pm 5.72 \pm 1.91)$ MeV, based on 2.9 fb$^{-1}$, 
is still statistics-limited.  

The Cabibbo-favored $D^+_s \to \mu \nu,\ \tau \nu$ process is easier 
in two respects.  
Unlike the $D^+$ case, where $\tau \nu$ is a relatively small effect, 
here it offers additional channels that enhance the utility of a dataset.  
In addition, $B$ factories possess enough tagging power in continuum 
charm production to make the best current single measurement.  
The one drawback is that $D_s$ production rates are smaller than $D^+$.  
Currently, the best measurement of $f_{D_s}$ is a preliminary result 
from Belle~\cite{Zupanc:2012cd}.  

Successful LQCD calculations of $D_{(s)}$ decay constants will give 
confidence in their results for $B$ decay constants.  
And while $f_B$ can be obtained from $B \to \tau \nu$, there is no 
analogous direct way to determine $f_{B_s}$.  
By contrast, in charm, both strange and non-strange decay constants 
are directly accessible.  

The key semileptonic modes are $D^0 \to K^- e^+ \nu,\ \pi^- e^+ \nu$.  
Additional statistical power may be obtained by including the isospin-related
$D^+$ decays, but both CKM matrix elements are accessible without  the need for
the more experimentally challenging $D_s$ decays.   The form factors, $f_K(q^2)$
and $f_\pi(q^2)$, are useful tests of LQCD.   One depends on similar LQCD
calculations to extract $|V_{ub}|$ from  $B \to \pi \ell \nu$ decays.  

For leptonic charm decays, $f_{D_{(s)}}$ parametrizes the probability that the 
heavy and light quarks ``find each other'' to annihilate.  Due to helicity
suppression the rate goes as $m_\ell^2$, and many NP models could have a
different parametric dependence on $m_\ell^2$.  New physics can be discussed in
terms  of generalized couplings~\cite{Kronfeld:2009cf}.  Models probed by this
decay include extended Higgs sectors, which contain new charged scalar states,
or models with broken left-right symmetry, which include heavy vector $W^\pm_R$
states.  

One can also search for new physics by testing relations that hold in the SM, 
but not necessarily in general.  
An example of such relation is a CKM ``charm unitarity triangle" relation:   
\begin{equation}\label{CharmTriangle}
V_{ud}^* V_{cd} + V_{us}^* V_{cs} + V_{ub}^* V_{cb} = 0 \,.
\end{equation}
Processes that are used to extract CKM parameters in Eq.~(\ref{CharmTriangle}) 
can be affected by new physics.  This can lead to disagreement between CKM 
elements extracted from different processes, or the triangle not closing.  
Finally, since all CP-violating effects in the flavor sector of the SM 
are related to the single phase of the CKM matrix, all of the CKM unitarity 
triangles, 
have the same area, $A = J/2$, where $J$ is the Jarlskog invariant.  
This fact could provide a non-trivial check of the standard model, 
given measurements of more than one triangle with sufficient accuracy.  
Unfortunately, the charm unitarity triangle will be harder to work with than the 
familiar $B$ physics triangle since it is rather ``squashed."  
In terms of the Wolfenstein parameter $\lambda=0.22$, the relation 
in Eq.~(\ref{CharmTriangle}) has one side ${\cal O}(\lambda^5)$ 
with the other two being ${\cal O}(\lambda)$.

\subsubsection*{\boldmath $D^0$ Oscillations (including CP Violation)}

The presence of $\Delta C = 2$ operators produce off-diagonal terms in the 
$D^0-\D0bar$ mass matrix, mixing the flavor eigenstates into the mass 
eigenstates  
\begin{equation}
|D_{1, 2} \rangle = p | D^0 \rangle \pm q | \D0bar \rangle \,.
\end{equation}
Neglecting CP violation leads to $|p|=|q|=1/\sqrt{2}$.
The mass and width splittings between the mass eigenstates are 
\begin{equation}\label{XandY}
x= \frac{m_1-m_2}{\Gamma_D}\,, \qquad y=\frac{\Gamma_1-\Gamma_2}{2 \Gamma_D}\,,
\end{equation}
where $\Gamma_D$ is the average width of the two mass eigenstates.  

The oscillation parameters $x$ and $y$ are both of order 1\% in the $D^0$ 
system.  These small values require the high statistics of $B$ factories  and
hadron machines.  Observations thus far have relied on the time-dependence of
several hadronic decays $K\pi,\ K\pi\pi^-,\ K_S\pi\pi$, etc., as well as
lifetime differences between CP-eigenstate decays ($KK,\ \pi\pi$)  and the
average lifetime (see the review in \cite{Beringer:1900zz}).  LHCb has made the
highest significance ($9\sigma$) observation of $D^0$ oscillations in a single
experiment~\cite{Aaij:2012nva}.  However, a non-zero value of $x$ has not yet
been established at $3\sigma$. LHCb and Belle II will be
able to pinpoint the value of $x$ in the next several years.  

Theoretical predictions for $x$ and $y$ in the SM are uncertain, although values
as high as 1\% had been expected~\cite{Falk:2001hx}. The predictions need to be
improved, and several groups are working to understand the problem using
technology such as the heavy-quark expansion, lattice QCD, and other
long-distance methods.

However, one can place an upper bound on the NP parameters by neglecting the  SM
contribution altogether and assuming that NP saturates the experimental 
result.   One subtlety is that the SM and NP contributions can have either the
same  or opposite signs.   While the sign of the SM contribution cannot be
calculated reliably due to  hadronic uncertainties, $x$ computed within a given
NP model can be  determined.  This stems from the fact that NP contributions
are generated by heavy degrees of freedom, making the short-distance
calculation reliable. 

Any NP degree of freedom will generally be associated with a generic heavy 
mass scale $\Lambda$,  
at which the NP interaction is most naturally described.  
At the scale $m_c$, this description must be modified by the effects of QCD.  
In order to see how NP might affect the mixing amplitude, it is instructive 
to consider off-diagonal terms in the neutral D mass matrix,
\begin{equation}\label{M12}
 M_{12} - \frac{i}{2}\, \Gamma_{12} =
 \frac{1}{2 M_{\rm D}} \langle \D0bar | 
{\cal H}_w^{\Delta C=-2} | D^0 \rangle 
+  \frac{1}{2 M_{\rm D}} \sum_n {\langle \D0bar | {\cal H}_w^{\Delta
  C=-1} | n \rangle\, \langle n | {\cal H}_w^{\Delta C=-1} 
| D^0 \rangle \over M_{\rm D}-E_n+i\epsilon}\,,
\end{equation}
where the first term contains ${\cal H}_w^{\Delta C=-2}$, which is an 
effective $|\Delta C| = 2$ Hamiltonian, represented by a set of operators 
that are local at the $\mu \sim m_D$ scale. 
This first term only affects $x$, but not $y$. 

As mentioned above, heavy BSM degrees of freedom cannot be produced in charm 
meson decays, but can nevertheless affect the effective $|\Delta C| = 2$ 
Hamiltonian by changing Wilson coefficients and introducing new operator 
structures.  By integrating out those new degrees of freedom associated 
with a high scale $\Lambda$,  
we are left with an effective Hamiltonian written 
in the form of a series of operators of increasing dimension.  
It turns out that a model-independent study of NP $|\Delta C| = 2$ 
contributions is possible, as any NP model will only modify Wilson 
coefficients of those operators~\cite{Golowich:2007ka,Gedalia:2009kh},
\begin{equation}
{\cal H}_{\rm NP}^{|\Delta C| = 2} = \frac{1}{\Lambda^2} 
\sum_{i=1}^8 C_i (\mu)\, Q_i (\mu) \,,
\qquad
\begin{array}{l}
Q_1 = (\overline{u}_L^\alpha \gamma_\mu c_L^\alpha) \,
(\overline{u}_L^\beta \gamma^\mu c_L^\beta)\,, \\
Q_2 = (\overline{u}_R^\alpha c_L^\alpha) \,
(\overline{u}_R^\beta c_L^\beta)\,, \\
Q_3 = (\overline{u}_R^\alpha c_L^\beta) \,
(\overline{u}_R^\beta c_L^\alpha) \,, \\
Q_4 = (\overline{u}_R^\alpha c_L^\alpha) \,
(\overline{u}_L^\beta c_R^\beta) \,,
\end{array}
\qquad
\begin{array}{l}
Q_5 = (\overline{u}_R^\alpha c_L^\beta) \,
(\overline{u}_L^\beta c_R^\alpha) \,, \\
Q_6 = (\overline{u}_R^\alpha \gamma_\mu c_R^\alpha) \,
(\overline{u}_R^\beta \gamma^\mu c_R^\beta)\,, \\
Q_7 = (\overline{u}_L^\alpha c_R^\alpha) \,
(\overline{u}_L^\beta c_R^\beta)\,, \\
Q_8 = (\overline{u}_L^\alpha c_R^\beta) \,
(\overline{u}_L^\beta c_R^\alpha) \,,
\end{array}
\label{SetOfOperators}
\end{equation}
where $C_i$ are dimensionless Wilson coefficients, and the $Q_i$ 
are the effective operators; $\alpha$ and $\beta$ are color indices. 
In total, there are eight possible operator structures contributing 
to $|\Delta C|=2$ transitions.  
Taking operator mixing into account, a set of constraints 
on the Wilson coefficients of Eq.~(\ref{SetOfOperators}) can be placed,
\begin{equation}
\big( |C_1|,\ |C_2|,\ |C_3|,\ |C_4|,\ |C_5| \big) \,\le\,
  \big( 57,\ 16,\ 58,\ 5.6,\ 16 \big)  
 \times 10^{-8} \left(\frac{\Lambda}{1~\mbox{TeV}} \right)^2.
\label{ConstraintsOnCoefficients}
\end{equation}

%
The constraints on $C_6-C_8$ are identical to those on 
$C_1-C_3$~\cite{Gedalia:2009kh}.
Note that Eq.~(\ref{ConstraintsOnCoefficients}) implies that new physics 
particles have highly suppressed couplings to charm quarks.  
Alternatively, the tight constraints of Eq.~(\ref{ConstraintsOnCoefficients}) 
probe NP at the very high scales:
$\Lambda \ge (4-10) \times 10^3$~TeV for tree-level NP-mediated charm mixing and 
$\Lambda \ge (1-3) \times 10^2$~TeV for loop-dominated mixing via new physics 
particles.

There is a beautiful effect at threshold, where the decay of the $\psi(3770)$
gives a quantum correlated $D^0-\D0bar$ pair, and like-sign $K^\pm\pi^\mp$
decays at equal times only arise from mixing, without the
doubly-Cabibbo-suppressed background.  However, this requires high-quality
particle identification,  and is very luminosity-intensive.  The event rate is of order
one  event per 5 fb$^{-1}$ (the current BESIII dataset is 2.9 fb$^{-1}$).  
Threshold does come into play in a different manner, however.   When mixing is
measured via the time dependence of hadronic decays,  one measures $x,y$ in a
rotated basis.   These parameters, denoted $x', y'$ in the case of
$K\pm\pi^\mp$,  can only be converted to the desired $x, y$ with knowledge of a 
strong final-state scattering phase, $\delta_{K\pi}$.   Threshold charm data
provide the only possibility to measure this  (and other related) phases.  

CP violation in $D^0-\D0bar$ mixing is an important area for future work.   In
Table \ref{Csection_tab_1}, we summarize the prospects for future  results on
these topics.  The entries related to $q/p$ parametrize CP violation; in the
absence of CP violation in mixing, $|q/p| =1$ and $\arg(q/p)=0$ in the phase
convention adopted.  

\begin{table}[t]
\centering
\begin{tabular}{c|c|c|c|c}   
\hline 
\hline
\multirow{2}{*}{Observable} & \multirow{2}{*}{Current Expt.}
             & LHCb          & Belle~II	        &  LHCb Upgrade      \\
  &          & (5 fb$^{-1}$) & (50 ab$^{-1}$) &  (50 fb$^{-1}$)    \\
\hline
$x$	     & (0.63 $\pm$ 0.20)\% & $\pm$0.06\% & $\pm$0.02\% & $\pm$0.02\% \\
$y$	     & (0.75 $\pm$ 0.12)\% & $\pm$0.03\% & $\pm$0.01\% & $\pm$0.01\% \\
$y_{\rm CP}$ & (1.11 $\pm$ 0.22)\% & $\pm$0.02\% & $\pm$0.03\% & $\pm$0.01\% \\
$|q/p|$      & 0.91 $\pm$ 0.17     &  $\pm$0.085 & $\pm$0.03   & $\pm0.03$   \\
${\rm arg}(q/p)$ & ($-10.2 \pm 9.2$)$^{\circ}$   & $\pm$4.4$^{\circ}$ 
                 & $\pm$1.4$^{\circ}$            & $\pm$2.0$^{\circ}$        \\
\hline 
\hline
\end{tabular}
\caption{Sensitivities of Belle II and LHCb to charm mixing related parameters,
along with the current results for these measurements; here ${\rm arg}(q/p)$
means ${\rm arg}\big[(q/p)(\overline A_{K^+K^-}/A_{K^+K^-})\big]$.  The second
column gives the 2011 world averages.  The remaining columns give the expected
accuracy at the indicated integrated luminosities. In the convention used in
HFAG fits, in the absence of CP violation $|q/p| =1$ and $\arg(q/p)=0$.}
\label{Csection_tab_1}
\end{table}

\subsubsection*{CP Violation in Decays}

A possible manifestation of new physics interactions in the charm
system is associated with the observation of CP violation~\cite{Artuso:2008vf,Bigi:2009jj}. 
This is due to the fact that all quarks that build up the hadronic states in weak 
decays of charm mesons belong to the first two generations. Since the $2\times2$ 
Cabibbo quark mixing matrix is real, no CP violation is possible in the
dominant tree-level diagrams which describe the decay amplitudes. 
CP-violating amplitudes can be introduced in the standard model by including 
penguin or box operators induced by virtual $b$ quarks. However, their 
contributions are strongly suppressed by the small combination of 
CKM matrix elements $V_{cb}V^*_{ub}$. Thus, it was believed that the 
observation of large CP violation in charm decays or mixing would be an 
unambiguous sign for new physics. The SM ``background" here is quite small,
giving CP-violating asymmetries of the order of $10^{-3}$. Hence, 
observation of CP-violating asymmetries larger than $1\%$ could indicate 
presence of new physics.

Recent measurements have indicated the possibility of direct CP violation 
in the decays $D^0 \to K^+K^-$ and $D^0 \to \pi^+\pi^-$~\cite{DirectCPexp}.  
The current world average is
\begin{equation}\label{CPviolationLHCb}
\Delta A_{\rm CP} = A_{\rm CP}(K^-K^+)-A_{\rm CP}(\pi^-\pi^+)
= -(0.33 \pm 0.12)\% \,,
\end{equation}
although the most recent LHCb result (included in this average) has a central
value with the opposite sign. These results triggered intense theoretical
discussions of the possible size of this quantity in the standard model and in
models of new physics~\cite{DirectCPtheory}. New measurements of
individual direct CP-violating asymmetries entering Eq.~(\ref{CPviolationLHCb})
and other asymmetries in the decays of neutral and charged $D$'s into $PP$,
$PV$, and $VV$ final states are needed to guide theoretical calculations (of
penguin amplitudes).

It is also important to measure CP-violating asymmetries in the decays of charmed 
baryon states, as those could have different theoretical and experimental systematics 
and could provide a better handle on theoretical uncertainties. 

No indirect CP violation has been observed in charm transitions yet. However, available 
experimental constraints can provide some tests of CP-violating NP models. 
For example, a set of constraints on the imaginary parts of Wilson coefficients of 
Eq.~(\ref{SetOfOperators}) can be placed,
\begin{equation}
\big( |\mbox{Im} C_1|,\ |\mbox{Im} C_2|,\ |\mbox{Im} C_3|,\ |\mbox{Im} C_4|,\ 
  |\mbox{Im} C_5| \big) \,<\,
\big( 11,\ 2.9,\ 11,\ 1.1,\ 3.0 \big) \times 10^{-8} 
 \left(\frac{\Lambda}{1~\mbox{TeV}} \right)^2.
\end{equation}
%
Just like the constraints of Eq.~(\ref{ConstraintsOnCoefficients}), 
they give a sense of how NP particles couple to the standard model.

\subsubsection*{Rare Decays}

The flavor-changing neutral current (FCNC) decay $D^0 \to \mu^+\mu^-$ 
is of renewed interest after the measurement of $B_s \to \mu^+ \mu^-$.  
While heavily GIM-suppressed, long-distance contributions
from $D^0 \to \gamma \gamma$, for example, also contribute.  
Direct knowledge of the decay $D^0 \to \gamma \gamma$ allows one to limit 
these contributions to the di-muon mode to below $6 \times 10^{-11}$.  

Decays $B \to K^{(*)} \ell^+ \ell^-$ have been the subject of great 
interest for many years, both rates and angular distributions offer 
the chance to see new physics effects.  
The analogous charm decays, $D_{(s)}^+ \to h^+ \mu^+\mu^-$, 
$D^0 \to h h' \mu^+\mu^-$ are likewise interesting.  
The former modes have long-distance contributions of order $10^{-6}$ 
from vector intermediaries ($\rho, \omega, \phi$) but these can be cut away.  
The standard model rate for the remaining decays is around $10^{-11}$.  
For the latter modes, one can form forward-backward and $T$-odd 
asymmetries with sensitivity to new physics.  

Experimentally, at present, there are only the upper 
limits on $D^0 \to \ell^+ \ell^-$ decays,
\begin{equation}
{\cal B}(D^0 \to \mu^+\mu^-) \le  1.1\times 10^{-8}, \qquad 
{\cal B}(D^0 \to e^+ e^-) \le 7.9\times 10^{-8}, \qquad
{\cal B}(D^0 \to \mu^\pm e^\mp) \le  2.6\times 10^{-7}.
\label{brs}
\end{equation}
Theoretically, just like in the case of mixing discussed above, 
all possible NP contributions to $c \to u \ell^+ \ell^-$ 
can also be summarized in an effective Hamiltonian,  
\begin{equation}
{\cal H}_{\rm NP}^{\rm rare}  = 
\sum_{i=1}^{10}  {\widetilde C}_i (\mu) ~ \widetilde Q_i,
\qquad 
\begin{array}{l}
\widetilde Q_1 = (\overline{\ell}_L \gamma_\mu \ell_L) \ 
(\overline{u}_L \gamma^\mu
c_L)\,, \\
\widetilde Q_2 = (\overline{\ell}_L \gamma_\mu \ell_L) \ 
(\overline{u}_R \gamma^\mu
c_R)\,, \\ 
\widetilde Q_3 = (\overline{\ell}_L \ell_R) \ (\overline{u}_R c_L) \,, 
\end{array}
\qquad 
\begin{array}{l}
\widetilde Q_4 = (\overline{\ell}_R \ell_L) \ 
(\overline{u}_R c_L) \,, \\
\widetilde Q_5 = (\overline{\ell}_R \sigma_{\mu\nu} \ell_L) \ 
( \overline{u}_R \sigma^{\mu\nu} c_L)\,,\\
\phantom{xxxxx} 
\end{array}
\label{SetOfOperatorsLL}
\end{equation}
where ${\widetilde C}_i$ are again Wilson coefficients, and the $ \widetilde Q_i$ 
are the effective operators. In this case, however, there are ten of them, 
with five additional operators $\widetilde Q_6, \dots, \widetilde Q_{10}$ 
that can be obtained from operators in Eq.~(\ref{SetOfOperatorsLL}) by 
the substitutions $L \to R$ and $R \to L$.
Further details may be found in Ref.~\cite{Golowich:2009ii}, 
where it is also noted that it might be advantageous to study correlations 
of new physics contributions to various processes, for instance $D^0-\D0bar$ mixing 
and rare decays.

\subsubsection*{Strong Phases}

Threshold data with correlated $D^0-\D0bar$ pairs may be used to 
extract strong phases in $D$ decays.  These phases enter into 
$B$ physics determinations of the CKM angle $\gamma$
from $B \to D^{(*)} K^{(*)}$ decays~\cite{Atwood:2003mj}.  
Without direct input from charm, these $B$ results suffer from 
ill-defined systematic uncertainties and lose precision.  
In addition, strong phases are needed to relate observables 
of $D^0-\D0bar$ oscillations measured with hadronic final states 
to the usual $x,\ y$ parameters.


\subsubsection*{Charmonium Spectroscopy}



In the last decade, the observations of the spin-singlet charmonium
states $h_c(1P)$ and $\eta_c(2S)$
have completed the charmonium multiplets below the open-charm threshold.
Experiments have also revealed a large number of unexpected $c
\overline{c}$ mesons
above the open-charm threshold, labelled $XYZ$ states.
These states may include tetraquarks, $c \overline{c}  g$ hybrids, meson
molecules, etc.
Many of the $XYZ$ states are narrow and some are manifestly exotic,
indicating a gap in our understanding of the QCD spectrum.
Understanding their spectrum and decays
are major challenges for lattice QCD and phenomenology.
Experimental data on the $XYZ$ states continues to accumulate,
and more can be expected from future  experiments.
The challenge of charmonium spectroscopy above the open-charm threshold
is discussed in detail in a Snowmass White Paper \cite{Bodwin:2013nua}.

\subsubsection*{Other Topics}

We finally list a few topics on ``engineering numbers."
Currently, charm lifetimes are dominated by FOCUS results; 
while the results are well-respected, a cross-check would 
be welcome.  These results serve to relate theoretical 
predictions for partial widths to the experimentally accessible 
quantities, branching fractions.  

Likewise, golden mode branching fractions for $D$ mesons are 
dominated by CLEO-c; a cross-check from BESIII is in order.  
For the baryons, where there are four weakly-decaying 
ground states, there are no absolute branching fraction results.  
For $\Lambda_c \to p K^- \pi^+$, the near-threshold enhancement 
of $\Lambda_c$ pairs measured by Belle in ISR~\cite{Pakhlova:2008vn} 
shows that BESIII should be able to provide a nice result 
with a modest-length run.  

In addition to topics discussed above, charm quarks will play a 
major role in the heavy-ion experimental programs at RHIC and LHC
for the next decade. Questions that will be addressed include identification
of the exact energy loss and hadronization mechanisms of charm 
(or beauty) quarks in propagation through Quark-Gluon Plasma (QGP), 
calculations of heavy quark transport coefficients, etc.

\subsection{Charm Physics Summary and Perspectives Beyond 2020}

Continued support of BESIII, LHCb, and Belle~II is critical to U.S.\ 
involvement in a vibrant charm program.  Investments in these
provide valuable access to exciting datasets.  
Attention should also be paid to possible opportunities at a future 
threshold experiment should one be built abroad.  

Theoretical calculations in charm physics are mainly driven by experimental
results. The challenges associated with nonperturbative QCD  dynamics are being
addressed by advances in lattice QCD and other  nonperturbative approaches.
While similar probes of the NP scale that  might reveal the ``grand design" of
flavor are available in the strange and beauty  systems, charm quarks furnish
unique access to processes involving  up quarks, more precise and complementary
to searches for FCNC top decays. Moreover, $D$ mesons are the only neutral
mesons composed of up-type quarks which have flavor oscillations, and thus
probe NP in the $\Delta F = 2$ transitions, providing complementary sensitivity
to $K$, $B$, and $B_s$ mixing.

\section{Report of the Lattice QCD Task Force}

The properties of the five least massive quarks offer a powerful tool
to indirectly study physics at energies many orders of magnitude above
those which are directly accessible to present or planned accelerators.  This
is made possible in large part by the quarks' strong interactions
which provide experimental physics with a host of bound states, common
and rare decay processes and mixings that enable clever and highly
sensitive studies of the properties of the underlying quarks.  Until
recently, the lack of predictive control of these same strong
interactions provided a large barrier to fully exploiting this
potential.  Ab initio lattice calculations are systematically 
removing this barrier, allowing us to fully exploit the strong
interactions of the quarks to search for physics beyond the standard
model.  In this section we describe the status and prospects for the
lattice QCD calculations needed for future
quark-flavor experiments.
Much of this material is drawn from a recent USQCD (the national U.S.\ lattice-QCD
collaboration) white paper~\cite{whitepaper13}.

Lattice QCD provides a first-principles method for  calculating low-energy
hadronic matrix elements  with reliable and systematically-improvable
uncertainties. Such matrix elements --- decay constants, form factors, mixing
matrix elements, etc.\ --- are needed  to determine the standard model (SM)
predictions for many processes and/or to extract CKM matrix elements.

In the last five years lattice QCD has matured into a precision tool.
Results with fully controlled errors are available
for nearly 20 matrix elements:
the decay constants
$f_\pi$, $f_K$, $f_D$, $f_{D_s}$, $f_B$ and $f_{B_s}$, 
semileptonic form factors for
$K\to \pi$, $D \to K$, $D\to\pi$, 
$B\to D$, $B\to D^{*}$, $B_s\to D_s$ and $B\to\pi$,
and the four-fermion mixing matrix elements
$B_K$, $f_B^2 B_B$ and $f_{B_s}^2 B_{B_s}$.
By contrast, in 2007 (when the previous USQCD white paper was 
written~\cite{whitepaper07}), 
only $f_K/f_\pi$ was fully controlled.
A sample of present errors is collected in Table~\ref{tab:error}.
For $K$ mesons, errors are at or below the percent level,
while for $D$ and $B$ mesons errors range from few to $\sim$10\%.

\begin{table}[t!]
\centering
\begin{tabular}{cccccc}
\hline\hline
Quantity   & CKM & Present & 2007 forecast & Present  & 2018  \\ 
& element & expt. error &\ lattice error\
& \ lattice error\ &\ lattice error\   \\  
\hline
$f_K/f_\pi$ & $|V_{us}|$ \rule[0mm]{0mm}{4mm} & 0.2\% &0.5\%&
{0.4\%} & 0.15\%  \\ 
$f_+^{K\pi}(0)$ & $|V_{us}|$ \rule[0mm]{0mm}{4mm} & 0.2\% & -- &
{0.4\%} & 0.2\% \\ 
$f_D$ \rule[0mm]{0mm}{4mm} & $|V_{cd}|$ & 4.3\% & 5\% &
2\% & $<1\%$ \\
$f_{D_s}$ \rule[0mm]{0mm}{4mm} & $|V_{cs}|$ & 2.1\% & 5\% &
2\% & $< 1\%$  \\
$D\to\pi\ell\nu$ \rule[0mm]{0mm}{4mm} & $|V_{cd}|$ & 2.6\% & -- &
4.4\% & 2\%  \\
$D\to K\ell\nu$ & $|V_{cs}|$ & 1.1\% \rule[0mm]{0mm}{4mm} & -- &
2.5\% & 1\%  \\ 
$B\to D^{*}\ell\nu$ \rule[0mm]{0mm}{4mm} & $|V_{cb}|$ & 1.3\% & -- &
1.8\% & $<1\%$  \\ 
$B\to \pi\ell\nu$ & $|V_{ub}|$ &4.1\% \rule[0mm]{0mm}{4mm} & -- &
8.7\% & 2\%  \\ 
$f_B$ \rule[0mm]{0mm}{4mm} & $|V_{ub}|$ & 9\% \rule[0mm]{0mm}{4mm} 
& -- & 2.5\% & $< 1\%$ \\
$\xi$ & $|V_{ts}/V_{td}|$ & 0.4\% & 2--4\% &
4\% & $< 1\%$ \\
$\Delta m_s$ & $|V_{ts}V_{tb}|^2$ & 0.24\% & 7--12\% &
11\% & 5\% \\
$B_K$ \rule[0mm]{0mm}{4mm} & ${\rm Im}(V_{td}^2)$
& 0.5\% &3.5--6\% & 1.3\% & $< 1\%$ \\
\hline\hline
\end{tabular}
\smallskip
\caption{History, status and future of selected lattice-QCD calculations
    needed for the determination of CKM matrix elements.
    2007 forecasts are from Ref.~\cite{whitepaper07}. 
    Most present lattice results are taken from 
    \href{latticeaverages.org}{latticeaverages.org}~\cite{Laiho:2009eu}.
    The quantity
    $\xi$ is $f_{B_s} \sqrt{B_{B_s}}/(f_{B}\sqrt{B_{B}})$.}
\label{tab:error}
\end{table}

The lattice community is embarking on
a three-pronged program of future calculations:
(i) steady but significant improvements in ``standard'' matrix elements
of the type just described,
leading to much improved results for CKM parameters (e.g., $V_{cb}$);
(ii) results for many additional matrix elements relevant for
searches for new physics and
(iii) the extension of lattice methods to more challenging matrix elements
which can both make use of old results and provide important information
for upcoming experiments.

Reducing errors in the standard matrix elements has been a major focus
of the lattice community over the last five years, and the improved
results illustrated in Table~\ref{tab:error} now play 
an important role in the determination of
the CKM parameters in the ``unitarity triangle fit.''
Lattice-QCD calculations involve various sources of systematic error
(the need for extrapolations to zero lattice spacing, infinite volume
and the physical light-quark masses, as well as fitting and operator
normalization)
and thus it is important to cross-check 
results using multiple discretizations of
the continuum {QCD} action.
(It is also important to check that  results for the hadron spectrum
agree with experiment. Examples of these checks are shown in
the 2013 whitepaper~\cite{whitepaper13}.)
This has been done for almost all the quantities noted above. 
This situation has spawned two lattice averaging efforts, 
{\tt latticeaverages.org}~\cite{Laiho:2009eu} and 
FLAG-1~\cite{Colangelo:2010et}, 
which have recently joined forces and expanded to form a worldwide
Flavor Lattice Averaging Group (FLAG-2), 
with first publication expected in mid-2013.

The ultimate aim of lattice-QCD calculations is to reduce errors in
hadronic quantities to the level
at which they become subdominant either to experimental errors or
other sources of error. As can be seen from Table~\ref{tab:error},
several kaon matrix elements are approaching this level, while
lattice errors remain dominant in most quantities involving
heavy quarks. Thus the most straightforward contribution of lattice QCD
to the future intensity frontier program will be the reduction
in errors for such quantities. Forecasts for the expected reductions
{by 2018} are shown in the table. These are based on a
Moore's law increase in computing power, and extrapolations using
existing algorithms. Past forecasts have been typically conservative
(as shown in the table) due to unanticipated
algorithmic or other improvements.
The major reasons for the expected 
reduction in errors are the use of $u$ and $d$
quarks with physical masses, the use of smaller lattice spacings
and improved heavy-quark actions, and the reduction in statistical errors.

Thus one key contribution of lattice QCD to the future flavor-physics
program will be a significant
reduction in the errors in CKM elements, most notably $V_{cb}$.
This feeds into the SM predictions for several of the rare decays
that are part of the proposed experimental program, 
e.g., $K\to\pi \nu\overline{\nu}$. For {these decays}, 
the parametric error {from}
$|V_{cb}|$, {which enters as the fourth power}, 
is the dominant source of uncertainty in the SM 
{predictions}.
The lattice-QCD improvements projected in 
Table~\ref{tab:error} will bring the theoretical uncertainties 
to a level commensurate with the projected experimental errors 
in time for the planned rare kaon-decay experiments at Fermilab.

The matrix elements discussed so far involve only a single hadron
and no quark-disconnected contractions. These are the most straightforward
to calculate (and are sometimes called ``gold-plated'').
The second part of the future lattice-QCD program for the intensity
frontier will be the extension of the calculations to other, similar,
matrix elements which are needed for the search for new physics.
This includes the mixing matrix elements for kaons, $D$ and $B$ mesons
arising from operators present in BSM theories but absent in the
SM, the form factors arising in $B\to K \ell^+\ell^-$ and
$\Lambda_b\to\Lambda \ell^+\ell^-$, non-SM form factors for
$K\to\pi$, $B\to\pi$ and $B\to K$ transitions.
We expect the precision attained for these quantities to
be similar to those for comparable quantities listed 
in Table~\ref{tab:error}. 

The third part of the lattice-QCD program is the least developed and
most exciting. This involves the development of new methods or
the deployment of known but challenging methods, and
allows a substantial increase in the repertoire of lattice calculations.
In particular, calculations involving two particles below
the inelastic threshold are now possible
(e.g., $K\to\pi\pi$ amplitudes~\cite{Blum:2011ng,Blum:2012uk,Boyle:2012ys}), 
quark-disconnected contractions are being controlled 
(e.g., $\eta'$ and $\eta$ masses~\cite{Christ:2010dd} and
the nucleon sigma term~\cite{Junnarkar:2013ac})
and processes involving two insertions of electroweak operators
are under pilot study (e.g., the long-distance part 
of $\Delta m_K$~\cite{Christ:2012se}).
During the next five years, we expect that these advances will lead
to a quantitative understanding of the $\Delta I=1/2$ rule, 
a prediction with $\sim5\%$ errors for the the SM contribution to
$\epsilon_K'$, and predictions with 10--20\% errors for the long-distance
contributions to $\Delta m_K$ and $\epsilon_K$.
This will finally allow us to use these hallowed experimental results 
in order to search for new physics.

These new methods should allow lattice QCD to contribute directly to the
proposed flavor-physics experiments. For example,
a calculation of the long-distance
contributions to $K\to\pi\nu\overline{\nu}$ decays should be possible, 
checking the present estimate that these contributions are small.
Similar methods should allow the calculation of 
the sign of the CP-conserving amplitude $K_S\to\pi^0 e^-e^+$,
thus resolving a major ambiguity in the SM prediction for
$K_L\to\pi^0 e^-e^+$.

We also expect progress on {even} more challenging calculations, for
which no method is yet known. An important example, in light of
recent evidence for CP violation in $D$ decays and for $D-\Dbar$ mixing,
is to develop a method for calculating the amplitudes
for $D\to\pi\pi,\ KK$ decays and $D-\Dbar$ mixing.
This requires dealing with four or more particles in a finite box,
as well as other technical details. 

These plans rely crucially on access to high-performance computing,
as well as support for algorithm and software development.
In the U.S., much of this infrastructure is coordinated by the
USQCD umbrella collaboration. Continued support for this effort
is essential for the program discussed here.

We also stress that there are substantial lattice-QCD efforts underway
to calculate the hadronic (vacuum polarization and light-by-light)
contributions to muonic $g-2$, {the light- and
strange-quark contents of the nucleon (which are needed to interpret
$\mu \to e$ conversion and dark-matter experiments), and the nucleon
axial form factor (which enters the determination of the neutrino flux
at many accelerator-based neutrino experiments).  Smaller-scale
lattice-QCD calculations of nucleon EDMs, proton- and neutron-decay
matrix elements, and neutron-antineutron oscillation matrix elements
are also in progress.} These are very important for the intensity
frontier as a whole, although not directly relevant to quark-flavor
physics.

In the remainder of this subsection, we describe the
major new efforts that are underway or envisaged for the
next 5 or so years,
considering in turn kaons, $D$ mesons and $B$ mesons,
and close with a 15-year vision.

\subsection{Future Lattice Calculations of Kaon Properties}

\underline{\boldmath\bf $K\to \pi \pi$ amplitudes:}~
These amplitudes are now active targets of lattice-QCD calculations.  
The final-state pions can be arranged to have physical,
energy-conserving relative momentum by imposing appropriate boundary
Corrections for the
effects of working in finite volume can be made following the analysis
of Ref.~\cite{Lellouch:2000pv}. 
A first calculation of the amplitude to the $I=2$ two-pion state, $A_2$,
has been performed~\cite{Blum:2011ng} with physical kinematics but 15\% finite
lattice spacing errors.  Calculations are now underway using two
ensembles with smaller lattice spacings which will
allow a continuum extrapolation, removing this error.  Results
with an overall systematic error of
$\approx 5\%$ are expected within the coming year.

The calculation of $A_0$ is much more difficult because of the overlap
between the $I=0$ $\pi\pi$ state and the vacuum, resulting in
disconnected diagrams and a noise to signal ratio that grows
exponentially with time separation.  In addition, for $I=0$,
$G$-parity boundary conditions must be employed and imposed on both the
valence and sea quarks.  These topics have been actively studied for the
past three years~\cite{Blum:2011pu} and $G$-parity boundary conditions
successfully implemented~\cite{Kelly:2012eh}.  
First results with physical kinematics are expected within two years 
from a relatively coarse, $32^3 \times 64$ ensemble.
Errors on $\epsilon_K'$ on the order of 15\% should be
achieved, with the dominant error coming from the finite lattice
spacing.  As in the case of the easier $A_2$ calculation, lessons
learned from this first, physical calculation will then be applied to
calculations using a pair of ensembles with two lattice spacings so
that a continuum limit can be obtained.  A five-year time-frame may be
realistic for this second phase of the calculation.  Essential to the
calculation of both $A_0$ and $A_2$ is the renormalization of the
lattice operators.  Significant efforts will be required in the next
2--3 years to extend the range of nonperturbative renormalization
methods up through the charm threshold and to a scale of 4--5 GeV where
perturbative matching to the conventional $\overline{\mbox{MS}}$
scheme will have small and controlled errors.

\underline{\boldmath\bf Long-distance contributions to 
$\Delta m_K$ \bf and $\epsilon_K$:}~ 
Promising techniques have been developed which allow the calculation
of the long-distance contribution to kaon mixing by lattice
methods.  By evaluating a four-point function including operators
which create and destroy the initial and final kaons and two effective
weak four-quark operators, the required second order amplitude can be
explicitly evaluated.  Integrating the space-time positions of the two
weak operators over a region of fixed time extent $T$ and extracting
the coefficient of the term which grows linearly with $T$ gives
precisely both $\Delta m_K$ and $\epsilon_K$.  This Euclidean space
treatment of such a second-order process contains unphysical
contributions which grow exponentially with $T$ and must be
subtracted.  The statistical noise remaining after this subtraction
gives even the connected diagrams the large-noise problems typical of
disconnected diagrams.  Preliminary results suggest that this problem
can be solved by variance reduction methods and large
statistics~\cite{Christ:2012se}. Given the central importance of GIM
cancellation in neutral kaon mixing, a lattice calculation that is not
burdened by multiple subtractions must include the charm quark mass
with consequent demands that the lattice spacing be small compared to
$1/m_c$~---~a substantial challenge for a calculation which should
also contain physical pions in an appropriately large volume.
Perturbative results~\cite{Brod:2010mj} as well as the first lattice
calculation~\cite{Christ:2012se} suggest that perturbation theory
works poorly at energies as low as the charm mass, making the
incorporation of charm in a lattice calculation a high priority.
Given the challenge of including both physical pions and active charm
quarks, the first calculation of $\Delta m_K$ may take 4--5 years.
Results for the long-distance part of $\epsilon_K$ may be obtained in
a similar time frame.  However, a more challenging subtraction
procedure must be employed for $\epsilon_K$.

\underline{\bf Rare kaon decays:}~ 
Given the promise of the first calculations of the long distance
contributions to $\Delta m_K$, a process that involves two $W^\pm$
exchanges, it is natural to consider similar calculations for the
second-order processes which enter important rare kaon decays such as
$K^0_L\to\pi^0\ell^+\ell^-$ and $K^+\to\pi^+\nu\overline{\nu}$.  While
in principle $K_L\to\ell^+\ell^-$ should also be accessible to lattice
methods, the appearance of three electroweak, hadronic vertices
suggests that this and similar processes involving $H_W^{\Delta S=1}$
and two photons, should be tackled only after success has been
achieved with more accessible, second order processes.

The processes $K^+\to\pi^+\nu\overline{\nu}$ and
$K_L\to\pi^0\nu\overline{\nu}$ may be the most straightforward
generalization of the current $\Delta m_K$ calculation.  Here the
dominant contribution comes from box and $Z$-penguin diagrams
involving top quarks, but with a 30\% component of the CP-conserving
process coming from the charm quark~\cite{Cirigliano:2011ny}.  While
the charm quark piece is traditionally referred to as ``short
distance,'' the experience with $\Delta m_K$ described above suggests
that a nonperturbative evaluation of this charm-quark contribution
may be a necessary check of the usual perturbative approach, which is
here believed to be reliable.  There are also ``longer distance''
contributions which are only accessible to lattice methods and will
become important when the accuracy of rare kaon decay experiments
reaches the 3\% level, or possibly sooner.  The long-distance
contributions to the decay $K_{L/S}\to\pi^0\ell^+\ell^-$ also appear
to be a natural target for a lattice-QCD calculation since the sign of
the CP-conserving process $K_S\to\pi^0\ell^+\ell^-$ may be only
determined this way.

\subsection[Future Lattice Calculations of $D$-meson Properties]{\boldmath
Future Lattice Calculations of $D$-meson Properties}

\underline{\boldmath\bf  $D\to \pi \pi,\ K K$ amplitudes:}~
Recent experimental evidence suggests that there may be CP violation
in $D\to\pi\pi$ and $D\to K K$ decays.  In order to interpret
these results, it is essential to be able to predict the CP violation
expected in the SM.  Even a result with a
large, but reliable, error could have a large impact.  This need will
become even more acute over the next five years as LHCb and Belle~II
improve the measurements.

This calculation is more challenging than that for
$K\to\pi\pi$ decays, which
represent the present frontier of lattice calculations.
In the kaon case, one must deal with the fact that two-pion
states in finite volume are not asymptotic states, and
the presence of multiple quark-disconnected contractions.
For $D$ decays, even when one has
fixed the strong-interaction quantum numbers of a final state
(say to $I=S=0$), the strong interactions necessarily bring in
multiple final states: $\pi\pi$ and
$K\bar K$ mix with $\eta\eta$, $4\pi$, $6\pi$, etc.
The finite-volume states used by lattice QCD are 
inevitably mixtures of all these possibilities, and one
must learn how, in principle and in practice, to disentangle
these states to obtain the desired matrix element.
Recently, a first step towards 
developing a complete method has been taken~\cite{Hansen:2012tf}, in which the
problem has been solved in principle for any number of two-particle
channels, and assuming that the scattering is dominantly $S$~wave.
This is encouraging, and this method may allow
one to  obtain semi-quantitative results for the amplitudes
of interest. We expect that turning this method into
practice will take $\sim 5$ years due to a number of numerical
challenges (in particular the need to calculate several
energy levels with good accuracy).

In the more distant future, we expect {that}
it will be possible to generalize the
methodology to include four-particle states; several groups
are actively working on the theoretical issues and much progress
has been made already for three particles. 

\underline{\boldmath\bf $D-\Dbar$ mixing:}~
Mixing occurs in the $D^0$--$\D0bar$ system, and there is no evidence yet
for CP violation in this mixing~\cite{Beringer:1900zz}.
The short-distance contributions can be
calculated for $D$ mesons using lattice QCD, as for kaons and $B$ mesons.
The challenge, however, is to calculate the
long-distance contributions.
As in the case of $\Delta m_K$ discussed above, there are two
insertions of the weak Hamiltonian, with many allowed states
propagating between them.
The $D$ system is much more challenging, however, since,
as for the decay amplitudes, there
are many strong-interaction channels having $E< m_D$.
Further theoretical work is needed to develop a practical method.

\subsection[Future Lattice Calculations of $B$-meson Properties]{\boldmath
Future Lattice Calculations of $B$-meson Properties}

\underline{\boldmath\bf $B \to D^{(*)} \ell\nu$ form factors at nonzero
recoil:}~
Lattice-QCD results for these form factors allow for the determination
of $|V_{cb}|$ from the measured decay rates. For the $B\to D^*\ell\nu$
form factor at zero recoil, the gap between experimental errors
($1.3\%$) and lattice errors (currently $\sim 1.8\%$) has narrowed
considerably over the last five years.  In the next five years, we
expect the lattice contribution to the error in $|V_{cb}|$ to drop
below the experimental one, as shown in Table~\ref{tab:error}.
Particularly important for this will be the extension of the 
$B\to D^{(*)} \ell\nu$ form-factor calculations to nonzero
recoil~\cite{Qiu:2011ur}.

\underline{\boldmath\bf Tauonic $B$-decay matrix elements:}~
Recently the \babar\ collaboration measured the ratios $R(D^{(*)}) =
{\cal B}(B\to D^{(*)} \tau\nu)/{\cal B}(B\to D^{(*)} \ell \nu)$ with
$\ell=e$ or $\mu$, and observed a combined excess over the existing
SM predictions of 3.4$\sigma$~\cite{Lees:2012xj}.  Those
SM predictions were based, however, on models of QCD, not
\emph{ab initio} QCD.
Realizing that it was much easier to obtain accurate results for these
ratios than for the form factors themselves, the Fermilab-MILC
collaboration responded quickly (using lattice data already
in hand), and provided the first lattice-QCD result for
$R(D)$~\cite{Bailey:2012jg}.  Their result slightly reduced the
discrepancy with experiment for $R(D)$ from $ 2.0 \to 1.7\sigma$.  At
present, the experimental errors in $R(D)$ ($\sim 16\%$) dominate over
lattice errors (4.3\%), so further lattice improvements are not needed
in the short run.  The experimental uncertainties will
shrink with the increased statistics available at Belle~II, and it
should be straightforward to reduce the corresponding lattice-QCD
error by a factor of two over the next five years.  Work is also in
progress to calculate $R(D^*)$, for which the uncertainties are
expected to be comparable to those of $R(D)$.

Belle~II will also reduce the uncertainty in the experimental
measurement of ${\cal B}(B\to\tau\nu)$ to the few-percent level with
its anticipated full data set.  In the next five years, lattice-QCD
calculations are expected to reduce the error in $f_B$ to the percent
level (see Table~\ref{tab:error}).  Particularly important for this
will be the use of finer lattice spacings that permit relativistic
$b$-quark actions~\cite{McNeile:2011ng}.  Combined with the
anticipated experimental precision, this will increase the reach of
new-physics searches in $B\to\tau\nu$; moreover, correlations between
$B\to\tau\nu$ and $B\to D^{(*)}\tau\nu$ decays can help distinguish
between new-physics models.

\underline{\boldmath\bf $B\to K\ell^+\ell^-$ and related decay form factors:}~ 
The branching ratio for $B\to K\ell^+\ell^-$ is now well measured,
and increasingly accurate results from LHCb,
and eventually Belle~II, are expected.
The SM prediction requires knowledge of
the vector and tensor $b\to s$ form factors across the kinematic range.
Present theoretical estimates use light-cone sum rules, but several
first-principles lattice-QCD calculations are nearing completion,
as reviewed in Ref.~\cite{Zhou:2013uu}.
The calculation is similar to that needed 
for the $B\to\pi\ell\nu$ form factor, and we expect similar
accuracy to be obtained over the next five years.

A related process is the baryonic decay
{$\Lambda_b\to\Lambda \ell^+\ell^-$}, recently measured 
by CDF. Here the extra spin degree of freedom can 
more easily distinguish
between SM and BSM contributions.
A lattice-QCD calculation of the required form factors has recently
been completed, using HQET to describe the $b$ quark~\cite{Detmold:2012vy}.
Errors of $\sim 10$--15\% in the form factors are obtained,
which are comparable to present experimental errors.
The latter errors will decrease with new results from LHCb,
and so improved LQCD calculations and cross-checks are needed.
Although the calculation is conceptually similar to
that for $B\to K\ell^+\ell^-$, given the presence of baryons
we expect the errors for {$\Lambda_b\to\Lambda\ell^+\ell^-$} to
lag somewhat behind.

\underline{\boldmath\bf Non-standard model form factors for 
$K\to\pi$ and $B\to\pi$ transitions:}~
The $B\to K$ vector and tensor form factors just discussed are also needed to
describe decays involving missing energy, 
$B\to K X$, in BSM theories~\cite{Kamenik:2011vy}.
Analogous form factors are needed for
$B\to\pi X$ and  $K\to\pi X$ decays.
The tensor form factors are also needed to evaluate some BSM contributions
to $K\to\pi\ell^+\ell^-$~\cite{Baum:2011rm}.
Thus it is of interest to extend the present calculations of
vector form factors in $K\to\pi$ and $B\to\pi$
to include the tensor matrix elements.
Since these are straightforward generalizations of present
calculations, we expect that comparable accuracy to the present errors
in Table~\ref{tab:error} can be obtained quickly, and that future
errors will continue to follow the projections for similar matrix
elements.

\subsection{Lattice QCD and Flavor Physics: 2018--2030}

The discussion above has laid out an ambitious vision for future
lattice-QCD calculations on a five-year timescale, 
explaining how they can provide essential
information for upcoming quark-flavor experiments. 
Also discussed are a number
of more challenging quantities which have become accessible to lattice
methods only recently. 
In this section we discuss
more generally the opportunities offered by lattice methods over the
extended time period covered by the Snowmass study.  However, we
should emphasize that these longer range forecasts are made difficult
by the very rapid evolution of this emerging field, which is driven by
both rapidly advancing commercial computer technology and continual,
difficult-to-anticipate advances in algorithms.

We begin with the conservative assumptions that exascale performance
($10^{18}$ floating point operations/second) will be achieved by 2022,
and that a further factor of 100 will be available by 2032.  These
represent factors of $10^2$ and $10^4$ over presently available
capability.  At fixed physical quark masses, the difficulty of modern
lattice-QCD algorithms scales with decreasing lattice spacing $a$ as
$1/a^{6}$ and with increasing physical linear problem size $L$ as
$L^5$.
Present large-scale lattice calculations at physical quark masses are
performed in volumes of linear size $L \approx 6$ fm and with inverse
lattice spacing $1/a$ as small as $\sim 2.5$ GeV.  Thus, these $10^2$
and $10^4$ advances in computer capability will allow an increase in
physical volume to 15 and 36 fm and in inverse lattice spacing to 5
and 10 GeV, respectively.  Statistical errors can be reduced from
their present percent-level for many quantities to 0.1\% or even
0.01\% as needed.

These three directions of substantial increase in capability translate
directly into physics opportunities.  The large increase in possible
Monte Carlo statistics is necessary if we are to decrease the errors
on many of the quantities in Table~\ref{tab:error} to the 0.1\% level.
Such increased statistics will also directly support perhaps 1\%
precision for results that depend on disconnected diagrams such as
$\epsilon_K'$ and the $K_L-K_S$ mass difference.  For most QCD
calculations, the non-zero pion mass implies that finite volume
effects decrease exponentially in the linear size of the system.
However, this situation changes dramatically when electromagnetic
effects are included.  Here the massless photon and related
difficulties of dealing with charged systems in finite volume result
in substantial finite volume errors which decrease only as a power of
$L$ as the linear system size $L$ becomes large.  The ability to work
on systems of linear size 20 or 30 fm will play an important role in
both better understanding electromagnetic effects using lattice
methods and achieving the 10\% errors in the computation of such
effects that are needed to attain 0.1\% errors in many of the
quantities in Table~\ref{tab:error}.

Finally the ability to work with an inverse lattice spacing as large
as 10 GeV will allow substantial improvements in the treatment of
heavy quarks.  Using $3~\mathrm{GeV} \le 1/a \le 5~\mathrm{GeV}$,
calculations involving charm quarks will have controlled finite
lattice spacing errors on the 1\% level or smaller.  As a result
calculation of the long-distance contributions, up to and including
the charm scale, will be possible for $\Delta m_K$, $\epsilon_K$ and
rare kaon decays yielding errors of order 1\% for these important
quantities.  The larger inverse lattice spacings in the range
$6~\mathrm{GeV} \le 1/a \le 10~\mathrm{GeV}$ will allow the present
estimates of the finite lattice spacing errors in bottom quark systems
to both be substantially reduced and to be 
refined using the new information provided by a 
larger range of lattice spacings.
This will allow many quantities involving
bottom quarks to be determined with errors well below 1\%.

While ever more difficult to forecast, a $10^4$ increase in
capability can be expected to significantly expand the range of
quantities that can be computed using lattice methods.  These include
the $D-\Dbar$ mixing and multi-particle $D$ decays discussed in
the previous section as well as even more challenging quantities such
as semileptonic $B$ decays with vector mesons in the final state.
These are relevant both for the extraction of CKM matrix elements
(e.g., $B\to\rho\ell\nu$ provides an alternative determination of
$|V_{ub}|$) and new-physics searches (e.g., measurements of $B\to
K^*\ell^+\ell^-$, $B\to K^* \gamma$ and $B_s\to\phi\gamma$).  
A second example is nonleptonic $B$ decays, such as $B\to D\pi(K)$, 
which can be used to obtain the CKM angle $\gamma$.

Clearly an enhanced computational capability of four orders of
magnitude, coupled with possibly equally large advances in numerical
algorithms, will have a dramatic effect on the phenomena that can be
analyzed and precision that can be achieved using lattice methods.
The possibility of making SM predictions with errors which
are an order of magnitude smaller than present experimental errors
will create an exciting challenge to identify quantities where
substantially increased experimental accuracy is possible and where
the impact of such measurements on the search for physics beyond the
SM most sensitive.  With the ability to make highly
accurate SM predictions for a growing range of quantities,
experiments can be designed that will achieve the greatest precision
for quantities sensitive to physics beyond the SM, rather
than being limited to those quantities which are least 
obscured by the effects of QCD.


\section{A U.S.\ Plan for Quark Flavor Physics}

Until recently, the U.S.\ had onshore accelerator facilities that supported a leadership role at both the
Energy and Intensity Frontiers.  With the successful start of the LHC and the termination of the
Tevatron program, the Energy Frontier has migrated offshore for the foreseeable future.  With 
choices summarized in Section~\ref{sec:intro}, the U.S.\  ceded leadership in much of 
quark-flavor physics. 
It is difficult to foresee a scenario that leads to the construction
of a facility in the U.S.\ that is capable of 
supporting $B$-physics or charm-physics experiments during the current decade
or even the next decade.  
Indeed, the only 
accelerator-based experiments currently
in the DOE pipeline are neutrino experiments and muon experiments at Fermilab, 
and to achieve their full potential these experiments depend on
Fermilab's \ProjectX\ facility, which has yet to achieve the first level of
DOE approval (``mission need").  It is under these rather dire 
circumstances, facing the
prospect that the U.S.\ accelerator-based HEP program may go the way of
the dodo bird, that we must contemplate the question of whether and how the U.S.
should pursue research in quark-flavor physics.

There is a strong physics case for quark-flavor physics that remains
robust in all LHC scenarios.  It rests, quite simply, on the potential of precision
quark-flavor experiments and studies of very rare decays 
to obseve the effect of high-mass virtual
particles.  If new physics is observed at LHC, tighter constraints from the flavor sector
will narrow the range of models that can account for the observed states.  If new physics
is not discovered at LHC, then the reach to mass scales beyond that of LHC will 
still offer the potential to find new physics and to estimate the scale needed for
direct observation.  International recognition of the 
importance of quark-flavor physics is evident from  the commitments  in Europe
and Asia to conduct the next-generation of $B$-physics, charm,
and kaon experiments.

In the U.S., 
the goal should be to construct an HEP program that has the breadth to 
assure meaningful participation in making the discoveries that will define
the future of particle physics.
The successful U.S.\ contributions to LHC have demonstrated
that physicists from U.S.\ laboratories and universities can play essential roles
in offshore experiments.   If this paradigm works at the Energy Frontier, it can
work at the Intensity Frontier as well.  Therefore, significant U.S.\ contributions
to offshore quark-flavor experiments such as LHCb and Belle~II should be
encouraged.  Also, in the one area where existing and foreseeable
facilities on U.S.\ soil can support a
world-leading program --- kaon physics --- the U.S.\ should embrace the opportunity.
The accelerator facilities required for kaon experiments are exactly those needed
for the neutrino program, so the costs are incremental and relatively modest.  Below,
we summarize the opportunities that exist now and those that will exist during the
next decade.

\subsection{Opportunities in This Decade}

The Task Force reports have described current, planned, and possible 
$B$-physics, charm, and kaon experiments in Europe and Asia.  There is a strong
and diverse international program.  The only U.S.\ entry in the discussion of
the immediate future for quark-flavor physics experiments is the ORKA proposal
at Fermilab, for an experiment which would make a precise measurement of the
$K^+ \to \pi^+ \nu \overline{\nu}$ branching fraction.

For the remainder of this decade, the plans in  Europe and Asia appear to be
set, and the experiments there (those already running or under construction)
will define the frontier of quark-flavor physics. These are LHCb and NA62 at
CERN, KLOE2 in Italy, PANDA in Germany, BESIII in China, and Belle~II, KOTO,
and TREK in Japan.  This is a rich program, and fortunately U.S.\ physicists
have some involvement in most of these experiments.  While they all have
important physics goals and capabilities, the scale of LHCb and Belle~II, and
their incredibly broad physics menus including both bottom and charm, means that
they will be the flagship experiments in quark-flavor physics. In view of that,
the U.S.\ should try to play a significant role in these experiments. 

The outstanding question is 
whether the ORKA experiment will go forward at Fermilab.  It 
received ``Stage~1" approval from  Fermilab in the fall of 2011, but has not been 
integrated into DOE's planned program thus far.  
A clear conclusion of this Snowmass working
group is that ORKA presents an extraordinary opportunity.  
If the U.S.\ HEP
program endeavors to achieve a leading role at the Intensity Frontier, 
ORKA should be pursued.

In short, the optimal U.S.\ plan in quark-flavor physics for the remainder of this decade has four elements. 
\begin{itemize}\vspace*{-6pt}
\item U.S.\ physicists should be supported to carry out significant
roles in LHCb and Belle II.
\item The ORKA experiment should move forward in a timely way at Fermilab.
\item Support for U.S.\ participation on other experiments 
that are in progress (e.g., KOTO, TREK, BESIII) should be maintained. 
\item Support for theory, and the computing facilities needed for 
lattice QCD, should be maintained.
\end{itemize}

\subsection{Opportunities in the Next Decade}

In the decade beginning around 2020, we can anticipate that LHCb will be well on
its path toward collecting $50 \, {\rm fb}^{-1}$ and Belle~II will be well on
its path toward  $50 \, {\rm ab}^{-1}$.   These will be complementary
data samples, overlapping in  some areas but providing different strengths in
others.  
We anticipate that the U.S.\ HEP program will be continuing its
emphasis on  Intensity Frontier experiments, with a commitment to providing
high-intensity proton sources for the production of neutrino
beams for neutrino experiments.  If so, the potential for such a high-intensity proton source
to support the next generation of rare kaon decay experiments is an opportunity unique to the
U.S.\ program.  In particular, \ProjectX\ at Fermilab can deliver more than an order of magnitude
increase in the beam power available for producing kaons compared to any other laboratory
in the world.  In addition, the CW-linac of \ProjectX\ can provide a time structure
that is programmable bunch-by-bunch.
That capability can be exploited in neutral kaon experiments to measure the momentum of
individual $K_L^0$'s via time-of-flight, opening the door to dramatic improvements in background rejection
for some challenging rare decays.

\ProjectX\ can be the leading facility in the world
for rare kaon decay experiments.

\subsection{Conclusions}

This report has described the physics case for precision studies of
flavor-changing interactions of bottom, charm, and strange quarks, and it has
described the experimental programs that are underway and foreseeable around the
world.  A substantial number of physicists in the U.S.\ are motivated to work in
this area, both theorists and experimentalists. Quark-flavor physics should be a
component in the plan for the future U.S.\ HEP program.

After enduring the full ``Snowmass process," the Quark Flavor Physics working
group has produced this report. It reflects a wide range of inputs. Its contents
and conclusions have been publicly vetted.   For instance, drafts of this
report were posted  for two rounds of public comment.

Our major conclusions can be summarized as follows:
\begin{itemize}\vspace*{-6pt}
\item Quark flavor physics is an essential element in the international high-energy physics program.
Experiments that study the properties of highly suppressed decays of strange, charm, and bottom
quarks have the potential to observe signatures of new physics at mass scales
well beyond those directly accessible by current or foreseeable accelerators.

\item The importance of quark flavor physics is recognized in Europe and Asia, as demonstrated
by the commitments to LHCb, NA62, KLOE-2, and Panda in Europe, and to Belle~II, BESIII,
KOTO, and TREK in Asia.  

\item In order for the U.S.\ HEP program to have the breadth to assure
meaningful participation in future discoveries, significant U.S.\ contributions
to offshore quark-flavor experiments is important, and continued support
for U.S.\ groups in these efforts is a sound investment.
In particular, 
U.S.\ contributions to LHCb and Belle~II should be
encouraged because of the richness of the physics menus of these experiments
and their reach for new physics.

\item Existing facilities at Fermilab are capable of mounting world-leading rare kaon decay 
experiments in this decade at modest incremental cost to running the Fermilab neutrino program.  
The proposed ORKA experiment, to measure the rare decay $K^+ \to \pi^+ \nu \overline{\nu}$
with high precision, provides such an opportunity.  This is a compelling opportunity that
should be exploited.

\item Longer term, \ProjectX\ 
at Fermilab can become the dominant facility in the world for rare kaon
decay experiments.  Its potential to provide ultra-high intensity kaon beams with tunable time structure 
is unprecedented.  While the physics case for \ProjectX\ is much broader than its capabilities 
for kaon experiments,  the power of a \ProjectX\ kaon program is a strong argument in
its favor.


\item Back-and-forth between theory and experiment 
has always led to unexpected progress in
quark-flavor physics, 
and this is expected to continue.  Therefore, stable support for
theorists working in this area is essential.   Lattice QCD plays a crucial role,
and support for the computing facilities needed for LQCD progress should be
maintained.

\end{itemize}

Quark flavor physics will be the source of future discoveries.  A healthy U.S.\
particle physics program will endeavor to be among the leaders in this research.


\newpage

\lhead[\fancyplain{}{\bf\thepage}]%
      {\fancyplain{}{\bf\bibname}}


\begin{thebibliography}{99}
\addcontentsline{toc}{section}{References}
\itemsep 0pt


\bibitem{Hewett:2012ns}
J.~L.~Hewett {\it et al.},
arXiv:1205.2671 [hep-ex].


\bibitem{Kobayashi:1973fv} 
  M.~Kobayashi and T.~Maskawa,
  Prog.\ Theor.\ Phys.\  {\bf 49}, 652 (1973).

\bibitem{Cabibbo:1963yz} 
  N.~Cabibbo,
  Phys.\ Rev.\ Lett.\  {\bf 10}, 531 (1963).

\bibitem{GIM}
S.L.~Glashow, J.~Iliopoulos, and L.~Maiani, Phys.\ Rev.\ D {\bf 2}, 1285 (1970).

\bibitem{Gaillard-Lee}
M.K.~Gaillard and B.W.~Lee, Phys.\ Rev.\ D {\bf 10}, 897 (1974).

\bibitem{Vainshtein}
A.I.~Vainshtein and I.B.~Khriplovich, 
Pisma Zh.\ Eksp.\ Theor.\ Fiz.\ {\bf 18}, 141 (1973) [JETP Lett.\ {\bf 18}, 83 (1973)].

\bibitem{Isidori:2010kg} 
  G.~Isidori, Y.~Nir and G.~Perez,
  Ann.\ Rev.\ Nucl.\ Part.\ Sci.\  {\bf 60}, 355 (2010)
  [arXiv:1002.0900 [hep-ph]];
and updates in
  G.~Isidori,
  arXiv:1302.0661 [hep-ph].


\bibitem{ckmfitter}
A.~H\"ocker, H.~Lacker, S.~Laplace and F.~Le Diberder,
Eur.\ Phys.\ J.\ C {\bf 21}, 225 (2001) 
[hep-ph/0104062];
and updates at \url{http://ckmfitter.in2p3.fr/}.

\bibitem{Charles:2004jd}
  J.~Charles {\it et al.},
  Eur.\ Phys.\ J.\ C {\bf 41}, 1  (2005) 
  [hep-ph/0406184].

\bibitem{Wolfenstein:1983yz}
  L.~Wolfenstein,
  Phys.\ Rev.\ Lett.\  {\bf 51}, 1945 (1983).

\bibitem{Hocker:2006xb}
  A.~Hocker and Z.~Ligeti,
  Ann.\ Rev.\ Nucl.\ Part.\ Sci.\  {\bf 56}, 501 (2006) 
  [hep-ph/0605217].

\bibitem{Ligeti:2004ak}
  Z.~Ligeti,
  Int.\ J.\ Mod.\ Phys.\  A {\bf 20}, 5105 (2005)
  [hep-ph/0408267].

\bibitem{Bona:2007vi}
  M.~Bona {\it et al.}  [UTfit Collaboration],
  JHEP {\bf 0803}, 049 (2008)
  [arXiv:0707.0636];
and updates at \url{http://utfit.org/}.

\bibitem{Lenz:2012az}
  A.~Lenz {\it et al.},
  Phys.\ Rev.\ D {\bf 86} (2012) 033008
  [arXiv:1203.0238 [hep-ph]].

\bibitem{Grossman:2009dw} 
  Y.~Grossman, Z.~Ligeti and Y.~Nir,
  Prog.\ Theor.\ Phys.\  {\bf 122}, 125 (2009)
  [arXiv:0904.4262 [hep-ph]].



\bibitem{brod-projectx}
J.~Brod, contribution to ``Kaon Physics with Project X'', 
in Ref.~\cite{Kronfeld:2013uoa}.

\bibitem{Kronfeld:2013uoa} 
  A.~S.~Kronfeld
  {\it et al.},
  arXiv:1306.5009 [hep-ex].

\bibitem{sebastian}
S.~J\"ager, talk given at the NA62 Physics Handbook Workshop, 
\href{http://indico.cern.ch/getFile.py/access?contribId=5&resId=0&materialId=slides&confId=65927}{http://indico.cern.ch/
getFile.py/access?contribId=5\&resId=0\&materialId=slides\&confId=65927}

\bibitem{Buras:2000qz} 
  A.~J.~Buras, P.~Gambino, M.~Gorbahn, S.~J\"ager and L.~Silvestrini, 
  Nucl.\ Phys.\ B {\bf 592}, 55 (2001)
  [hep-ph/0007313].

\bibitem{Bauer:2009cf} 
  M.~Bauer, S.~Casagrande, U.~Haisch and M.~Neubert,
  JHEP {\bf 1009}, 017 (2010)
  [arXiv:0912.1625 [hep-ph]].

\bibitem{Blanke:2007wr} 
  M.~Blanke, A.~J.~Buras, S.~Recksiegel, C.~Tarantino and S.~Uhlig,
  JHEP {\bf 0706}, 082 (2007)
  [arXiv:0704.3329 [hep-ph]].

\bibitem{Buras:1998ed} 
  A.~J.~Buras and L.~Silvestrini,
  Nucl.\ Phys.\ B {\bf 546}, 299 (1999)
  [hep-ph/9811471].
  
\bibitem{Buras:1999da} 
  A.~J.~Buras, G.~Colangelo, G.~Isidori, A.~Romanino and L.~Silvestrini,
  Nucl.\ Phys.\ B {\bf 566}, 3 (2000)
  [hep-ph/9908371].

\bibitem{uli-projectx}
U.~Haisch,  contribution to ``Kaon Physics with Project X'', 
in Ref.~\cite{Kronfeld:2013uoa}.

\bibitem{Adler:2008zza}
S.~Adler {\it et al.} (E949 \& E787 Collaborations), 
      Phys. Rev.\ D {\bf 77}:052003 (2008) [arXiv:0709.1000]

\bibitem{Grossman:1997sk} 
  Y.~Grossman and Y.~Nir,
  Phys.\ Lett.\ B {\bf 398}, 163 (1997)
  [hep-ph/9701313].

\bibitem{Blanke:2009pq}
M.~Blanke, 
  Acta Phys. Polon. B {\bf  41}, 127 (2010),  [arXiv:0904.2528]

\bibitem{Mescia:2006jd} 
  F.~Mescia, C.~Smith and S.~Trine,
  JHEP {\bf 0608}, 088 (2006)
  [hep-ph/0606081].

\bibitem{Blanke:2008yr} 
  M.~Blanke, A.~J.~Buras, B.~Duling, K.~Gemmler and S.~Gori,
  JHEP {\bf 0903}, 108 (2009)
  [arXiv:0812.3803 [hep-ph]].

\bibitem{Blanke:2009am}
M.~Blanke, A.~J.~Buras, B.~Duling, S.~Recksiegel, and C.~Tarantino,
Acta Phys. Polon B {\bf 41}, 657 (2010)
[arXiv:0906.5454 [hep-ph]].

\bibitem{Buras:2012jb} 
  A.~J.~Buras, F.~De Fazio and J.~Girrbach,
  JHEP {\bf 1302}, 116 (2013)
  [arXiv:1211.1896 [hep-ph]].

\bibitem{Buras:2004qb} 
  A.~J.~Buras, T.~Ewerth, S.~J\"ager and J.~Rosiek,
  Nucl.\ Phys.\ B {\bf 714}, 103 (2005)
  [hep-ph/0408142].

\bibitem{Kamenik:2011vy} 
  J.~F.~Kamenik and C.~Smith,
  JHEP {\bf 1203}, 090 (2012)
  [arXiv:1111.6402 [hep-ph]].

\bibitem{Cirigliano:2007xi} 
  V.~Cirigliano and I.~Rosell,
  Phys.\ Rev.\ Lett.\  {\bf 99}, 231801 (2007)
  [arXiv:0707.3439 [hep-ph]].

\bibitem{Efrosinin:2000yv} 
  V.~P.~Efrosinin, I.~B.~Khriplovich, G.~G.~Kirilin and Y.~G.~Kudenko,
  Phys.\ Lett.\ B {\bf 493}, 293 (2000)
  [hep-ph/0008199].

\bibitem{NA62}
\url{http://na62.web.cern.ch/na62/Documents/ReferenceDocuments.html}.

\bibitem{KOTO}
\url{http://koto.kek.jp/}.

\bibitem{TREK}
\url{http://trek.kek.jp/}.

\bibitem{KLOE2}
\url{http://www.lnf.infn.it/kloe2/}.

\bibitem{Comfort:2011zz}
J.~Comfort {\it et al.},
FERMILAB-PROPOSAL-1021 (2011).

\bibitem{KOPIO}
KOPIO Experiment Proposal (2005),
\url{http://www.bnl.gov/rsvp/KOPIO.htm}.



\bibitem{LatticeWhitePaper}
Lattice QCD at the intensity frontier, 
T.~Blum {\it et al.}\ [USQCD Collaboration], available at 
\url{http://www.usqcd.org/documents/13flavor.pdf}.

\bibitem{Zupan:2011mn}
J.~Zupan,
arXiv:1101.0134 [hep-ph].

\bibitem{LHCb:2013oba} 
R.~Aaij {\it et al.}  [LHCb Collaboration],
arXiv:1304.2600 [hep-ex].

\bibitem{Aaij:2012nna}
R.~Aaij {\it et al.} [LHCb Collaboration],
Phys.\ Rev.\ Lett.\ {\bf 110} (2013) 021801
[arXiv:1211.2674 [hep-ex]].

\bibitem{Buras:2012ru}
A.~J.~Buras, J.~Girrbach, D.~Guadagnoli and G.~Isidori,
Eur.\ Phys.\ J.\ C {\bf 72} (2012) 2172
[arXiv:1208.0934 [hep-ph]].

\bibitem{DeBruyn:2012wk}
K.~De~Bruyn {\it et al.}, Phys.\ Rev.\ Lett.\ {\bf 109}, 041801 (2012) [arXiv:1204.1737].

\bibitem{Chatrchyan:2013bka}
S.~Chatrchyan {\it et al.}  [CMS Collaboration], arXiv:1307.5025 [hep-ex], submitted to Phys.\ Rev.\ Lett.

\bibitem{Aaij:2013aka}
A.~Aaij {\it et al.}  [LHCb Collaboration], arXiv:1307.5024 [hep-ex], submitted to Phys.\ Rev.\ Lett.

\bibitem{Abazov:2011yk}
V.~M.~Abazov {\it et al.} [D0 Collaboration],
Phys.\ Rev.\ D {\bf 84} (2011) 052007
[arXiv:1106.6308 [hep-ex]].

\bibitem{Laplace:2002ik}
  S.~Laplace, Z.~Ligeti, Y.~Nir and G.~Perez,
  Phys.\ Rev.\ D {\bf 65} (2002) 094040
  [hep-ph/0202010].

\bibitem{Xing:2012mt} 
Z.~Xing {\it et al.} [LHCb Collaboration],
arXiv:1212.1175 [hep-ex].

\bibitem{Lees:2013uzd}
J.~P.~Lees {\it et al.} [BaBar Collaboration],
arXiv:1303.0571 [hep-ex].

\bibitem{Fajfer:2012jt} 
  S.~Fajfer, J.~F.~Kamenik, I.~Nisandzic and J.~Zupan,
Phys.\ Rev.\ Lett.\  {\bf 109}, 161801 (2012)
  [arXiv:1206.1872 [hep-ph]].

\bibitem{Bodwin:2013nua}
   G.~T.~Bodwin, E.~Braaten, E.~Eichten, S.~L.~Olsen, T.~K.~Pedlar and
J.~Russ,
   arXiv:1307.7425 [hep-ph].

\bibitem{Aushev:2010bq} 
T.~Aushev {\it et al.} [Belle II Collaboration],
arXiv:1002.5012 [hep-ex].

\bibitem{Bona:2007qt}
M.~Bona {\it et al.} [SuperB Collaboration],
[arXiv:0709.0451 [hep-ex]].

\bibitem{Adachi:2012et}
I.~Adachi {\it et al.} [Belle Collaboration],
Phys.\ Rev.\ Lett.\ {\bf 108} (2012) 171802
[arXiv:1201.4643 [hep-ex]].

\bibitem{Aubert:2009aw}
B.~Aubert {\it et al.} [BaBar Collaboration],
Phys.\ Rev.\ D {\bf 79} (2009) 072009
[arXiv:0902.1708 [hep-ph]].

\bibitem{Ishino:2006if}
H.~Ishino {\it et al.} [Belle Collaboration],
Phys.\ Rev.\ Lett.\ {\bf 98} (2007) 211801
[arXiv:hep-ex/0608035].

\bibitem{Lees:2012kx}
J.~P.~Lees {\it et al.} [BaBar Collaboration],
Phys.\ Rev.\ D {\bf 87} (2012) 052009
[arXiv:1206.3525 [hep-ph]].

\bibitem{Beringer:1900zz} 
  J.~Beringer {\it et al.}  [Particle Data Group Collaboration],
  Phys.\ Rev.\ D {\bf 86}, 010001 (2012).

\bibitem{Amhis:2012bh}
Y.~Amhis {\it et al.} [Heavy Flavor Averaging Group],
[arXiv:1207.1158 [hep-ex]], and updates at
\url{http://www.slac.stanford.edu/xorg/hfag/}.

\bibitem{Gronau:2005kz}
M.~Gronau, 
Phys.\ Lett.\ B {\bf 627} (2005) 82
[arXiv:hep-ph/0508047].

\bibitem{Fujikawa:2008pk}
M.~Fujikawa {\it et al.} [Belle Collaboration],
Phys.\ Rev.\ D {\bf 81} (2010) 011101
[arXiv:0809.4366 [hep-ex]].

\bibitem{Charles:2011va}
J.~Charles {\it et al.} [CKM Fitter Group],
Phys.\ Rev.\ D {\bf 84} (2011) 033005
[arXiv:1106.4041 [hep-ph]].

\bibitem{Aubert:2009wt}
B.~Aubert {\it et al.} [BaBar Collaboration],
Phys.\ Rev.\ D {\bf 81} (2010) 051101
[arXiv:0912.2453 [hep-ex]].

\bibitem{Hara:2010dk}
K.~Hara {\it et al.} [Belle Collaboration],
Phys.\ Rev.\ D {\bf 82} (2010) 071101
[arXiv:1006.4201 [hep-ex]].

\bibitem{Lees:2012ju}
J.~P.~Lees {\it et al.} [BaBar Collaboration],
[arXiv:1207.0698 [hep-ex]].

\bibitem{Adachi:2012mm}
I.~Adachi {\it et al.} [Belle Collaboration],
Phys.\ Rev.\ Lett.\ {\bf 110} (2013) 131801
[arXiv:1208.4678 [hep-ex]].

\bibitem{cdf_b_results}
CDF $B$-physics results may be found at 
\url{http://www-cdf.fnal.gov/physics/new/bottom/bottom.html}.

\bibitem{d0_b_results}
D\O\ $B$-physics results may be found at 
\url{http://www-d0.fnal.gov/Run2Physics/WWW/results/b.htm}.

\bibitem{Alves:2008zz} 
  A.~A.~Alves, Jr. {\it et al.}  [LHCb Collaboration],
  JINST {\bf 3}, S08005 (2008).

\bibitem{Aaij:2012me} 
  R.~Aaij {\it et al.},
  JINST {\bf 8}, P04022 (2013)
  [arXiv:1211.3055 [hep-ex]].

\bibitem{Bediaga:2012py} 
  R.~Aaij {\it et al.}  [LHCb Collaboration],
  LHCb-PUB-2012-006,
  Eur.\ Phys.\ J.\ C {\bf 73}, 2373 (2013)
  [arXiv:1208.3355 [hep-ex]];
  See also a summary in LHCb-PUB-2012-009.

\bibitem{lhcb_upgrade_tdr}
 I.~Bediaga {\it et al.} [LHCb Collaboration]
 CERN-LHCC-2012-007 ; LHCb-TDR-12.

\bibitem{Stone:2008ak} 
  S.~Stone and L.~Zhang,
  Phys.\ Rev.\ D {\bf 79}, 074024 (2009)
  [arXiv:0812.2832 [hep-ph]].

\bibitem{LHCb:2012ae} 
  R.~Aaij {\it et al.}  [LHCb Collaboration],
  Phys.\ Rev.\ D {\bf 86}, 052006 (2012)
  [arXiv:1204.5643 [hep-ex]].

\bibitem{Chatrchyan:2012rga} 
  S.~Chatrchyan {\it et al.}  [CMS Collaboration],
  JHEP {\bf 1204}, 033 (2012)
  [arXiv:1203.3976 [hep-ex]].

\bibitem{Aad:2012pn} 
  G.~Aad {\it et al.}  [ATLAS Collaboration],
  Phys.\ Lett.\ B {\bf 713}, 387 (2012)
  [arXiv:1204.0735 [hep-ex]].

\bibitem{Straub:2012jb} 
  D.~M.~Straub,
  arXiv:1205.6094 [hep-ph].

\bibitem{Aaij:2013iag} 
  R.~Aaij {\it et al.}  [LHCb Collaboration],
  arXiv:1304.6325 [hep-ex].

\bibitem{CMSKstarmumu}
[CMS Collaboration],
CMS-PAS-BPH-11-009, 2013.

\bibitem{ATLASKstarmumu}
[ATLAS Collaboration],
ATLAS-CONF-2013-038, 2013.

\bibitem{Aaij:2013qha} 
  R.~Aaij {\it et al.}  [LHCb Collaboration],
  Phys.\  Rev.\  Lett.\  110, {\bf 241802} (2013)
  [arXiv:1303.7125 [hep-ex]].

\bibitem{Aaij:2012vn} 
  R.~Aaij {\it et al.}  [LHCb Collaboration],
  JHEP {\bf 1206}, 058 (2012)
  [arXiv:1204.1620 [hep-ex]].

\bibitem{Aaij:2012dz} 
  R.~Aaij {\it et al.}  [LHCb Collaboration],
  JHEP {\bf 1206}, 141 (2012)
  [arXiv:1205.0975 [hep-ex]].

\bibitem{Aaij:2012pz}
 R.~Aaij {\it et al.}  [LHCb Collaboration],
 Phys.\ Rev.\ D {\bf 85}, 091103 (2012)
  [arXiv:1202.5087 [hep-ex]].

\bibitem{Aaij:2013zoa}
  R.~Aaij {\it et al.}  [LHCb Collaboration],
  Phys.\  Rev.\  Lett.\  110, {\bf 222001} (2013)
  [arXiv:1302.6269 [hep-ex]].

\bibitem{Aaij:2012zr} 
  R.~Aaij {\it et al.}  [LHCb Collaboration],
  Phys.\ Rev.\ D {\bf 85}, 112004 (2012)
  [arXiv:1201.5600 [hep-ex]].

\bibitem{Aaij:2013fia} 
  R.~Aaij {\it et al.}  [LHCb Collaboration],
  Phys.\ Lett.\ B {\bf 724} (2013)
  [arXiv:1304.4518 [hep-ex]].

\bibitem{Heijne:2012qia} 
  V.~A.~M.~Heijne {\it et al.} [LHCb Collaboration], LHCb-CONF-2012-014,
  Frascati Phys.\ Ser.\  {\bf 56}, 162 (2012).


\bibitem{Asner:2008nq} 
  D.~M.~Asner, T.~Barnes, J.~M.~Bian, I.~I.~Bigi, N.~Brambilla, I.~R.~Boyko, V.~Bytev and K.~T.~Chao {\it et al.},
  Int.\ J.\ Mod.\ Phys.\ A {\bf 24}, S1 (2009)
  [arXiv:0809.1869 [hep-ex]].

\bibitem{Gersabeck:2012hz} 
  M.~Gersabeck {\it et al.} [LHCb Collaboration],
  arXiv:1209.5878 [hep-ex].

\bibitem{panda_fair}
U. Wiedner, 
Prog.\ Part.\ Nucl.\ Phys.\ {\bf 66}, 477 (2011).


\bibitem{Eisenstein:2008aa} 
  B.~I.~Eisenstein {\it et al.}  [CLEO Collaboration],
  Phys.\ Rev.\ D {\bf 78}, 052003 (2008)
  [arXiv:0806.2112 [hep-ex]].

\bibitem{Rong:2012pb} 
  G.~Rong,
  arXiv:1209.0085 [hep-ex].

\bibitem{Zupanc:2012cd} 
  A.~Zupanc {\it et al.} [Belle Collaboration],
  arXiv:1212.3942 [hep-ex].

\bibitem{Kronfeld:2009cf} 
  A.~S.~Kronfeld,
  arXiv:0912.0543 [hep-ph].

\bibitem{Aaij:2012nva} 
 R.~Aaij  {\it et al.}  [LHCb Collaboration],
  Phys.\ Rev.\ Lett.\  {\bf 110}, 101802 (2013)
  [arXiv:1211.1230 [hep-ex]].

\bibitem{Falk:2001hx} 
  A.~F.~Falk, Y.~Grossman, Z.~Ligeti and A.~A.~Petrov,
  Phys.\ Rev.\ D {\bf 65}, 054034 (2002)
  [hep-ph/0110317];
  A.~F.~Falk, Y.~Grossman, Z.~Ligeti, Y.~Nir and A.~A.~Petrov,
  Phys.\ Rev.\ D {\bf 69}, 114021 (2004)
  [hep-ph/0402204].

\bibitem{Golowich:2007ka}
  E.~Golowich, J.~Hewett, S.~Pakvasa and A.~A.~Petrov,
  Phys.\ Rev.\  D {\bf 76}, 095009 (2007)
  [arXiv:0705.3650 [hep-ph]].

\bibitem{Gedalia:2009kh}
  O.~Gedalia, Y.~Grossman, Y.~Nir and G.~Perez,
  Phys.\ Rev.\  D {\bf 80}, 055024 (2009)
  [arXiv:0906.1879 [hep-ph]].
   M.~Ciuchini et al.,
  Phys.\ Lett.\  B {\bf 655}, 162 (2007).
  [arXiv:hep-ph/0703204].
  
\bibitem{Artuso:2008vf}
  M.~Artuso, B.~Meadows and A.~A.~Petrov,
  Ann.\ Rev.\ Nucl.\ Part.\ Sci.\  {\bf 58}, 249 (2008);
  A.~Ryd and A.~A.~Petrov,
  Rev.\ Mod.\ Phys.\  {\bf 84}, 65 (2012);
  S.~Bianco, F.~L.~Fabbri, D.~Benson and I.~Bigi,
  Riv.\ Nuovo Cim.\  {\bf 26N7}, 1 (2003);
  [arXiv:hep-ex/0309021].
 G.~Burdman and I.~Shipsey,
  Ann.\ Rev.\ Nucl.\ Part.\ Sci.\  {\bf 53}, 431 (2003);
    X.~Q.~Li, X.~Liu and Z.~T.~Wei,
  Front.\ Phys.\ China {\bf 4}, 49 (2009).

\bibitem{Bigi:2009jj} 
  I.~I.~Bigi,
  arXiv:0902.3048 [hep-ph].

\bibitem{DirectCPexp}
  R.~Aaij {\it et al.}  [LHCb Collaboration],
  Phys.\ Rev.\ Lett.\  {\bf 108}, 111602 (2012);
  T.~Aaltonen {\it et al.}  [CDF Collaboration],
  Phys.\ Rev.\ Lett.\  {\bf 109}, 111801 (2012);
  New measurements do not confirm those results: 
  R.~Aaij {\it et al.}  [LHCb Collaboration],
  Phys.\ Lett.\ B {\bf 723}, 33 (2013);

\bibitem{DirectCPtheory} 
  M.~Golden and B.~Grinstein,
  Phys.\ Lett.\ B {\bf 222}, 501 (1989);
  J.~Brod, A.~L.~Kagan and J.~Zupan,
  Phys.\ Rev.\ D {\bf 86}, 014023 (2012);
  B.~Bhattacharya, M.~Gronau and J.~L.~Rosner,
  Phys.\ Rev.\ D {\bf 85}, 054014 (2012);
  I.~I.~Bigi and A.~Paul,
  JHEP {\bf 1203}, 021 (2012);
 G.~Isidori, J.~F.~Kamenik, Z.~Ligeti and G.~Perez,
  Phys.\ Lett.\ B {\bf 711}, 46 (2012);
   J.~Brod, Y.~Grossman, A.~L.~Kagan and J.~Zupan,
  JHEP {\bf 1210}, 161 (2012);
  W.~Altmannshofer, R.~Primulando, C.~-T.~Yu and F.~Yu,
  JHEP {\bf 1204}, 049 (2012);
 Y.~Grossman, A.~L.~Kagan and J.~Zupan,
  Phys.\ Rev.\ D {\bf 85}, 114036 (2012);
  H.~-Y.~Cheng and C.~-W.~Chiang,
  Phys.\ Rev.\ D {\bf 85}, 034036 (2012)
  [Erratum-ibid.\ D {\bf 85}, 079903 (2012)]; 
  G.~Hiller, Y.~Hochberg and Y.~Nir,
  Phys.\ Rev.\ D {\bf 85}, 116008 (2012);
  T.~Feldmann, S.~Nandi and A.~Soni,
  JHEP {\bf 1206}, 007 (2012).

\bibitem{Golowich:2009ii}
  E.~Golowich, J.~Hewett, S.~Pakvasa and A.~A.~Petrov,
  Phys.\ Rev.\  D {\bf 79}, 114030 (2009).

\bibitem{Atwood:2003mj} 
  D.~Atwood and A.~Soni,
  Phys.\ Rev.\ D {\bf 68}, 033003 (2003)
  [hep-ph/0304085].



\bibitem{Pakhlova:2008vn} 
  G.~Pakhlova {\it et al.}  [Belle Collaboration],
  Phys.\ Rev.\ Lett.\  {\bf 101}, 172001 (2008)
  [arXiv:0807.4458 [hep-ex]].



\bibitem{whitepaper13}
T.~Blum {\em et al.} [USQCD Collaboration],
{\em Lattice QCD at the Intensity Frontier},
\url{http://www.usqcd.org/documents/13flavor.pdf} (2013).

\bibitem{whitepaper07}
R.~Brower {\em et al.} [USQCD Collaboration],
{\em Fundamental parameters from future lattice  calculations},
\url{http://www.usqcd.org/documents/fundamental.pdf} (2007).

\bibitem{Laiho:2009eu} 
  J.~Laiho, E.~Lunghi and R.~S.~Van de Water,
  Phys.\ Rev.\ D {\bf 81}, 034503 (2010)
  [arXiv:0910.2928 [hep-ph]].

\bibitem{Colangelo:2010et}
  G.~Colangelo {\it et al.} [FLAG],
  Eur.\ Phys.\ J.\ C {\bf 71}, 1695 (2011)
  [arXiv:1011.4408 [hep-lat]].

\bibitem{Blum:2011ng} 
  T.~Blum {\it et al.} [RBC and UKQCD Collaborations],
  Phys.\ Rev.\ Lett.\  {\bf 108}, 141601 (2012)
  [arXiv:1111.1699 [hep-lat]].

\bibitem{Blum:2012uk} 
  T.~Blum {\it et al.} [RBC and UKQCD Collaborations],
  Phys.\ Rev.\ D {\bf 86}, 074513 (2012)
  [arXiv:1206.5142 [hep-lat]].

\bibitem{Boyle:2012ys}
  P.~A.~Boyle {\it et al.}  [RBC and UKQCD Collaborations],
  arXiv:1212.1474 [hep-lat].

\bibitem{Christ:2010dd} 
  N.~H.~Christ {\it et al.} [RBC and UKQCD Collaborations],
  Phys.\ Rev.\ Lett.\  {\bf 105}, 241601 (2010)
  [arXiv:1002.2999 [hep-lat]].

\bibitem{Junnarkar:2013ac} 
  P.~Junnarkar and A.~Walker-Loud,
  arXiv:1301.1114 [hep-lat].

\bibitem{Christ:2012se} 
  N.~H.~Christ, T.~Izubuchi, C.~T.~Sachrajda, A.~Soni and J.~Yu [RBC and UKQCD Collaborations],
  arXiv:1212.5931 [hep-lat].

\bibitem{Lellouch:2000pv} 
  L.~Lellouch and M.~Luscher,
  Commun.\ Math.\ Phys.\  {\bf 219}, 31 (2001)
  [hep-lat/0003023].

\bibitem{Blum:2011pu} 
  T.~Blum {\it et al.} [RBC and UKQCD Collaborations],
  Phys.\ Rev.\ D {\bf 84}, 114503 (2011)
  [arXiv:1106.2714 [hep-lat]].

\bibitem{Kelly:2012eh} 
  C.~Kelly {\it et al.} [RBC and UKQCD Collaborations],
  PoS LATTICE {\bf 2012}, 130 (2012).

\bibitem{Brod:2010mj} 
  J.~Brod and M.~Gorbahn,
  Phys.\ Rev.\ D {\bf 82}, 094026 (2010)
  [arXiv:1007.0684 [hep-ph]].

\bibitem{Cirigliano:2011ny} 
  V.~Cirigliano, G.~Ecker, H.~Neufeld, A.~Pich and J.~Portoles,
  Rev.\ Mod.\ Phys.\  {\bf 84}, 399 (2012)
  [arXiv:1107.6001 [hep-ph]].

\bibitem{Hansen:2012tf} 
  M.~T.~Hansen and S.~R.~Sharpe,
  Phys.\ Rev.\ D {\bf 86}, 016007 (2012)
  [arXiv:1204.0826 [hep-lat]].

\bibitem{Qiu:2011ur}
  S.~-W.~Qiu {\it et al.}  [Fermilab Lattice and MILC Collaborations],
  PoS LATTICE {\bf 2011} (2011) 289
  [arXiv:1111.0677 [hep-lat]].

\bibitem{Lees:2012xj} 
  J.~P.~Lees {\it et al.}  [BaBar Collaboration],
  Phys.\ Rev.\ Lett.\  {\bf 109}, 101802 (2012)
  [arXiv:1205.5442 [hep-ex]].

\bibitem{Bailey:2012jg} 
  J.~A.~Bailey {\it et al.} [Fermilab Lattice and MILC Collaborations],
  Phys.\ Rev.\ Lett.\  {\bf 109}, 071802 (2012)
  [arXiv:1206.4992 [hep-ph]].

\bibitem{McNeile:2011ng} 
  C.~McNeile, C.~T.~H.~Davies, E.~Follana, K.~Hornbostel and G.~P.~Lepage [HPQCD Collaboration],
  Phys.\ Rev.\ D {\bf 85}, 031503 (2012)
  [arXiv:1110.4510 [hep-lat]].

\bibitem{Zhou:2013uu} 
  R.~Zhou,
  arXiv:1301.0666 [hep-lat].

\bibitem{Detmold:2012vy} 
  W.~Detmold, C.~-J.~D.~Lin, S.~Meinel and M.~Wingate,
  Phys.\ Rev.\ D {\bf 87}, 074502 (2013)
  [arXiv:1212.4827 [hep-lat]].

\bibitem{Baum:2011rm} 
  I.~Baum, V.~Lubicz, G.~Martinelli, L.~Orifici and S.~Simula,
  Phys.\ Rev.\ D {\bf 84}, 074503 (2011)
  [arXiv:1108.1021 [hep-lat]].

\end{thebibliography}
\end{document}